\documentclass[prd, onecolumn, nofootinbib, floatfix]{revtex4}

\usepackage{epsfig} 
\usepackage{amsmath}
\usepackage{amsbsy,color}

\newcommand{\beq}{\begin{equation}}
\newcommand{\eeq}{\end{equation}}
\newcommand{\beqa}{\begin{eqnarray}}
\newcommand{\eeqa}{\end{eqnarray}}

\newcommand{\ls}{\lesssim}

\def\ha{\frac{1}{2}}
\def\imag{\dot \imath}

\def \sn2{\left(S/N\right)^2}

\def \dk{\delta_{\bf k}}
\def \dkp{\delta_{\bf k'}}
\def \k{{\bf k}}
\def \kp{{\bf k'}}
\def \kpp{{\bf k''}}

\begin{document} 

\title{Thinking Outside the Box:\\
Effects of Modes Larger than the Survey on
Matter Power Spectrum Covariance
}
\author{Roland de Putter$^{1,2}$, Christian Wagner$^1$, Olga Mena$^2$, Licia Verde$^1$, Will J.~Percival$^3$
\vspace{0.1cm}
}

\affiliation{$^1$ICC, University of Barcelona (IEEC-UB), Marti i Franques 1, Barcelona 08028, Spain\\
$^2$Instituto de Fisica Corpuscular, Universidad de Valencia-CSIC, Spain\\
$^3$Institute of Cosmology \& Gravitation, University of Portsmouth, Dennis Sciama Bldg., Portsmouth, PO1 3FX, UK} 
\date{\today}

\begin{abstract}
Accurate power spectrum (or correlation function) covariance matrices
are a crucial requirement for cosmological parameter estimation
from large scale structure surveys. In order to minimize reliance on
computationally expensive mock catalogs,
it is important to have a solid analytic understanding of the different
components that make up a covariance matrix.
Considering the matter power spectrum covariance matrix,
it has recently been found
that there is a potentially dominant effect on mildly non-linear scales
due to power in modes of size equal to and larger than the survey volume.
This {\it beat coupling} effect has been derived analytically in perturbation
theory and while it has been tested with simulations, some questions remain unanswered.
Moreover, there is an additional effect of
these large modes, which has so far not been included in
analytic studies, namely the effect on the estimated {\it average}
density which enters the power spectrum estimate. In this article,
we work out analytic, perturbation theory based expressions
including both the beat coupling and this {\it local average effect}
and we show that while, when isolated, beat coupling indeed causes large excess covariance
in agreement with the literature,
in a realistic scenario
this is compensated almost entirely by the local average effect, leaving only $\sim 10 \%$
of the excess.
We test our analytic expressions by comparison to a suite of large N-body simulations,
using both full simulation boxes and subboxes thereof to study
cases without beat coupling, with beat coupling and with both beat coupling {\it and}
the local average effect.
For the variances, we find excellent agreement with the analytic expressions
for $k < 0.2 h$Mpc$^{-1}$ at $z=0.5$, while the correlation coefficients
agree to beyond $k=0.4 h$Mpc$^{-1}$.
As expected, the range of agreement increases towards higher redshift and decreases slightly
towards $z=0$.
We finish by including the large-mode effects in a full covariance matrix
description for arbitrary survey geometry and confirming its validity using simulations.
This may be useful as a stepping stone towards building an actual galaxy (or other tracer's) power spectrum
covariance matrix.

\end{abstract} 

\maketitle

\section{Introduction \label{sec:intro}}

Galaxy surveys (and surveys of other dark matter tracers) are an
important tool for constraining cosmological
parameters and the tracer's power spectrum (or its Fourier transform, the correlation function) is the most valuable observable.
It can be used to measure the scale of baryon acoustic oscillations
(BAO, see e.g.~\cite{seoeis03,eisetal03,coleetal05,blakeetal11}), which is a particularly robust feature
and a strong probe of dark energy through the measurement's dependence on the expansion history,
but it is also common to use the full shape of the power spectrum to constrain cosmology (e.g.~\cite{efstetal02,reidetal10}).
In order for these measurements to be useful for cosmology however, it is crucial
to have an accurate estimate of the observable's covariance matrix.
As surveys get larger, calculating such a matrix directly from simulations
becomes ever more challenging.
In fact, large numbers of mock catalogs are required to estimate
covariance matrices directly from simulations and it is computationally very
costly to create such numbers with the required volume and resolution.
Moreover, the covariance matrix is cosmology dependent so ideally one would
either be able to generate a large number of matrices for different cosmologies, or
show that this cosmology dependence can be safely neglected.
This requirement increases the number of mocks needed
in the brute force method even further.
It is therefore important to develop a good analytical
understanding in order to eventually rely less on simulations.

While the density of tracers may have a non-trivial relation with the dark matter density,
a good first step towards understanding the true covariance matrix is to study that of the
dark matter power spectrum. The two main complications in understanding the dark matter matrix
are the mode mixing due to the finite survey volume and the non-Gaussian nature of the matter overdensity
field on small scales ($k \gtrsim 0.1 h$Mpc$^{-1}$) due to non-linear evolution.
While the former is relatively straightforward to quantify (\cite{FKP}),
the latter effects are the subject of a large number of
studies~\cite{scoccetal99,meikwhite99,coorayhu01,pthal,sirko05,sefusattietal06,neyrincketal06,neyszap07,smith08,satoetal09,neyrinck11,harpen11,nganetal11}, involving N-body simulations, perturbation theory and the halo model.
For covariances on non-linear scales, an important (but often overlooked)
role is played by modes of wavelength comparable to and larger than the survey size.
In particular, \cite{HRS06,rimesham06} found that the interplay of mode mixing by the survey window
with non-linear correlations between a pair of small-scale modes and a large mode,
causes a covariance contribution proportional to the power in super-survey modes,
and that, in fact, this is the dominant contribution to the covariance matrix
on small scales. The added covariance affects both the variances
and the off-diagonal covariances.
This effect is usually called {\it beat coupling} (see also \cite{sefusattietal06, takjain08, reidsperbode09,takahashietal09}).
A second consequence of the presence of modes larger than the survey is that
the average density (which is needed to construct the overdensity)
estimated from the survey has a variance itself, leading to an additional
covariance contribution proportional to the large scale power.
We will refer to this as the {\it local average} effect.

In the standard approach to estimating covariances
from N-body simulations, the power spectrum is calculated using the full,
periodic simulation volume. However, this completely misses the large effects
discussed above because there are no modes larger than the ``survey'' volume.
In order to capture these effects, one thus needs to consider either a subvolume
of a (much) larger simulation, or implement a varying zero mode (\cite{sirko05}) in the simulations.
While some such studies have been performed (see e.g.~\cite{takahashietal09, reidsperbode09}),
a systematic study, leading to a complete and well tested analytic description
has until now been lacking.

In this article, we aim to present such a study. We use a set of large ($L = 2400 h^{-1}$Mpc)
N-body simulations and study the aforementioned effects of large modes by considering
significantly smaller subboxes ($L = 600 h^{-1}$Mpc), thus imitating the real world scenario of a finite
survey embedded in an infinite universe. We build an analytic model for the covariance matrix,
properly including the effect of the window function and describing non-linear effects using perturbation
theory. This description
includes the beat coupling model from \cite{HRS06}, but adds to this an analytic estimate
of the local average effect. The latter effect needs to be included when building a
covariance matrix for a realistically estimated power spectrum and has to our knowledge not been
calculated analytically previously.

We compare our model to simulations at five redshifts in the range $z=0-2$
and generally find excellent agreement
until well into the non-linear regime.
Since we test several combinations of covariance matrix contributions
individually by employing different approaches to estimating spectra from simulations,
we are confident that our theoretical understanding of the dark matter
covariance is correct and that we have captured the main effects.

One of our main results is that, while beat coupling indeed
causes excess covariance which soon dominates, the contribution
from the local average effect cancels most of this out, leaving only
a reduced beat coupling contribution of about $10 \%$ of the original.

In addition to the effects on non-linear scales discussed above,
the mode mixing due to the window function also has a well known effect on linear scales,
correlating neighboring power spectrum estimators and reducing the variances.
This effect needs to be taken into account for a realistic survey.
We calculate these covariances directly using analytic expressions from \cite{FKP}
and show good agreement with simulations. Thus, while the main goal of this paper is to study
the beat coupling and local average effects related to super-survey modes, we also provide
a complete expression that can be used to calculate a (dark matter) covariance matrix
for an arbitrary survey geometry and can serve as a stepping stone for calculating a galaxy
(or other dark matter tracer) spectrum covariance matrix.
We also study the cosmology dependence of the covariance matrix and find that for cosmological parameter variations
relevant to current cosmological constraints, the covariance matrix undergoes changes at the $\sim 30 \%$ level.

The outline of this article is as follows. We explain the analytic description
of the covariance matrix in section \ref{sec:form}, using a simple formalism apt for the description
of simulation results. This will lead to three
expressions that can be tested against simulations (without beat coupling, with beat coupling,
and with beat coupling {\it and} local average effect). We then describe our N-body simulations
in section \ref{sec:nbody} and compare the results to theory in section \ref{sec:results}.
While for the main results in this paper, we stick to a simple description
in terms of a discrete set of modes,
we discuss a more rigorous and complete treatment of the window function
in section \ref{sec:gen}, leading to expressions that can be applied to
arbitrary survey geometry.
We conclude and summarize our work in
section \ref{sec:disc}. Finally, we use our analytic expressions
to briefly discuss the cosmology dependence of the covariance matrix in the Appendix.

As this article is rather long and technical, we provide some recommendations for a reader on a tight schedule.
On a first reading, this reader may want to skip the mathematical discussion in section \ref{sec:form} and jump straight to
section \ref{sec:outlook}, where the main equations are summarized. Moreover, in section \ref{sec:gen},
one could jump to Eq.~(\ref{eq:c3 geom}), which presents the final covariance matrix expression for
general geometry, and then focus on the results in Figs \ref{fig:varszp}-\ref{fig:corrik14zp},
where the green points and lines show the covariance matrix corresponding to said equation.

\section{Covariance Matrix Formalism}
\label{sec:form}

In this section, we work out the formalism for calculating a covariance matrix
for a power spectrum measured from a cubic volume with uniform selection function,
i.e.~the geometry of N-body simulations. We describe the density field in terms
of a discrete set of modes, which significantly simplifies notation. We
will provide a more general description, which can be applied to
an arbitrary survey geometry (following \cite{FKP}), in section \ref{sec:gen}. The more technical discussion there
also serves to motivate the simpler approach taken in this section.

\subsection{N-body Simulation Geometry (Cubic Box)}

We consider a cubic subvolume $V = L^3$ of the universe (which later in this article
will be modeled by taking a subvolume of a much larger periodic simulation box).
The matter overdensity field $\delta({\bf x}) \equiv (\rho({\bf x}) - \bar{\rho})/\bar{\rho}$,
where $\rho({\bf x})$ is the matter density and $\bar{\rho}$ its mean, can be expanded as
\beq
\delta({\bf x}) = \int \frac{d^3 {\bf k}}{(2\pi)^3}\, e^{- \imag {\bf k}\cdot {\bf x}}\,\delta({\bf k}).
\eeq
To significantly simplify notation, we choose to describe the density field in the volume $V$
in terms of discrete Fourier modes in analogy to the usual description of the density field
in a periodic box, i.e. we effectively describe the field in the subvolume by
\beq
\label{eq:periodicFT}
\delta({\bf x}) \sim \sum_{\bf k} e^{- \imag {\bf k}\cdot {\bf x}}\,\delta_{\bf k},
\quad k_i = \frac{2\pi}{L}\,n_i \quad {\rm for} \quad n_i = 0, \pm 1,\dots.
\eeq
To be more exact, we define the discrete modes as
\beq
\delta_{\bf k} \equiv \frac{1}{V}\, \int_V d^3 {\bf x} \, e^{\imag {\bf k}\cdot {\bf x}}\,\delta({\bf x}).
\eeq
In other words, $\delta_{\bf k}$ is equal to the FKP estimator (\cite{FKP}) $F({\bf k})$ (up to a factor $V^{-\ha}$)
applied to the volume $V$, and is therefore a weighted average over a range of continuum Fourier modes. In particular,
while there is no true ${\bf k} = {\bf 0}$ mode, the subvolume sees an effective zero mode (see also \cite{sirko05})
$\delta_{\bf 0} = V^{-1}\, \int_V d^3 {\bf x} \,\delta({\bf x})$ which gets its main contribution
from continuum Fourier modes $|{\bf k}| < 2\pi/L$. Note that Eq.~(\ref{eq:periodicFT}) is not to be taken
literally, as the field is not actually periodic with respect to the volume $V$. We discuss the effect of taking
a subvolume more rigorously in section \ref{sec:gen}.

The statistics of the discrete overdensity modes can be characterized by a sequence of connected $n-$point functions,
\beqa
\langle \dk \, \dkp \rangle &=& \frac{P_k}{V} \, \delta^{\rm K}_{{\bf k} + {\bf k'}} \nonumber \\
\langle \dk \, \dkp \, \delta_{\bf k''} \rangle_c &=& \frac{B_{{\bf k}, {\bf k'}, {\bf k''}}}{V^2} \, \delta^{\rm K}_{{\bf k} + {\bf k'} + {\bf k''}} \nonumber \\
\langle \dk \, \dkp \, \delta_{\bf k''} \, \delta_{\bf k'''}\rangle_c &=& \frac{T_{{\bf k}, {\bf k'}, {\bf k''}, {\bf k'''}}}{V^3} \, \delta^{\rm K}_{{\bf k} + {\bf k'} + {\bf k''} + {\bf k'''}} \nonumber \\
&\dots& \nonumber,
\eeqa
where we have defined the power spectrum $P_k$, bispectrum $B_{{\bf k}, {\bf k'}, {\bf k''}}$
an the trispectrum $T_{{\bf k}, {\bf k'}, {\bf k''}, {\bf k''}}$.
Note that these are technically
weighted averages of the true, continuum power, bi- and trispectrum, for example (see also section \ref{sec:gen})
\beq
P_{0} \equiv V \, \langle |\delta_{\bf 0}|^2 \rangle = V \, \int \frac{d^3 {\bf k}}{(2\pi)^3} \, P(k) \, \prod_{i=x,y,x} \, j^2_0(L k_i/2),
\eeq
where $j_0(x) = \sin x/x$ is the zeroth spherical Bessel function.

The power of an individual mode can be estimated as
\beq
\hat{P}_{\bf k} \equiv V \, |\delta_{\bf k}|^2,
\eeq
such that $\langle \hat{P}_{\bf k} \rangle = P_k$ as desired, from which one can define a bin averaged estimator in order
to maximize signal to noise,
\beq
\hat{P}_{i} \equiv \frac{1}{N_i}\, \sum_{{\bf k} \in i} \hat{P}_{\bf k}.
\eeq
Here, the sum is over all ${\bf k} = \frac{2\pi}{L}\, {\bf n}$,
with integer components $n_i$, such that $k$ lies in some small range defining the $i$-th bin. $N_i$ is
the number of modes in $i$. The expectation value
$P_i \equiv \langle \hat{P}_{i} \rangle \approx P_{k_i}$, with $k_i$ a typical mode inside the bin.

We are interested in the covariance matrix of this bin averaged estimator,
\beq
{\bf C}_{ij} \equiv \langle \delta \hat{P}_{i} \, \delta \hat{P}_{j} \rangle.
\eeq
Starting from the covariances in the individual mode estimators,
\beq
\label{eq:covindiv}
\langle \delta \hat{P}_{\bf k} \, \delta \hat{P}_{\bf k'} \rangle = P_k^2\, \left( \delta^{\rm K}_{{\bf k}+{\bf k'}} + \delta^{\rm K}_{{\bf k}-{\bf k'}} \right)
+ \frac{1}{V} \, T_{{\bf k}, -{\bf k}, {\bf k'}, -{\bf k'}},
\eeq
we obtain
\beq
\label{eq:covgen}
{\bf C}_{ij} = \frac{1}{N_i}\, \sum_{{\bf k} \in i}  \frac{1}{N_j}\, \sum_{{\bf k'} \in j} \, \langle \delta \hat{P}_{\bf k} \, \delta \hat{P}_{\bf k'}\rangle
= \frac{2 P_{ i}^2}{N_i} \, \delta^{\rm K}_{ij} + \frac{\bar{T}_{ij}}{V},
\eeq
where we have introduced the shorthand notation $\bar{T}_{ij}$ for the bin averaged trispectrum.
The term proportional to the power spectrum squared is diagonal and given by simply counting modes because the binning in $k$-space only
includes pairs of estimators that have zero covariance. As we will discuss in detail in section \ref{sec:gen}, this is not
in general the case as the window function may correlate different bins.

\subsection{N-body Simulation Geometry: Perturbation Theory}
\label{subsec:PT}

The power spectrum appearing in the first term on the right hand side of Eq.~(\ref{eq:covgen}) is the full non-linear
power spectrum, which can be estimated directly from simulations.
The trispectrum is harder to get from simulations (but see \cite{harpen11}),
but can be modeled using a variety of perturbation theory schemes or using the halo model. In this work,
we will use the simple framework of (Eulerian) standard perturbation theory (SPT; \cite{Bernetal02,fry84, goroffetal86, scoccetal99}).
Anticipating angle averaging, i.e.~modulo transformations $\k \leftrightarrow -\k $ and $\kp \leftrightarrow -\kp $,
the leading order trispectrum for the configuration of interest is given as
(see\footnote{Note that the factor $16$ in Eq.~(7)
of \cite{scoccetal99} should be a factor $8$.} \cite{fry84,scoccetal99})
\beqa
\label{eq:tri}
T_{{\bf k}, -{\bf k}, {\bf k'}, -{\bf k'}} &=& 12\, P^{\rm lin}_k \, P^{\rm lin}_{k'} \, \left[ F_3(\k, -\k, \kp ) P^{\rm lin}_k + (\k \leftrightarrow \kp ) \right] \nonumber \\
+ 8 \, &P^{\rm lin}_{|\k - \kp|}& \, \left[ F_2(\k - \kp, \kp) \, P^{\rm lin}_{k'} + (\k \leftrightarrow \kp )\right]^2 \nonumber \\
+ 16 \, &P^{\rm lin}_k& \, P^{\rm lin}_{k'} \, P^{\rm lin}_{0} \, F_2(-{\bf 0}, \k) \, F_2(\kp, {\bf 0}),
\eeqa
where $F_2$ and $F_3$ are the kernels for the second and third order contributions to the density field, as
given in the appendix of \cite{goroffetal86}. We can now distinguish two types of terms. The first two lines
represent the ``standard'' trispectrum contributions, that have always been included in PT studies of the trispectrum
(see e.g.~\cite{scoccetal99}). When no modes larger than the survey are present, like in the artificial case
of a power spectrum estimated from the full volume of a periodic (N-body simulation) box,
these are the only terms entering the trispectrum, giving
\beq
\label{eq:tri0}
\bar{T}^0_{ij} = \frac{1}{N_i}\, \sum_{{\bf k} \in i}  \frac{1}{N_j}\, \sum_{{\bf k'} \in j} \, [ P^{\rm lin}_k \, P^{\rm lin}_{k'} \, \left[ F_3(\k, -\k, \kp ) P^{\rm lin}_k
+ (\k \leftrightarrow \kp ) \right]
+ 8 \, P^{\rm lin}_{|\k - \kp|} \, \left[ F_2(\k - \kp, \kp) \, P^{\rm lin}_{k'} + (\k \leftrightarrow \kp )\right]^2  ].
\eeq
We will evaluate $\bar{T}^0_{ij}$ numerically as no further analytic simplifications are possible.

\subsection{The Beat Coupling Effect}
\label{sec:bc}

The third line in Eq.~(\ref{eq:tri}) is the beat coupling contribution, the importance of which has been realized
only more recently (\cite{HRS06, takahashietal09, takjain08, reidsperbode09, rimesham06,sefusattietal06}). Our notation here requires some explanation (see also section \ref{sec:gen}).
Consistent with our discussion in the previous section, quantities evaluated at ${\bf k} = {\bf 0}$
should really be interpreted in terms of an {\it effective} zero mode arising from contributions at $k < 2\pi/L$,
i.e.~$P^{\rm lin}_{0} = V \, \langle|\delta^{\rm lin}_{\bf 0}|^2\rangle \approx P^{\rm lin}(k \sim \pi/L)$,
generated by continuum contributions $T(\k + {\bf \epsilon}, -\k + {\bf \epsilon'}, \kp + {\bf \epsilon''}, -\kp + {\bf \epsilon'''})$
with ${\bf \epsilon}+{\bf \epsilon'}=-({\bf \epsilon''}+{\bf \epsilon'''}) \sim \pi/L$.
Moreover, $F_2$, given by
\beq
F_2({\bf k_1}, {\bf k_2}) = \frac{5}{7} + \frac{{\bf \hat{k}_1} \cdot {\bf \hat{k}_2}}{2} \left( \frac{k_1}{k_2} + \frac{k_2}{k_1}\right)+ \frac{2}{7} \, ({\bf \hat{k}_1} \cdot {\bf \hat{k}_2})^2,
\eeq
is at first sight not well defined if one of the arguments equals ${\bf 0}$. $F_2(-{\bf 0}, \k)$ (and the other affected term) should
thus be interpreted as a limit
\beq
F_2(-{\bf 0}, \k) = \int \frac{d\Omega_{\hat{\bf \epsilon}}}{4\pi} F_2(-{\bf \epsilon}, \k),
\eeq
with $|{\bf \epsilon}| \sim \pi/L$ (see section \ref{sec:gen} for a more rigorous treatment of the beat coupling term with the same result).
The integral over the direction of ${\bf \epsilon}$ makes this quantity well defined.

Using the fact that
\beq
\label{eq:avg F2}
\int \frac{d\Omega_{\bf k_1}}{4\pi} \, F_2({\bf k_1}, {\bf k_2}) = \frac{17}{21},
\eeq
the beat coupling term can be angle (or bin) averaged analytically so that we end up with
\beq
\label{eq:cov}
\bar{T}_{ij} = \bar{T}^0_{ij} 
+ 16 \, \left(\frac{17}{21}\right)^2 \, P^{\rm lin}_{k_i} \, P^{\rm lin}_{k_j} \, P^{\rm lin}_{0}.
\eeq

Physically, the presence of the beat coupling term is an interesting interplay of mode mixing due to the window functions
with correlations between pairs of non-linear modes and one larger mode (see \cite{HRS06}).
Due to the finite volume from which the power spectrum is measured, the estimator $\hat{P}(\k)$ really consists of
a weighted average of pairs of density modes $\delta(\k + {\bf \epsilon}) \, \delta(-(\k + {\bf \epsilon'}))$,
with $|{\bf \epsilon}|$ of order of the fundamental mode $2\pi/L$. For $k$ in the non-linear regime,
such pairs correlate with the large scale perturbation $\delta({\bf \epsilon'}-{\bf \epsilon})$,
which in turn causes the covariance between the power spectrum estimators to be proportional to the power in these
large modes. In our description, this is captured by the power in the effective zero mode.
Note that the beat coupling creates an excess both in the variance and in the off-diagonal
elements of the covariance matrix.

\subsection{The Local Average Effect}

While the beat coupling term derived in the previous section adds significant covariance on non-linear scales,
there is an additional effect coupling small scale covariance to power in the zero-mode that plays a role in
the power spectrum estimated from an actual survey. This second effect is caused by the fact that to obtain the overdensity
$\delta$, one needs an estimate of the average density $\bar{\rho}$. In a realistic survey, one does not know the true average number density
of, say,  galaxies, but instead has to rely on an estimate of the average density within the survey volume, which is modulated
by the zero mode $\delta_{\bf 0}$. This results in a decrease in covariance, partially canceling the beat coupling
effect\footnote{The difference between using the true and the local average
was also commented on briefly in \cite{neyszasza09}}.
We will refer to this contribution as the {\it local average effect}.

The local average effect causes the true overdensity estimator to be given
by
\beq
\label{eq:dk d0}
\tilde{\delta}_{\bf k} \equiv \frac{\delta_{\bf k}}{1 + \delta_{\bf 0}}.
\eeq
Since $\bar{\rho}$ also appears in the numerator of $\delta$, there technically
is also a $\delta_{\bf 0}$ contribution there. However, the Fourier transform of the zero mode
for non-zero wave vector $\k$ vanishes so this term can be omitted.
Eq.~(\ref{eq:dk d0}) can be expanded in powers of $\delta_{\bf 0}$ to derive expressions
for the $n$-point functions to the desired order in perturbation theory.
For example, to next-to-leading order, the expectation value of the
power spectrum estimator becomes
\beqa
\langle \hat{\tilde{P}}_{\bf k} \rangle &\equiv& \langle \tilde{\delta}_{\bf k} \, \tilde{\delta}_{\bf -k} \rangle \nonumber \\
&=& V \, \langle \delta_{\bf k} \, \delta_{\bf -k} \, \left(1 - 2 \delta_{\bf 0} + 3 \delta_{\bf 0}^2 + \mathcal{O}\left(\delta_{\bf 0}^3\right)\right) \rangle \nonumber \\
&=& P_{k} - 2 \frac{B_{\k, -\k, {\bf 0}}}{V} + 3 \frac{P_{\bf 0}}{V} \, P_{k} + \mathcal{O}\left(P^3/V^2\right)~.
\eeqa
The expectation value of the angle-averaged estimator is then
\beqa
\langle \hat{\tilde{P}}_{i} \rangle &=& P_{k_i} - \frac{2}{V} \, \frac{68}{21} \, P_{k_i} \, P_0 + \frac{3}{V} P_{k_i} \, P_0 + \mathcal{O}\left(P^3/V^2\right) \nonumber \\
&=& P_{k_i} \, \left( 1 - \frac{73}{21} \, \frac{P_0}{V}\right) + \mathcal{O}\left(P^3/V^2\right)~.
\eeqa
Hence, the power spectrum receives a small bias due to the local average effect.
The relative correction is $\lesssim 10^{-4}$ for a 1$h^{-3}$Gpc$^3$ survey
so it can be safely ignored for a realistic survey.
Here,
we have used that the bispectrum (\cite{sefusattietal06})
\beq
B_{\k, \kp, {\bf k''}} = 2 \, P_{k} \, P_{k'} \, F_2(\k, \kp) \,  +  \, ({\rm cyclic})
\eeq
and we have applied the identity (\ref{eq:avg F2}) to carry out the angle averaging.

Using the same approach for the covariance matrix, one gets to next to leading order
\beq
\label{eq:cov bc+la}
\tilde{\bf C}_{ij} = \langle \hat{\tilde{P}}_i \, \hat{\tilde{P}}_j \rangle - \langle \hat{\tilde{P}}_i \rangle \, \langle \hat{\tilde{P}}_j \rangle
= \frac{2 P_{k_i}^2}{N_i} \, \delta^{\rm K}_{ij}  + \frac{\bar{T}^0_{ij}}{V}
+ 16 \, \left(\frac{17}{21}\right)^2 \, P^{\rm lin}_{k_i} \, P^{\rm lin}_{k_j} \, \frac{P^{\rm lin}_{0}}{V}
- \frac{188}{21} \, P^{\rm lin}_{k_i} \, P^{\rm lin}_{k_j} \, \frac{P^{\rm lin}_0}{V},
\eeq
where we have ignored a small relative correction to the diagonal term of order $P^{\rm lin}_0/V$.
The local average effect thus introduces a term of the same form as the beat coupling term from the previous section,
proportional to $P_{k_i} \, P_{k_j} \, P_0/V$, but with opposite sign. Comparing the coefficients, one finds
that the two effects almost entirely cancel out, leaving only a small positive coefficient,
$16\left(\frac{17}{21}\right)^2 - \frac{188}{21} \approx 1.5$ $\sim 10 \%$ of the original beat coupling
coefficient. Note that the local average effect to this order consists of contributions
from the bispectrum and trispectrum, but also from terms that would even be there had the field remained
completely Gaussian (but that happen to be of the same order as the leading non-Gaussian corrections).

Eq.~(\ref{eq:cov bc+la}) thus gives us an expression for the most realistic case we will consider,
where modes larger than the survey are present {\it and} the spectrum is estimated using the local
average in the survey volume.

\subsection{Theory Summary and Outlook}
\label{sec:outlook}

In the remainder of this article, we will use N-body simulations to test
the expressions derived above.
We will consider three different cases which allow us to separately constrain different combinations of contributions to the total covariance:
\begin{itemize}
\item {\bf Case 1:} {\it periodic box}\\
The spectrum is estimated from the full, periodic simulation volume.
In this case, there is no beat coupling nor a local average effect and the prediction for the covariance is
to leading order
\beq
\label{eq:c1}
{\bf C}_{ij} =  \frac{2 P_{k_i}^2}{N_i} \, \delta^{\rm K}_{ij} + \frac{\bar{T}_{ij}^0}{V}.
\eeq
We obtain the (non-linear) power spectrum appearing above by applying the
Halofit\footnote{We have checked that the difference between using the Halofit spectrum and the average
simulated spectrum is small for $k < 0.4 h/$Mpc.} prescription \cite{Smithetal03}
to the linear power spectrum calculated using CAMB \cite{LewChalLas00}.
The bin-averaged trispectrum $\bar{T}_{ij}^0$ is given by Eq.~(\ref{eq:tri0}), where we make one modification. Instead of using the linear
power spectra, we use the non-linear spectra. This is consistent to the desired order in perturbation theory and
turns out to slightly improve the accuracy of the model on strongly non-linear scales.

\item
{\bf Case 2:} {\it subbox of periodic box}\\
Since modes larger than the ``survey volume'' are now present, there is a beat coupling effect. The leading order covariance
prediction is
\beq
\label{eq:c2}
{\bf C}_{ij} =  \frac{2 P_{k_i}^2}{N_i} \, \delta^{\rm K}_{ij} + \frac{\bar{T}_{ij}^0}{V}
+ 16 \, \left(\frac{17}{21}\right)^2 \, P_{k_i} \, P_{k_j} \, \frac{P_{0}}{V}.
\eeq
Note that also for the additional trispectrum terms, we choose to use the non-linear power spectrum
as opposed to the linear one.

\item
{\bf Case 3:} {\it subbox of periodic box, using subbox mean}\\
This is the most realistic case, where not only there are modes larger than the survey, but the overdensity
$\delta = (\rho({\bf x}) - \bar{\rho})/\bar{\rho}$ is calculated in terms of the average density $\bar{\rho}$
of the subbox, as opposed to the ``true'' average density of the full box.
The leading order part of the covariance matrix is given by
\beqa
\label{eq:c3}
\tilde{\bf C}_{ij} = \frac{2 P_{k_i}^2}{N_i} \, \delta^{\rm K}_{ij}  + \frac{\bar{T}^0_{ij}}{V}
+ \frac{676}{441} \, P_{k_i} \, P_{k_j} \, \frac{P_0}{V}.
\eeqa

\end{itemize}

In the next section, we first describe the details of our simulations. We will discuss the results of our comparison in
section \ref{sec:results}.

\section{N-body simulations}
\label{sec:nbody}

In order to test our analytic predictions we use a large suite of N-body simulations.
We have two sets of simulations consisting of 160 runs of a $2400h^{-1}$Mpc box with $768^3$ particles
and 1024 runs of a $600h^{-1}$Mpc box with $192^3$ particles. The initial conditions of the 160 and 1024 simulations
were set up using different realizations of a Gaussian random field with the power spectrum given by CAMB.
We adopted a flat $\Lambda$CDM cosmology consistent with the current observational constraints \cite{Komatsuetal10}.
The cosmological parameters are the present-day matter fraction $\Omega_m=0.27$, Hubble constant $h=0.7$,
baryon fraction $\Omega_b h^2 = 0.023$, spectral index $n_s=0.95$, and present-day normalization $\sigma_8=0.7913$.

The particles were displaced from their initial grid points according to second-order Lagrangian perturbation
theory using an initial redshift $z_i=19$.

The simulations were performed with the Tree-PM code Gadget-2 \cite{springel05} taking only the gravitational
force into account. We applied a force softening of $70h^{-1}$kpc and used a particle mesh
of $2048^3$ and $512^3$ for the $2400h^{-1}$Mpc and $600h^{-1}$Mpc runs, respectively.
Using a much higher resolution simulation, we checked that with these simulation settings the power
spectrum derived from the simulation data is accurate at the 1\% level up to $k<0.2h{\rm Mpc}^{-1}$ and
remains accurate within 4\% up to $k<0.4h{\rm Mpc}^{-1}$ at all redshifts.

To compute the power spectrum from the simulation data, we assign the particles using the
cloud-in-cell (CIC) scheme to a regular grid with a fixed grid spacing of $1.5625h^{-1}$Mpc in
all cases (full box, subbox, subbox with zero-padding). Hence, the Nyquist frequency of the
grid is the same in all cases, $k_{\rm Ny}\approx 2h^{-1}$Mpc, which is 5 times larger than
the scales we consider in this paper $k<0.4h{\rm Mpc}^{-1}$. Therefore we expect the effects
due to finite grid size (smearing and aliasing effects) to be very small. Nevertheless,
we do correct for the smearing due to the CIC assignment \cite{jing05}.

The simulations and the power spectrum computation needed about $300,000$ CPU hours on our in-house cluster Hipatia
at ICCUB.

\section{Results}
\label{sec:results}

\begin{figure*}
  \includegraphics[width=0.48\columnwidth]{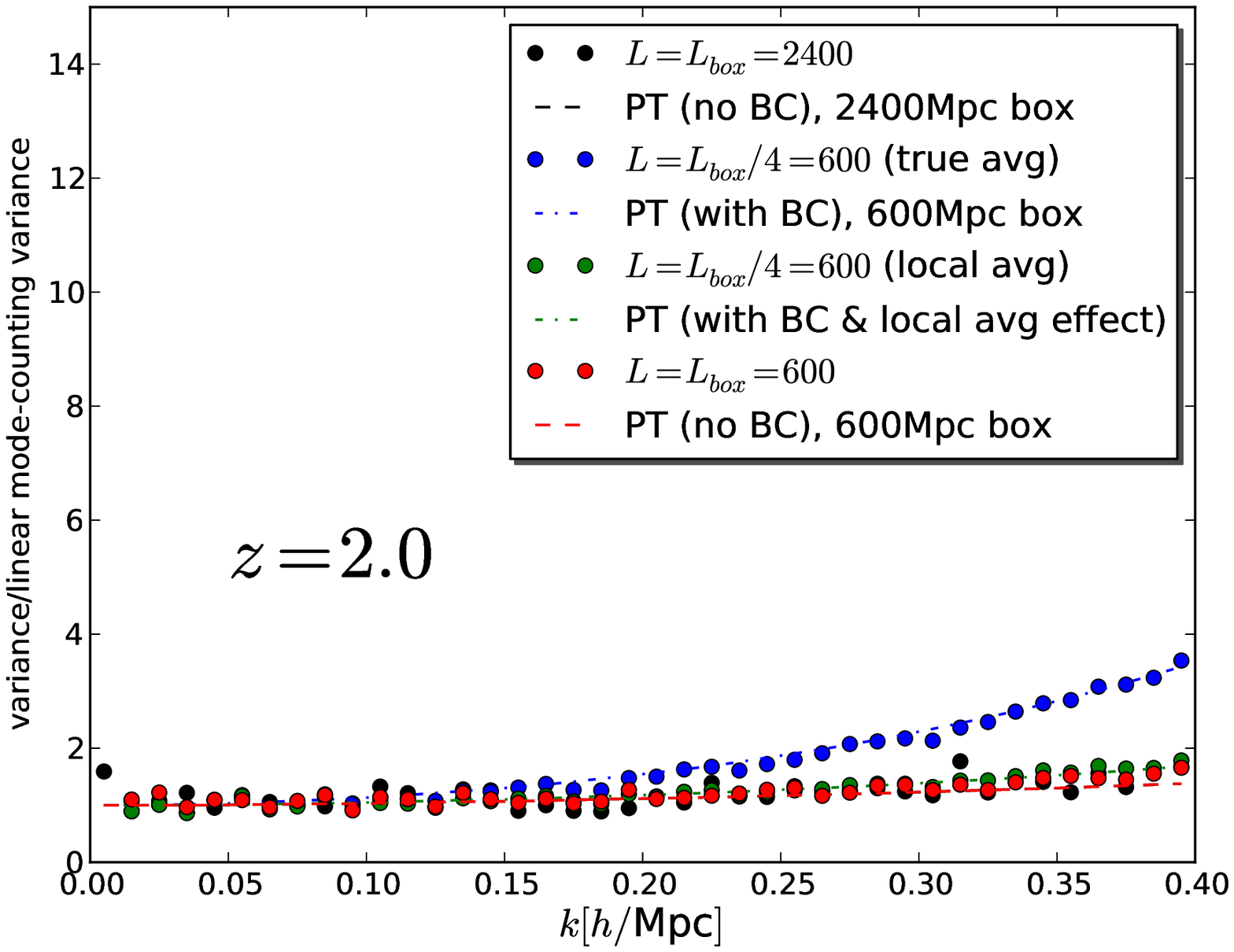}
  \includegraphics[width=0.48\columnwidth]{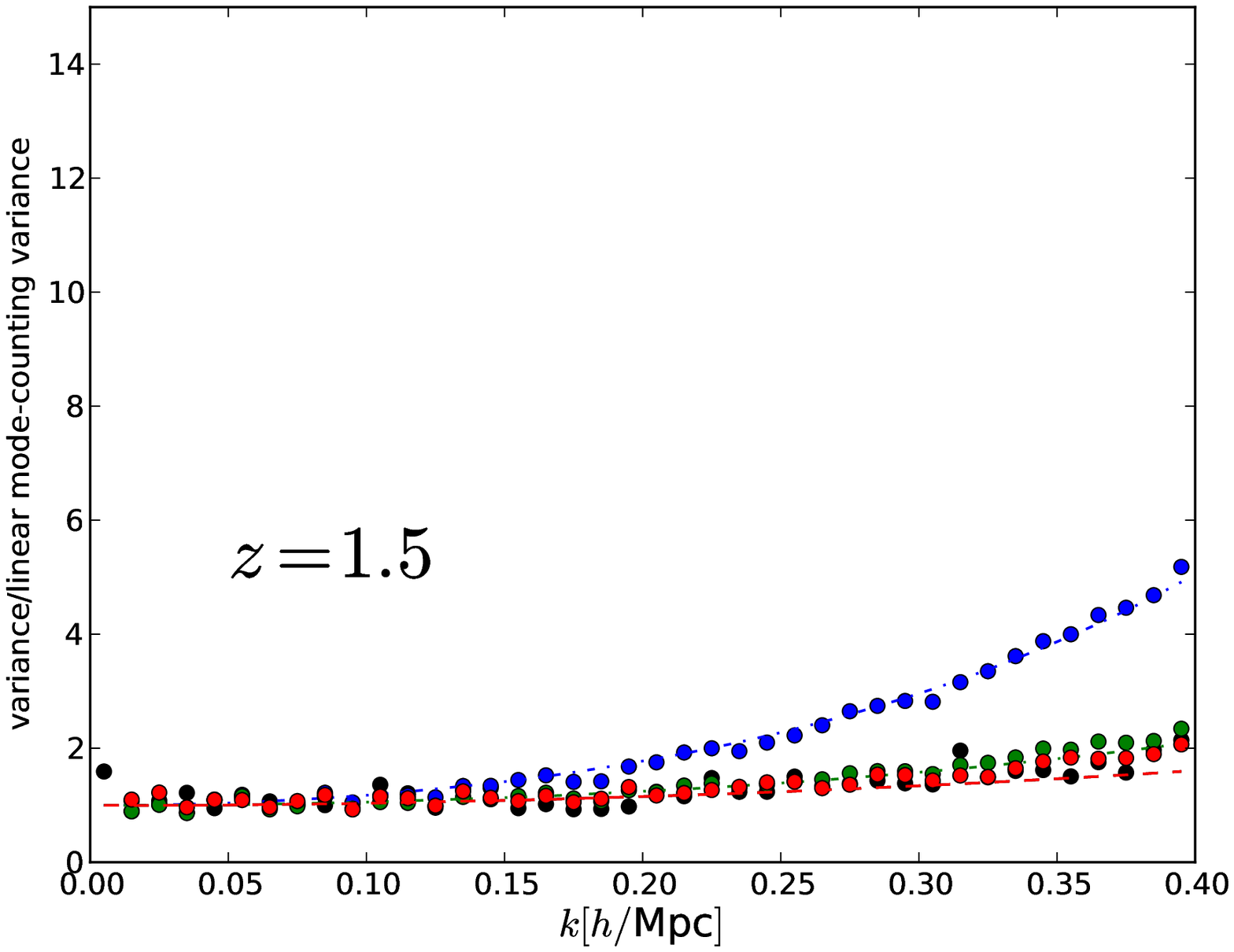}
  \includegraphics[width=0.48\columnwidth]{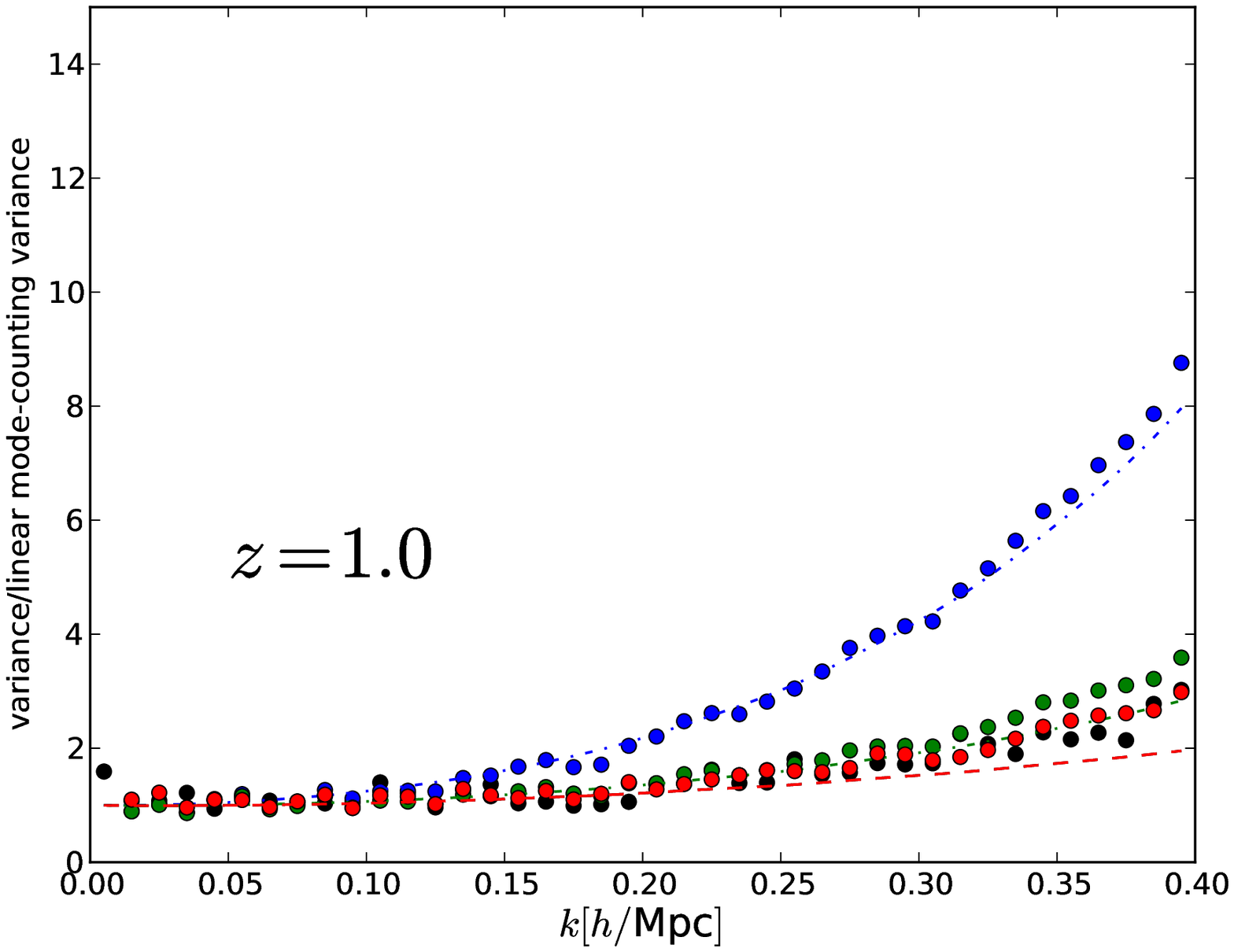}
  \includegraphics[width=0.48\columnwidth]{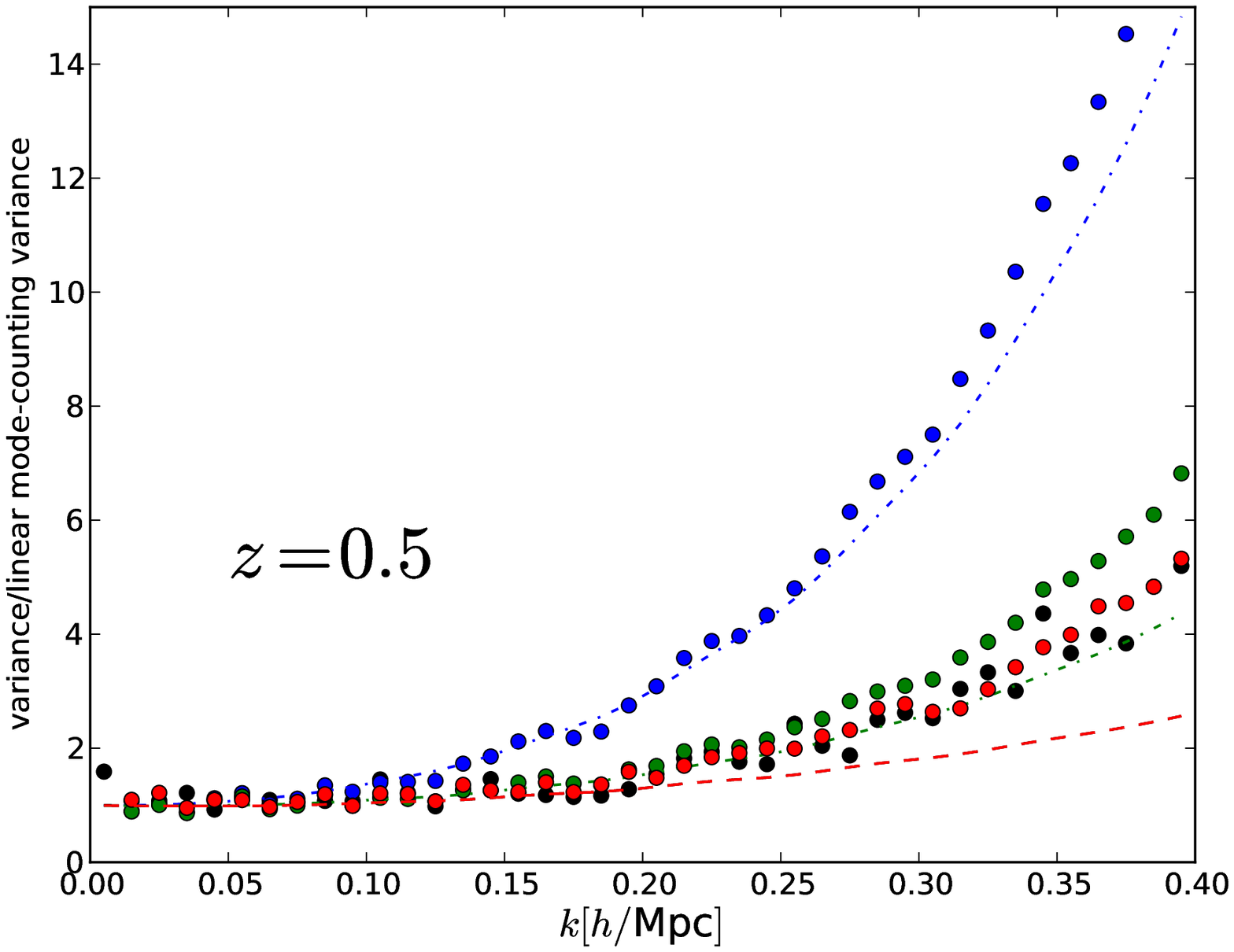}
  \includegraphics[width=0.48\columnwidth]{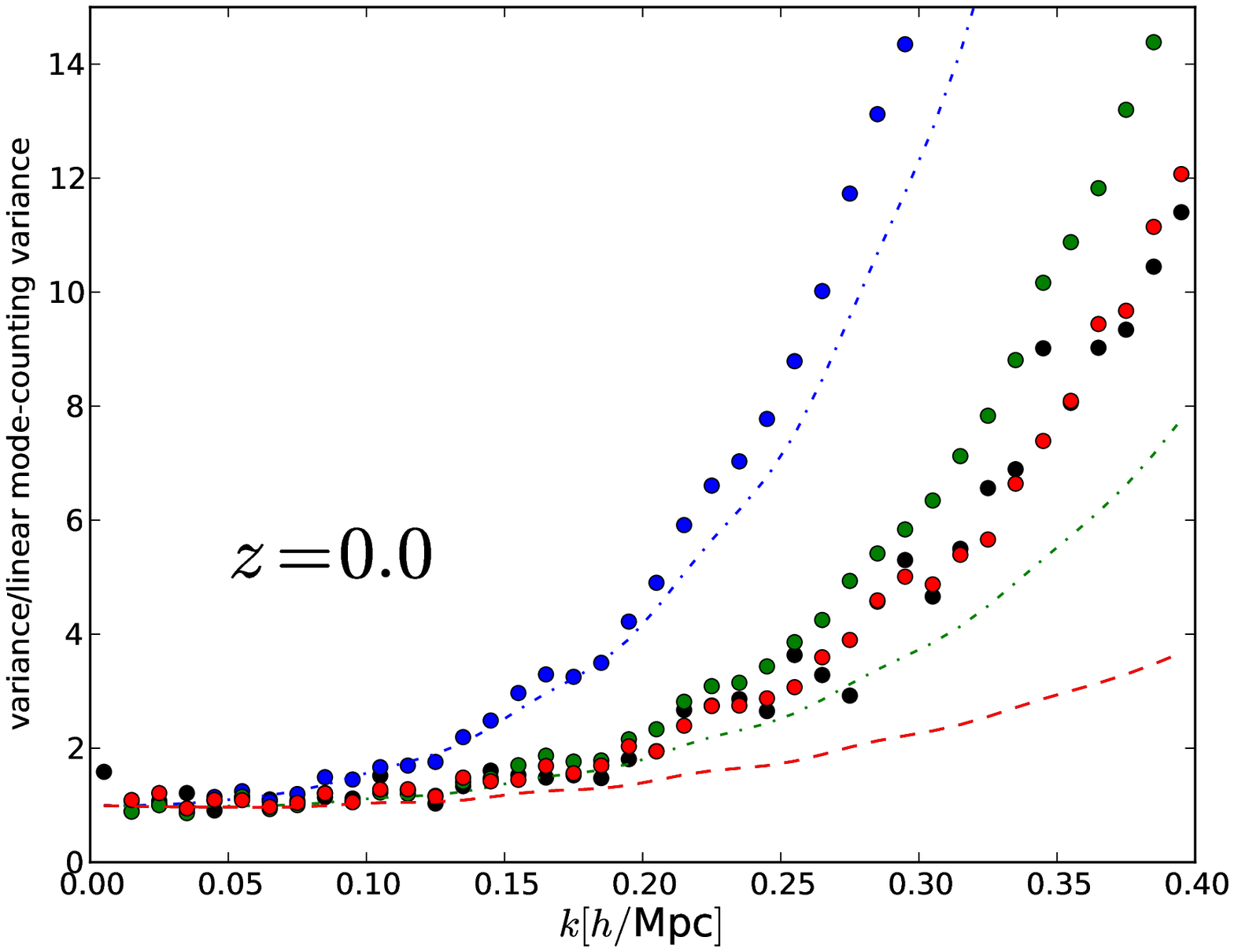}
  \caption{Power spectrum variance relative to variance based on linear spectrum and mode
  counting ($2 P_{\rm lin}^2/N_k$ with $N_k$ the number of modes per bin) for redshifts
  $z = 0 - 2$. The dots are results from N-body simulations and the (dashed) lines
  theory predictions from Eqs (\ref{eq:c1})-(\ref{eq:c3}), using perturbation theory (PT). The results in black and
  red represent the case where no modes larger than the survey are present (but other non-linear contributions
  are included). The blue results
  represent the case where
  these modes {\it are} present and thus display large excess covariance due to beat coupling. In
  the most realistic case (green), the large modes also affect the power spectrum through the estimated
  average density and this {\it local average effect} reduces the excess covariance by $\sim 90 \%$.}
  \label{fig:vars}
\end{figure*}

We now compare the simulation results to the analytic predictions summarized in section \ref{sec:outlook}.
For the covariances of the full periodic box (Case 1), we compare to the 1024 simulations of the $600 h^{-1}$Mpc cubed volume
and to the 160 simulations of the $2400 h^{-1}$Mpc cubed volume.
To study the {\it subbox} Cases 2 \& 3, we divide the $L = 2400 h^{-1}$Mpc volume into 64 smaller boxes with side $L/4=600 h^{-1}$Mpc each,
which provides us with $160 \times 64 = 10,240$ different subbox realizations. 

For each volume, we estimate the power spectrum $\hat{P}_i$
in isotropic bins of width $\Delta k = 0.01 h {\rm Mpc}^{-1}$ in the range $ k = 0 - 0.4 h {\rm Mpc}^{-1}$.
We then estimate the covariance matrix for the full box, or for a particular subbox of it, by
\beq
\label{eq:covest}
{\bf \hat{C}}_{i j} = \frac{1}{N_r - 1} \, \sum_r (\hat{P}_i - \langle\hat{P}_i\rangle) (\hat{P}_j - \langle\hat{P}_j\rangle),
\eeq
where the sum runs over all $N_r$ simulation realizations and $\langle \hat{P}_i \rangle$ is the average power spectrum over simulations.
In the subbox case, we then improve the accuracy by averaging the subbox covariance estimate in Eq.~(\ref{eq:covest})
over all 64 subboxes. Note that, since the subboxes living in the same simulation volume are not independent, it would
be wrong to directly apply Eq.~(\ref{eq:covest}) to all $N_r = 10,240$ subboxes.

Since the non-linear effects discussed in section \ref{sec:form} become stronger with time,
it is interesting to consider the covariance matrix for a range of redshifts from $z = 2 - 0$. We expect our analytic predictions
to have the largest range of validity at $z=2$, as non-linear effects there are smallest.

In Fig.~\ref{fig:vars}, we show results for the variances in the power spectrum, normalized by the variance based
on mode counting and the linear power spectrum. Case 1 is tested by both the full $L=2400 h^{-1}$Mpc box case (black dots - N-body results, black dashed - analytic)
and the full $L=600 h^{-1}$Mpc box case (red). The blue dots and dashed lines test Case 2 and the green ones test Case 3.
For all cases and at all redshifts, we find good agreement between theory and simulation
for bins up to a maximum value $k_{\rm max}$ that lies in the non-linear regime.
The values for $k_{\rm max}$ are reasonable given the range of scales over which perturbation
theory is expected to be applicable, 
$k_{\rm max} \sim 0.4 h{\rm Mpc}^{-1}$ at $z = 2$ to $k_{\rm max} \sim 0.15 - 0.2 h{\rm Mpc}^{-1}$
at $z = 0$.

\begin{figure*}
  \begin{center}{
  \includegraphics[width=0.48\columnwidth]{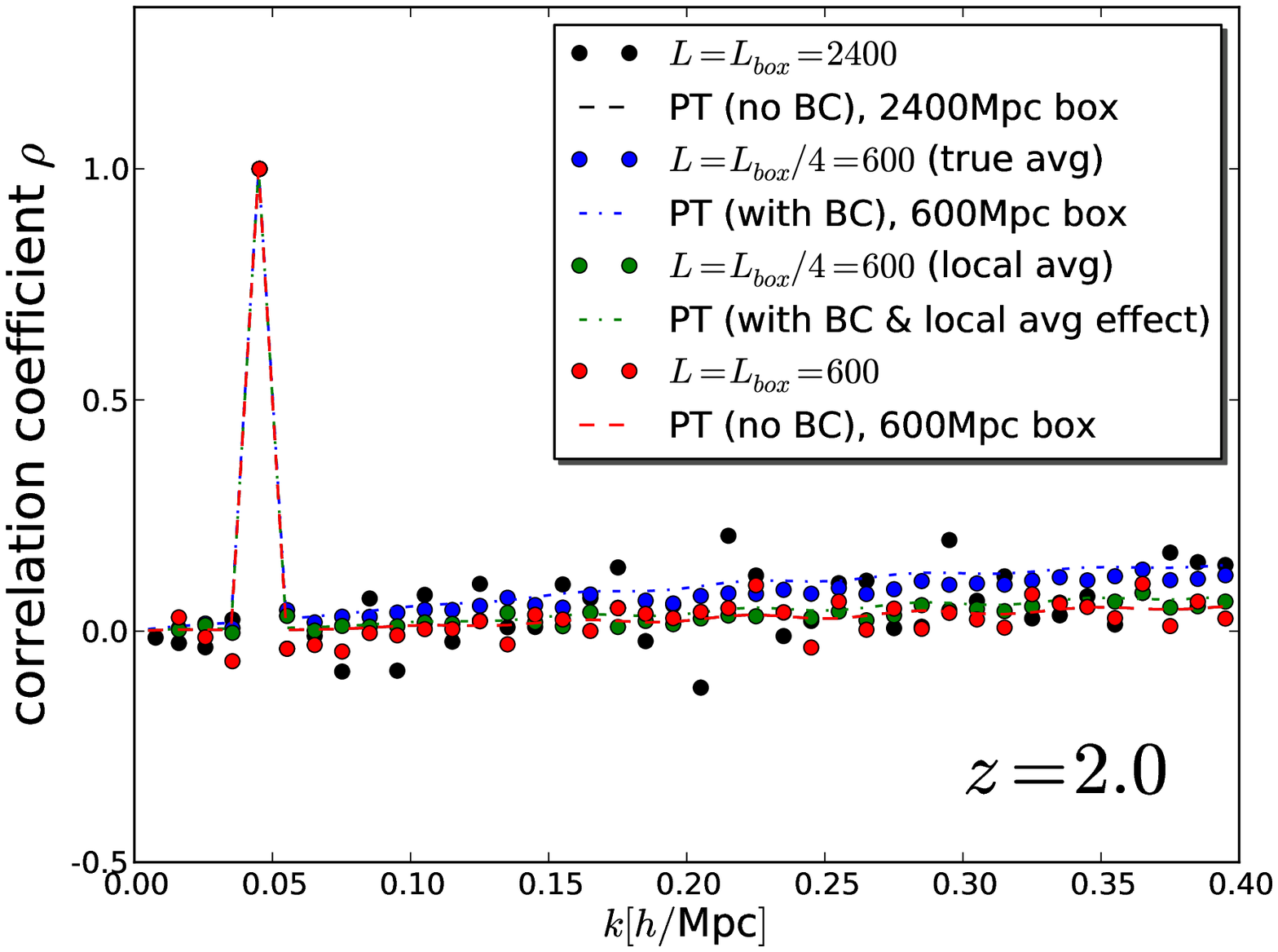}
  \includegraphics[width=0.48\columnwidth]{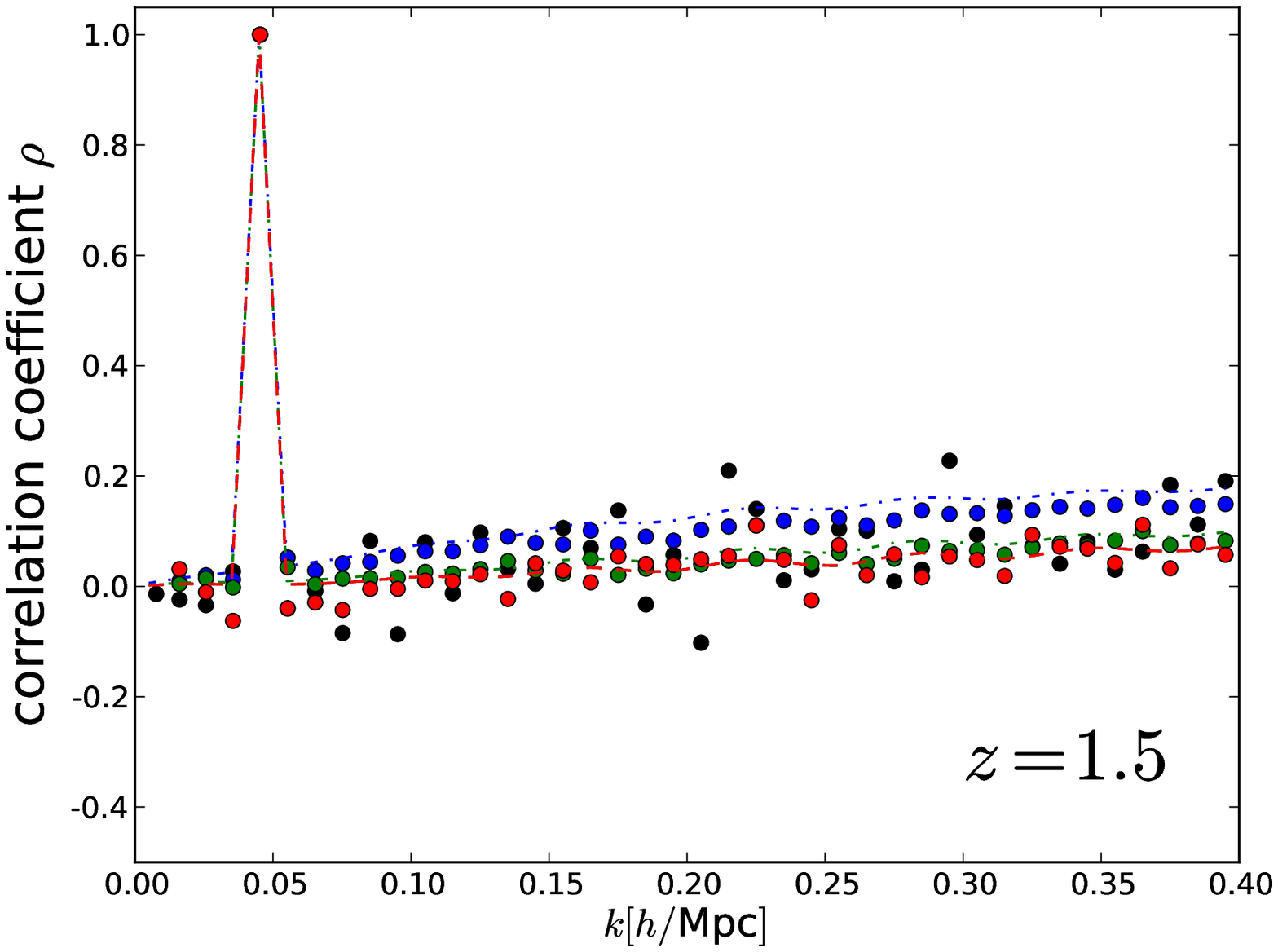}
  \includegraphics[width=0.48\columnwidth]{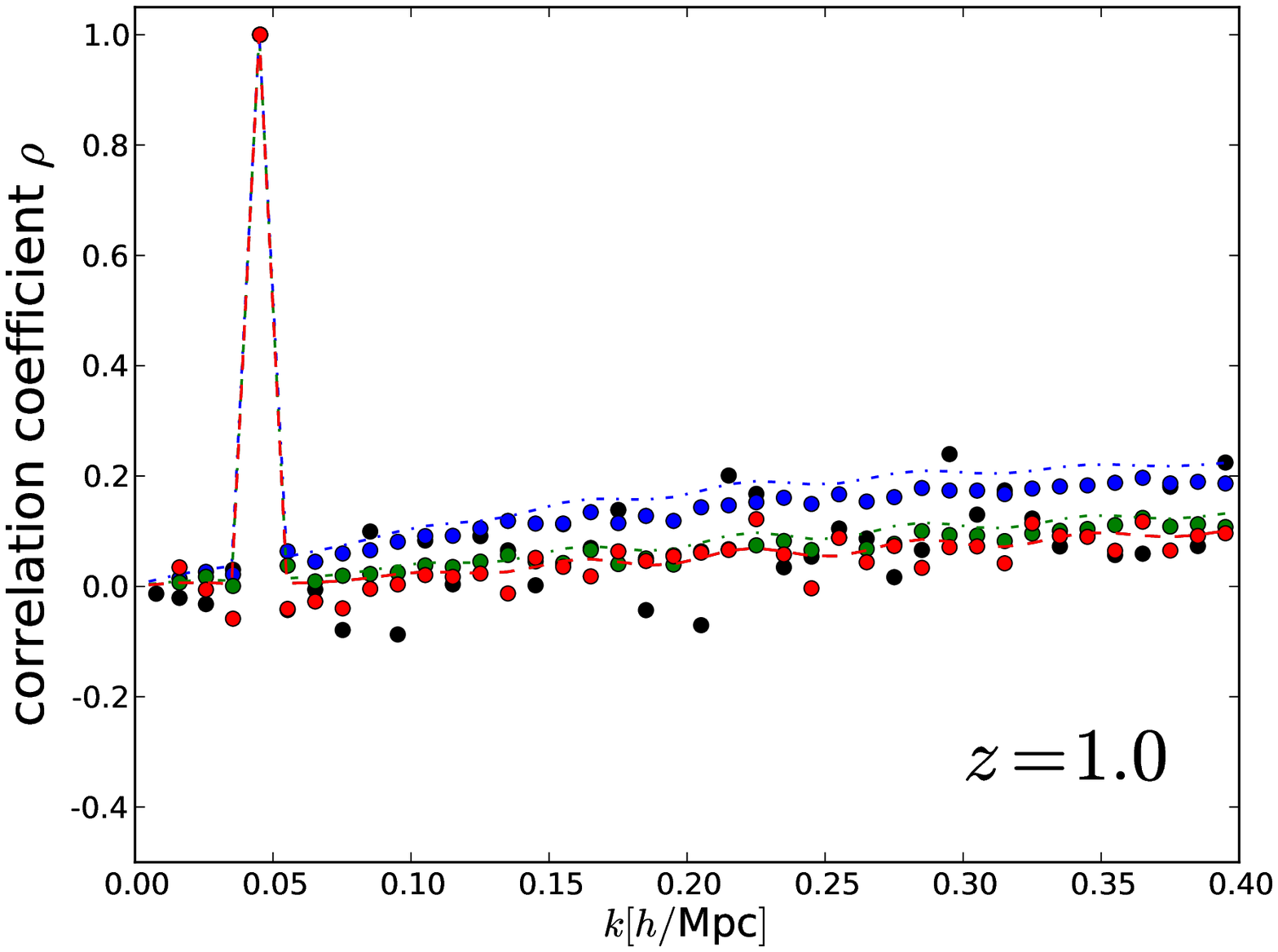}
  \includegraphics[width=0.48\columnwidth]{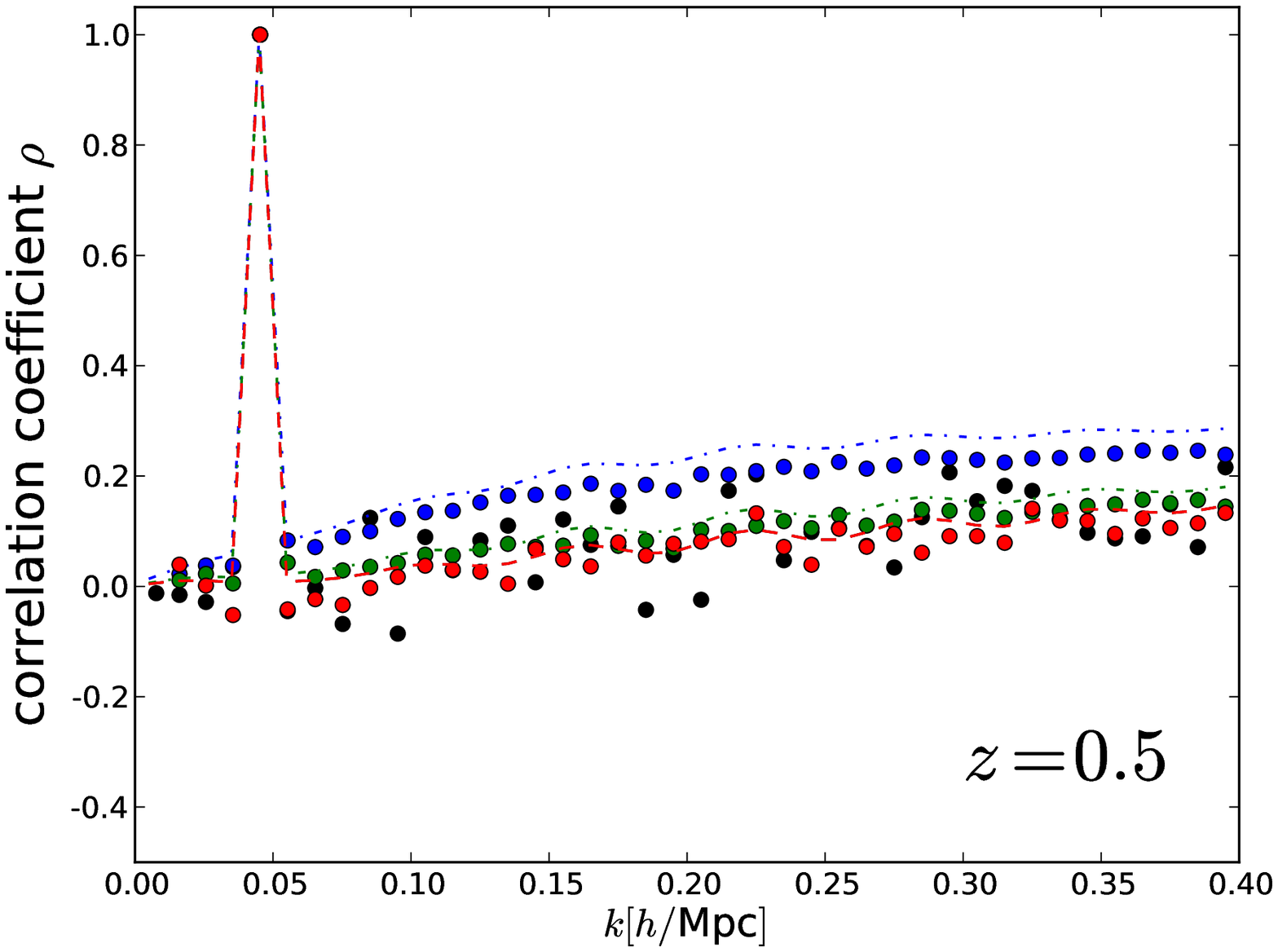}
  \includegraphics[width=0.48\columnwidth]{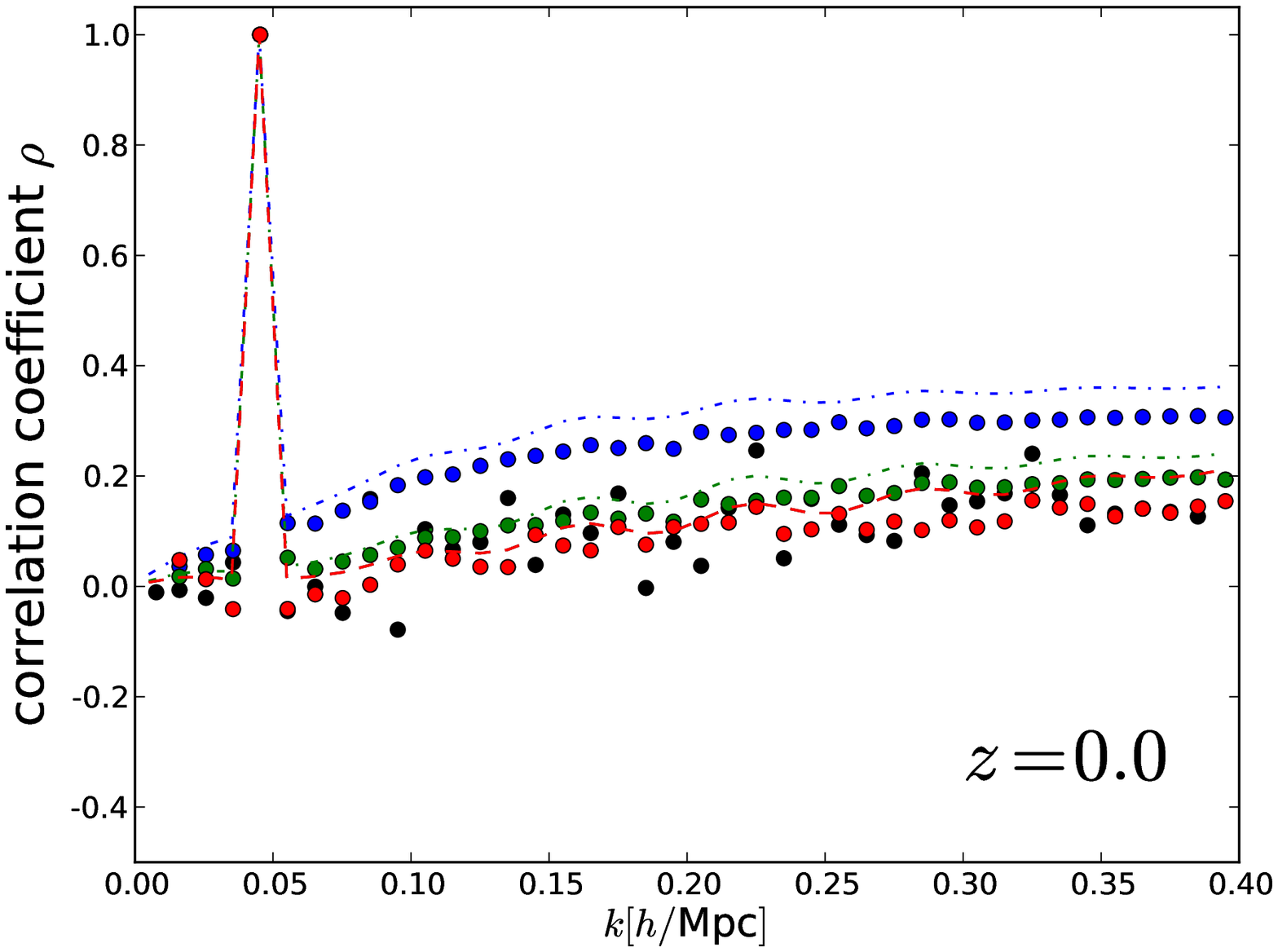}
  }
  \end{center}
  \caption{Correlation coefficients $\rho_{ij} \equiv {\bf C}_{ij}/\sqrt{{\bf C}_{ii} \, {\bf C}_{jj}}$
  of $i$-th bin at $k$ on x-axis, relative to $j$-th bin at $k=0.04-0.05 h$Mpc$^{-1}$.
  Same comparisons and color coding as in Fig.~\ref{fig:vars}. The beat coupling
  causes a significant increase in the off diagonal correlations (blue), but this is again undone
  by the local average effect (green). The effect is described well by Eqs (\ref{eq:c1})-(\ref{eq:c3}).
  }
  \label{fig:corrik4}
\end{figure*}

\begin{figure*}
  \begin{center}{
  \includegraphics[width=0.48\columnwidth]{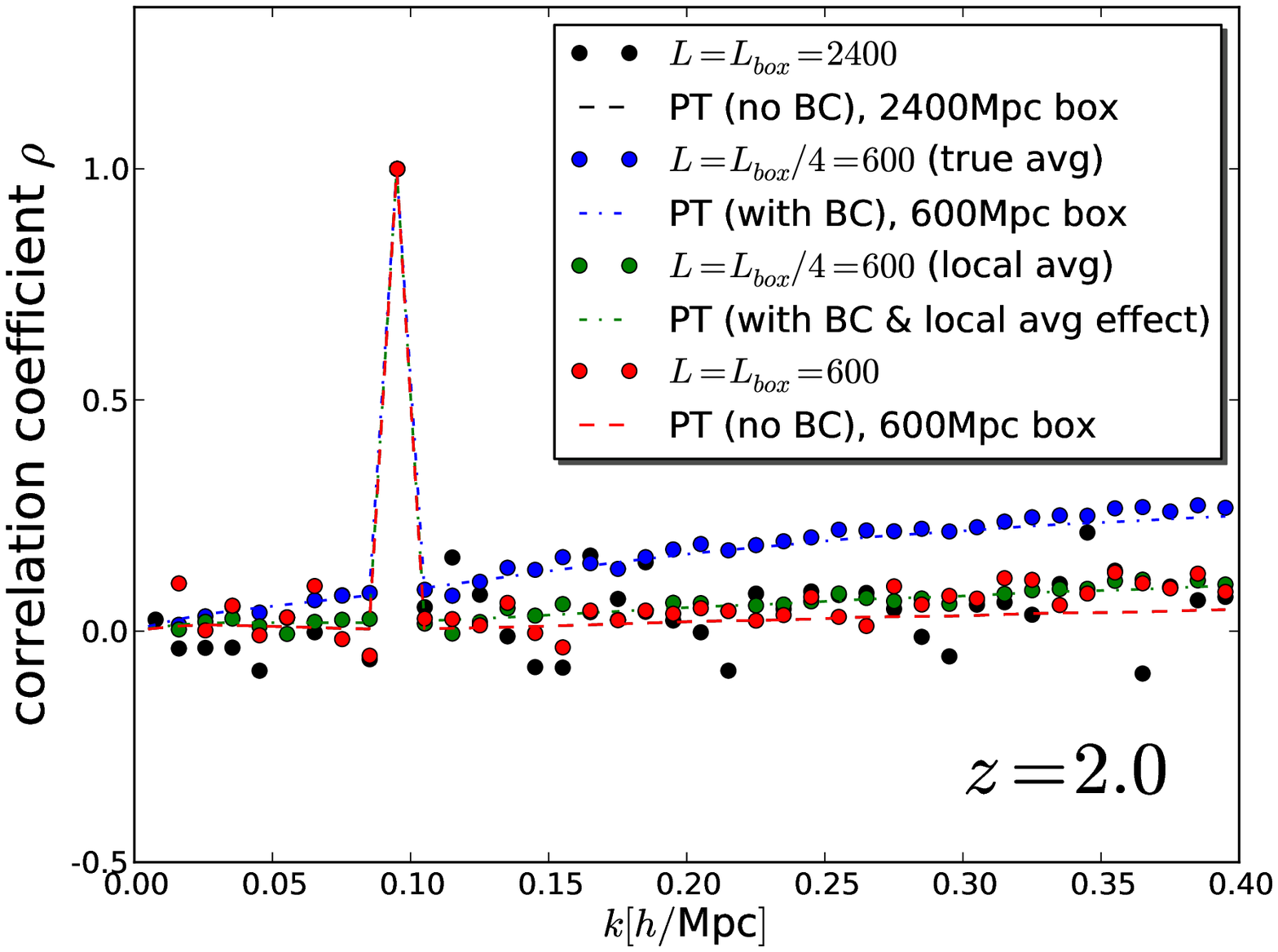}
  \includegraphics[width=0.48\columnwidth]{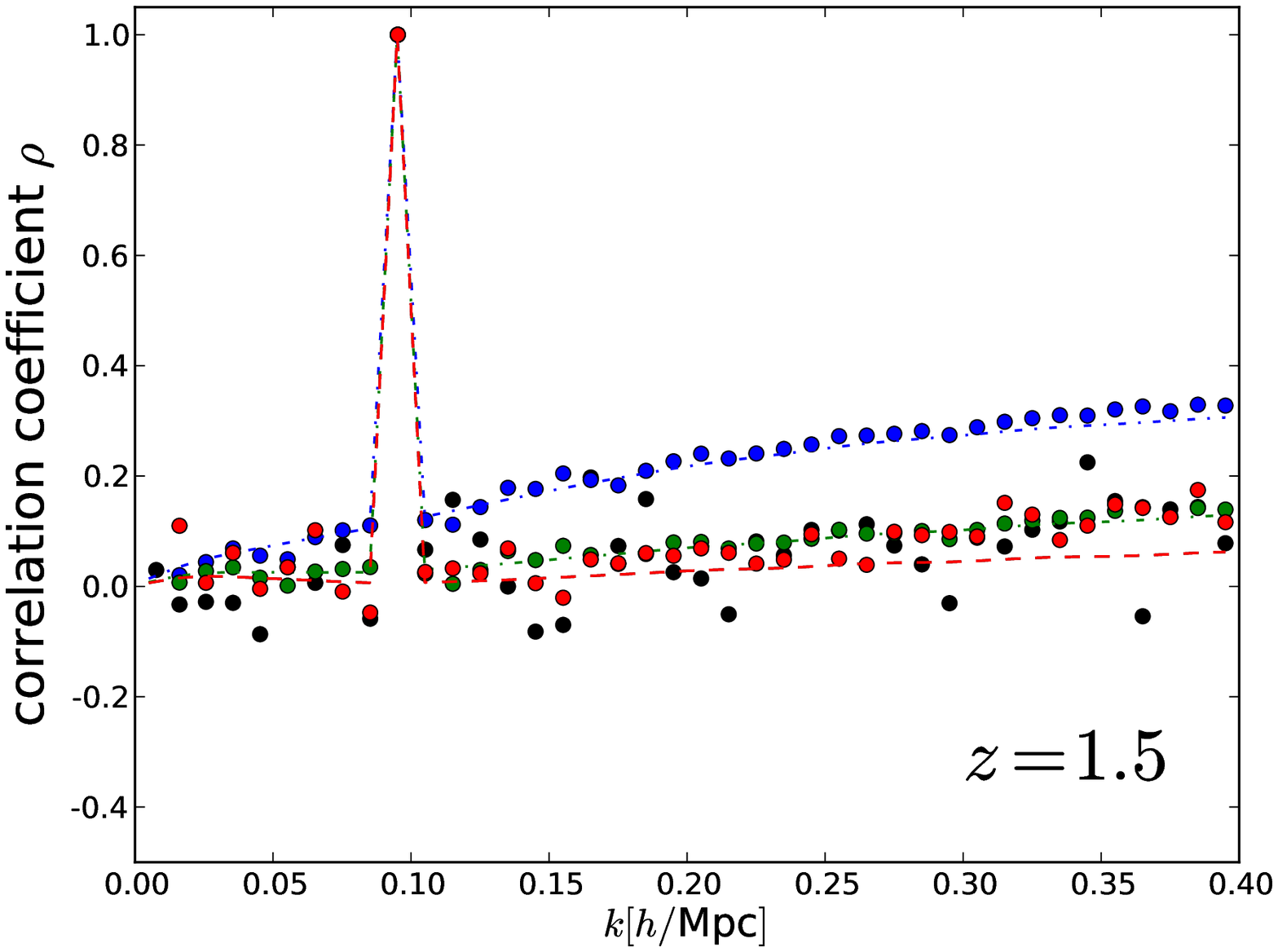}
  \includegraphics[width=0.48\columnwidth]{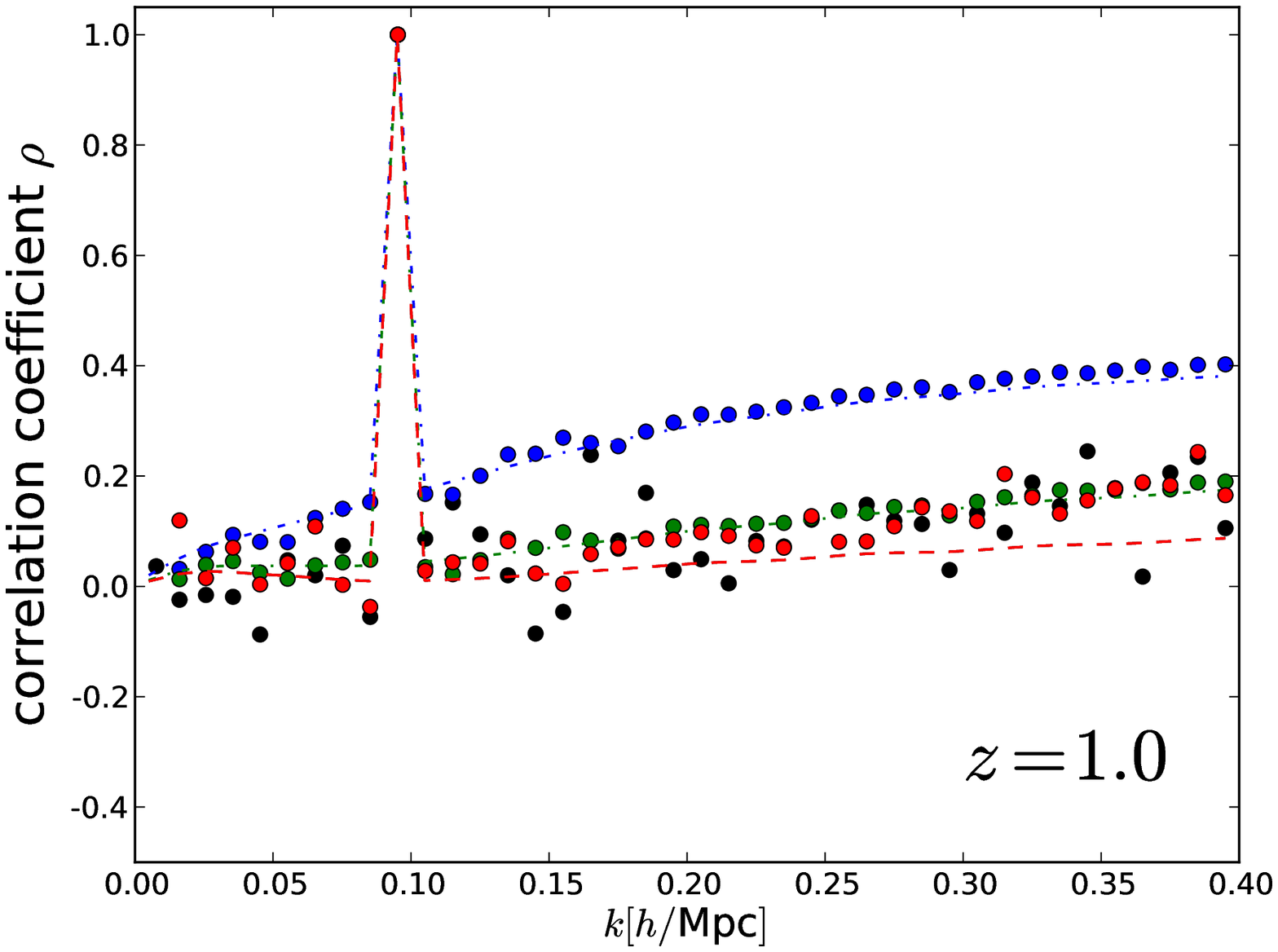}
  \includegraphics[width=0.48\columnwidth]{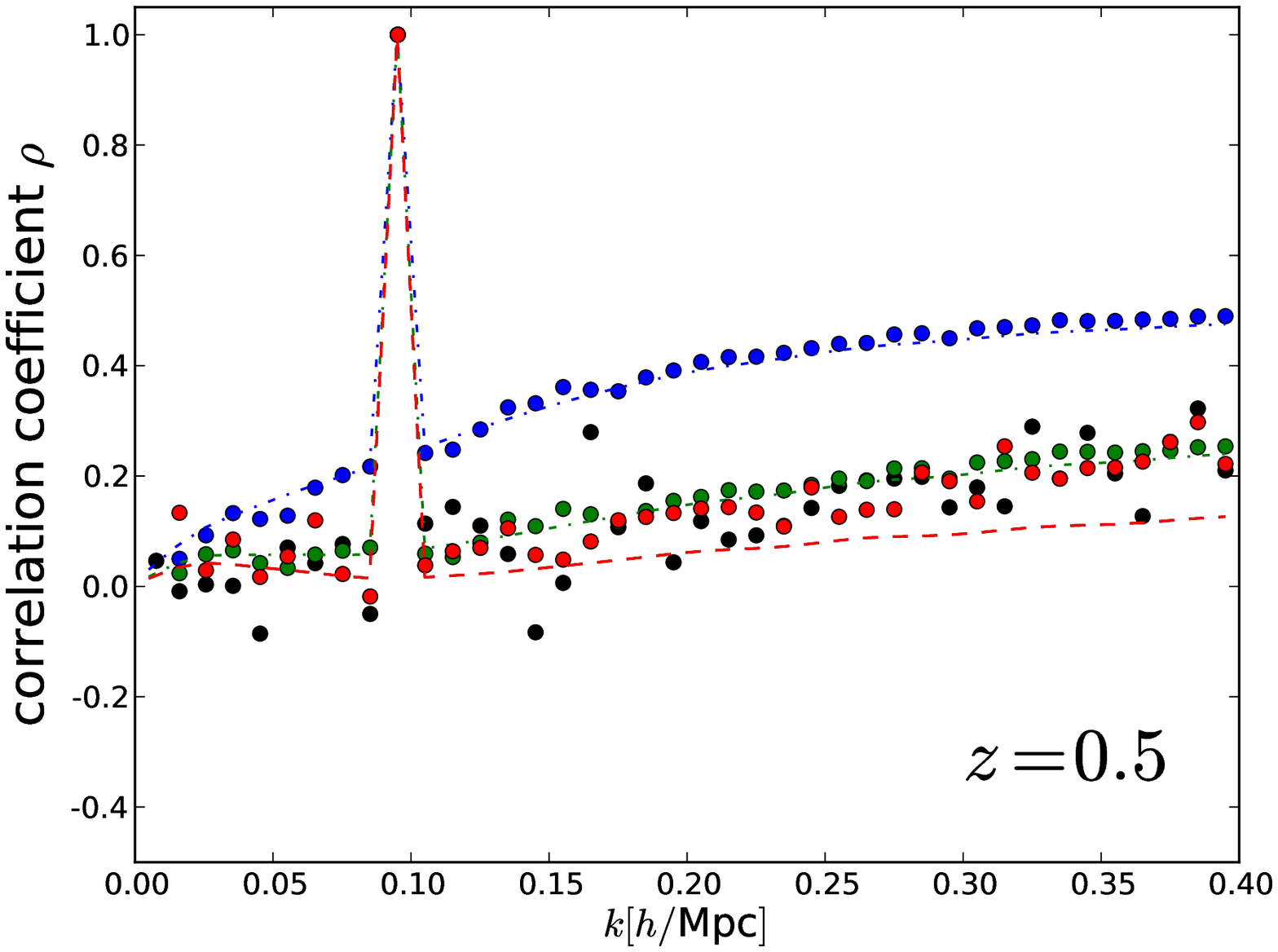}
  \includegraphics[width=0.48\columnwidth]{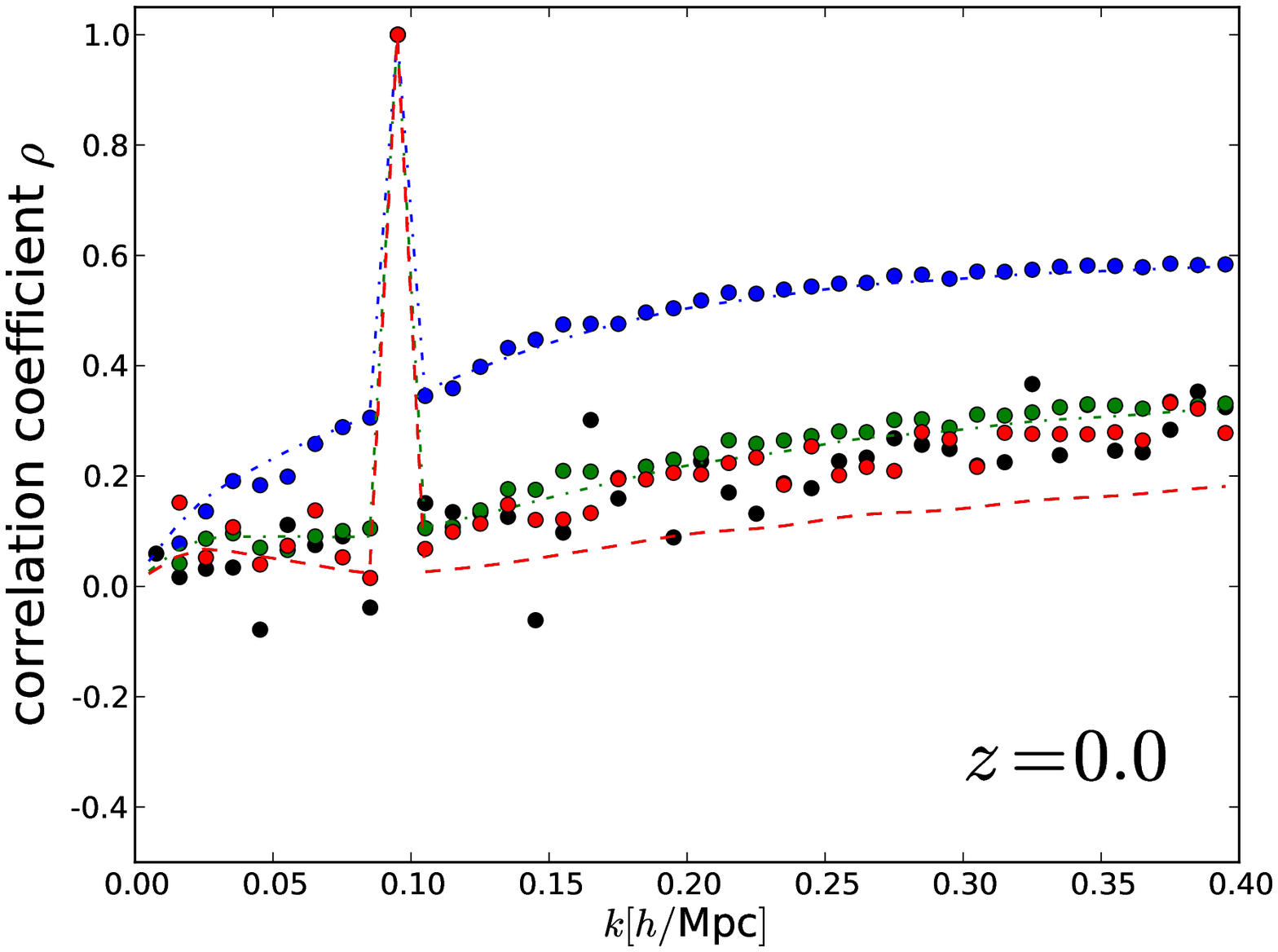}
  }
  \end{center}
  \caption{As in Fig.~\ref{fig:corrik4}, but here correlation coefficients
  relative to bin at $k=0.09-0.1 h$Mpc$^{-1}$ are shown.}
  \label{fig:corrik9}
\end{figure*}

\begin{figure*}
  \begin{center}{
  \includegraphics[width=0.48\columnwidth]{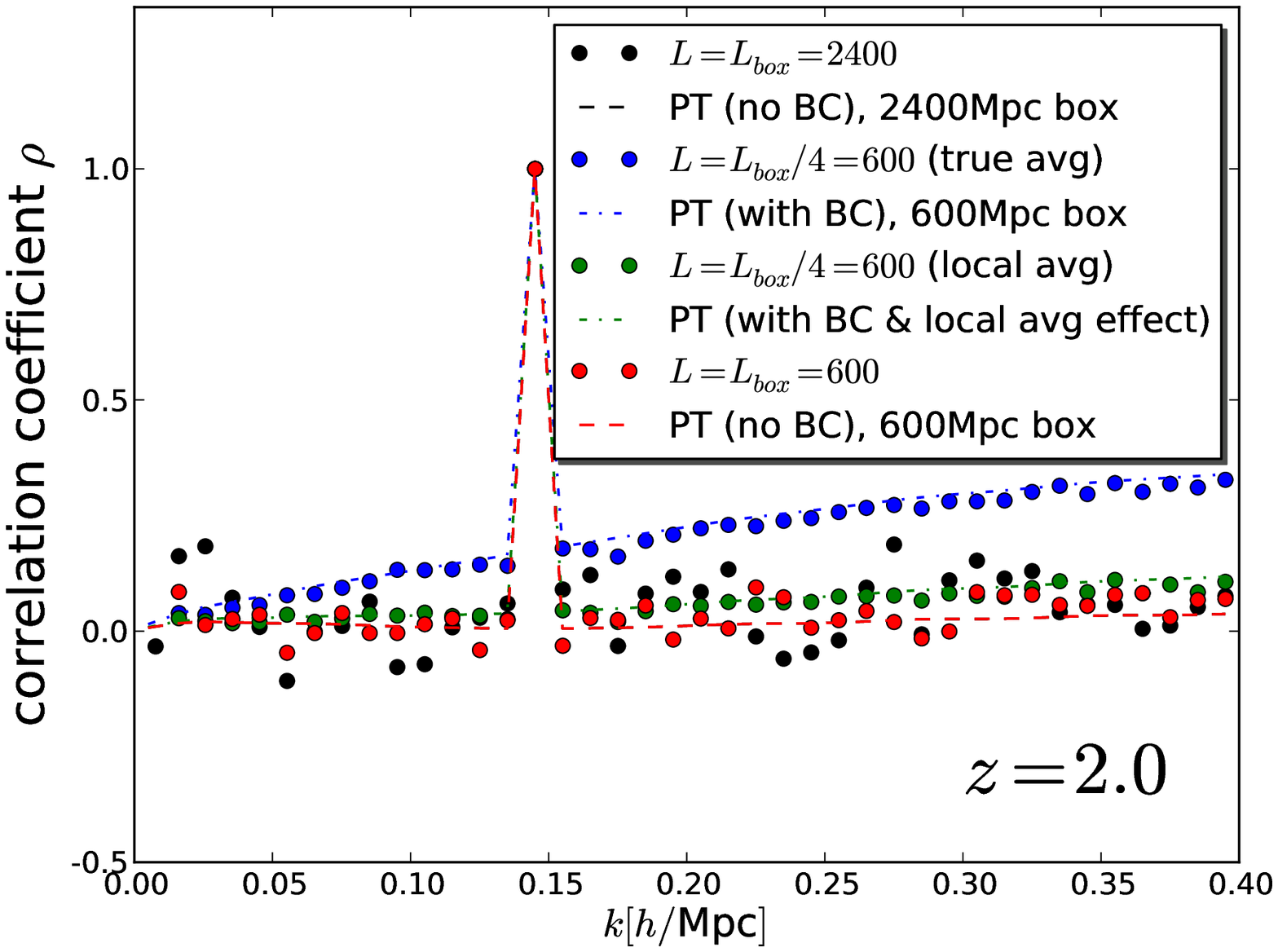}
  \includegraphics[width=0.48\columnwidth]{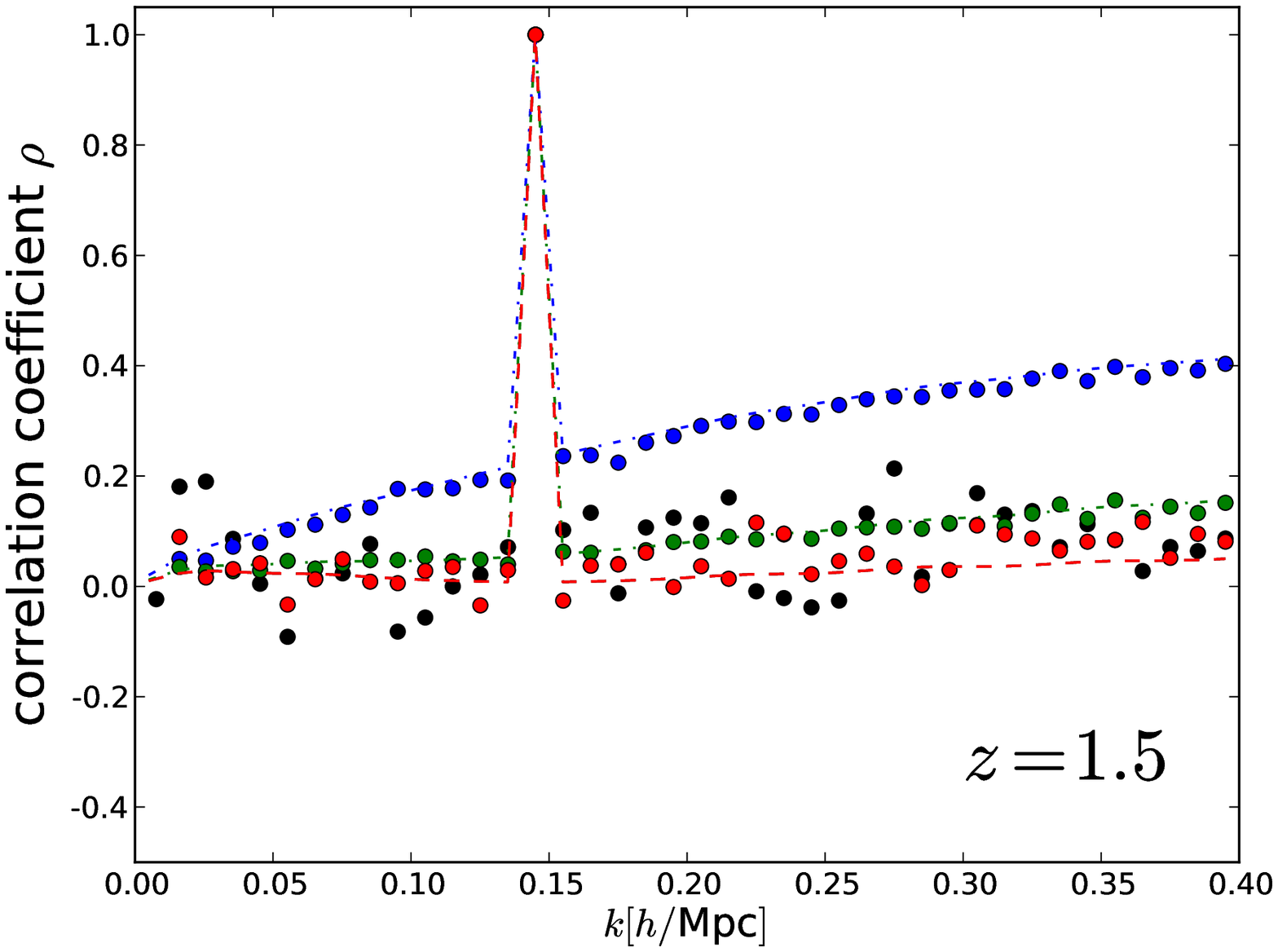}
  \includegraphics[width=0.48\columnwidth]{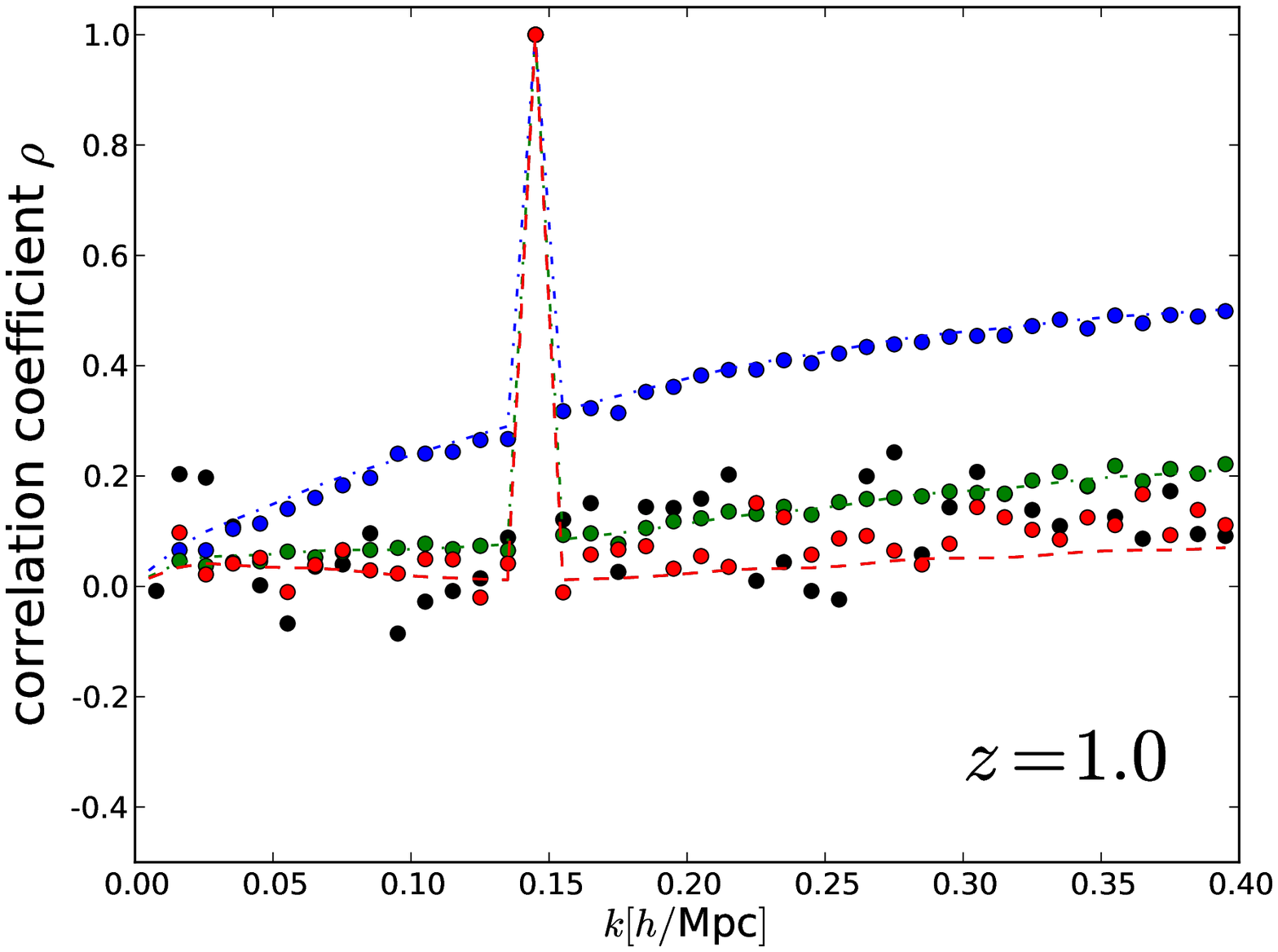}
  \includegraphics[width=0.48\columnwidth]{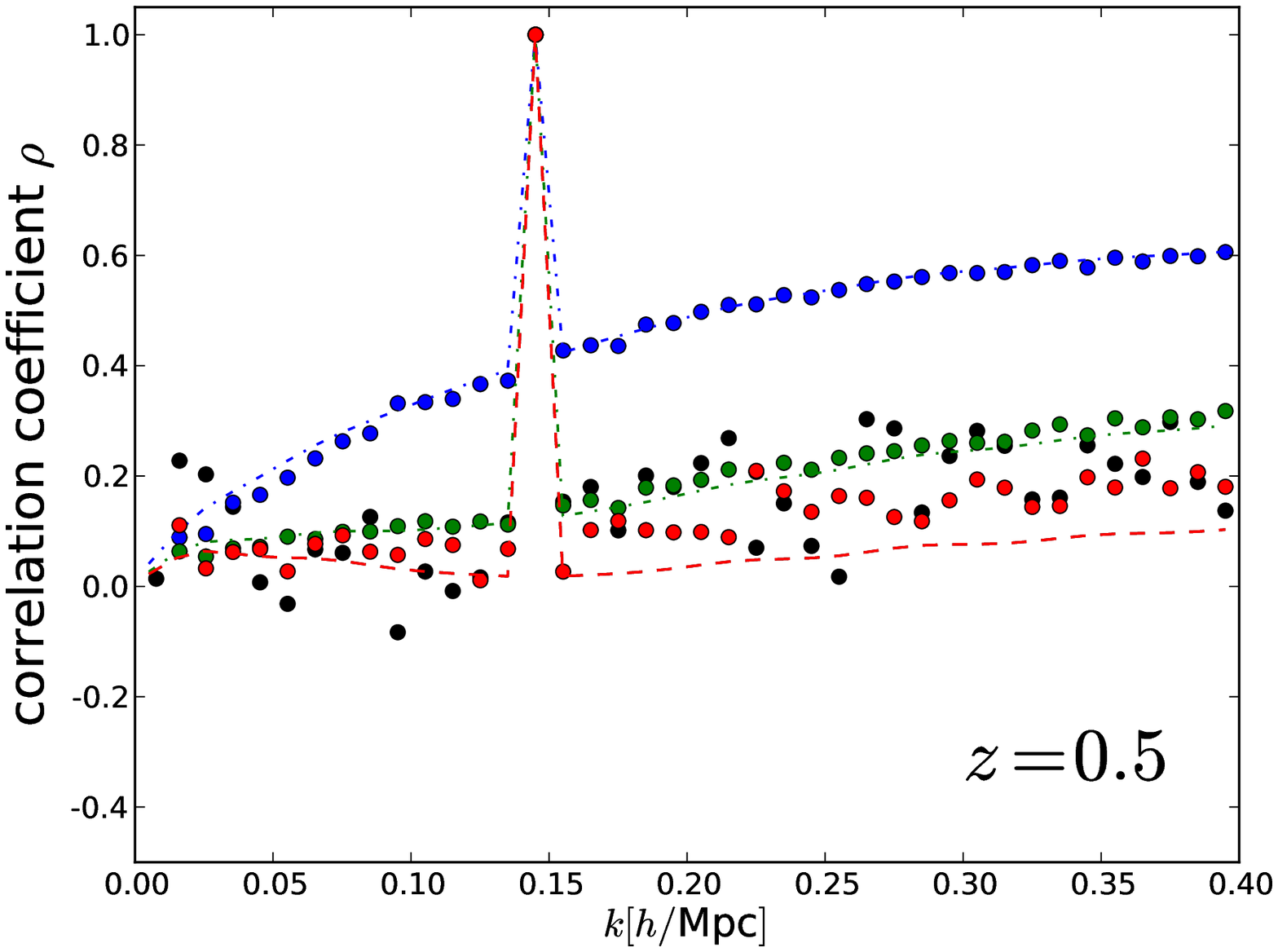}
  \includegraphics[width=0.48\columnwidth]{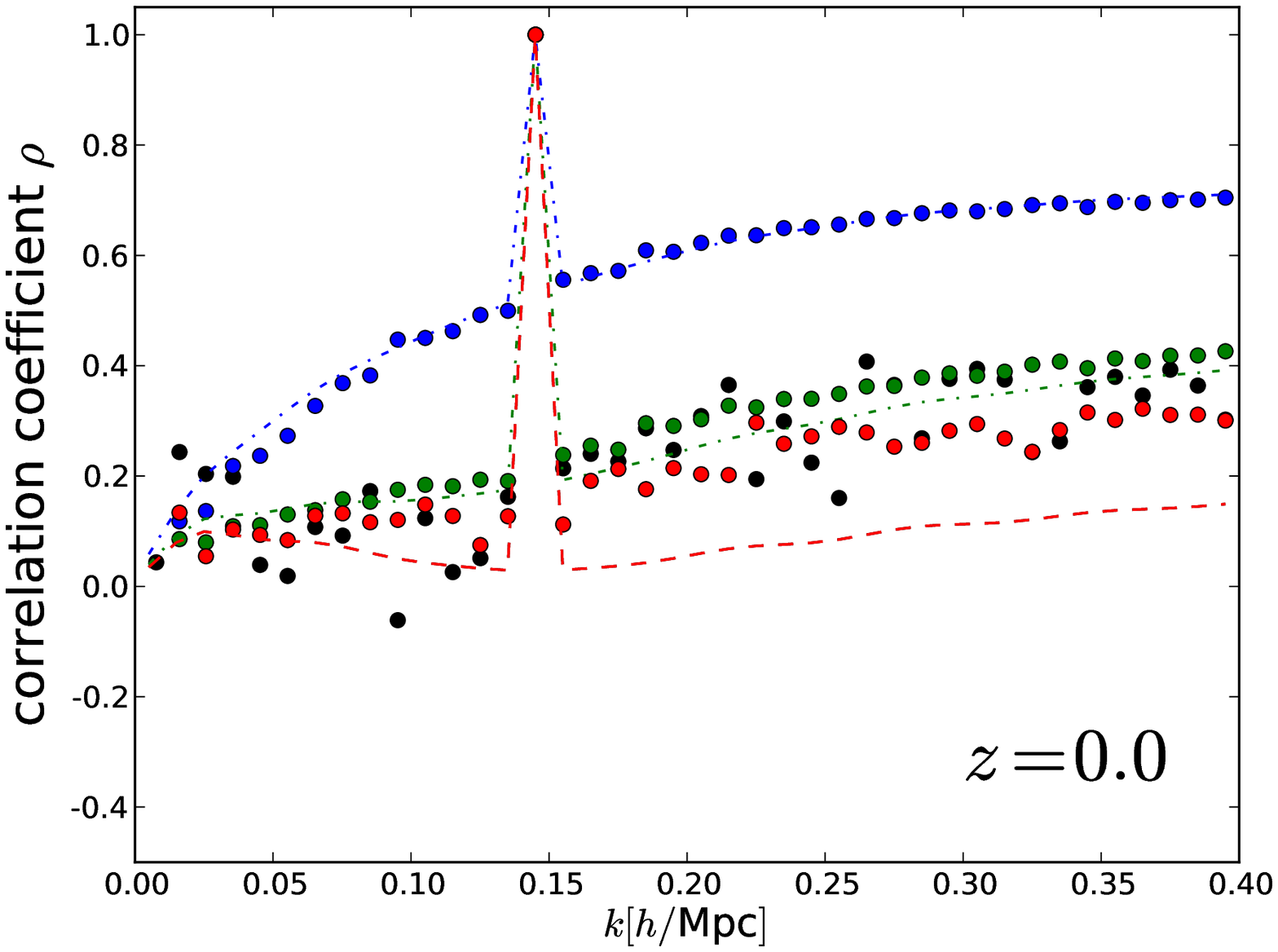}
  }
  \end{center}
  \caption{As in Fig.~\ref{fig:corrik4}, but here correlation coefficients
  relative to bin at $k=0.14-0.15 h$Mpc$^{-1}$ are shown.}
  \label{fig:corrik14}
\end{figure*}

We now comment on the three scenarios of interest individually.
\begin{itemize}

\item {\bf Case 1:} {\it periodic box}\\
The $L = 600 h^{-1}$Mpc and $L = 2400 h^{-1}$Mpc full-box simulation
results agree very well with each other, as expected because they are both described by Case 1 and have no beat coupling.
They also agree well with the analytic prediction. This confirms that excess covariance is not simply caused by the mere presence of
large modes. The following two cases show that what matters is the presence of modes larger than the ``survey'' volume.

\item
{\bf Case 2:} {\it subbox of periodic box}\\
Beat coupling introduces a large amount of excess variance and is well described by perturbation theory (for the range of
scales discussed above).

\item
{\bf Case 3:} {\it subbox of periodic box, using subbox mean}\\
As expected from theory, the local average effect undoes most of the excess variance due to beat coupling. Given the noise in
the simulation results, the Case 3 simulation results do not distinguish with much significance
the Case 3 theory prediction from the Case 1 prediction (but certainly from the Case 2 prediction).
However, the Case 3 simulation results do clearly display
a larger variance than the Case 1 results at large $k$, looking consistent with having $\sim 10 \%$ of the beat coupling
excess variance.

\end{itemize}

In Figs.~\ref{fig:corrik4} - \ref{fig:corrik14}, we quantify the off diagonal correlations
by plotting the correlation coefficients between bins,
\beq
\label{eq:defrho}
\rho_{i j} = \frac{{\bf C}_{i j}}{\sqrt{{\bf C}_{ii} \, {\bf C}_{jj}}},
\eeq
where we keep one bin fixed and let the value of $k$ corresponding to the other one
vary on the horizontal axis. The fixed bins are at $k \approx 0.05 h{\rm Mpc}^{-1}$ (Fig.~\ref{fig:corrik4}),
$k \approx 0.1 h{\rm Mpc}^{-1}$ (Fig.~\ref{fig:corrik9}) and $k \approx 0.15 h{\rm Mpc}^{-1}$ (Fig.~\ref{fig:corrik14}).

The conclusions for the correlation coefficients are similar to those drawn about the variances and again
the analytic and numerical results agree well. In fact, for Cases 2 \& 3, there now is good agreement up
to $k = 0.4 h$Mpc$^{-1}$ for all redshifts.
We do note that
the correlations relative to the $k \approx 0.05 h{\rm Mpc}^{-1}$ bin appear to be overestimated
somewhat by the analytic expressions. However, this is because the variance from simulations
in the $k = 0.04 - 0.05 h{\rm Mpc}^{-1}$
bin is somewhat high due to noise. Since this variance appears in the denominator of Eq.~(\ref{eq:defrho}),
it brings the simulation correlation coefficients down, thus explaining the slight disagreement with
the analytic result.
The other deviation is that the correlation coefficients relative to the bins at wave number
($k \approx 0.1 h{\rm Mpc}^{-1}$ and $k \approx 0.15 h{\rm Mpc}^{-1}$)
are higher than expected for Case 1 at low redshift. Whereas they should theoretically not
have a beat coupling contribution, these correlation coefficients behave as if they do get such a contribution
similar to the one in the scenario with beat coupling and the local average effect (Case 3).
This may simply be due to non-linear effects beyond the order included in our theoretical expressions.
Other than this, the agreement is very good.

\begin{figure*}
  \begin{center}{
  \includegraphics[width=0.48\columnwidth]{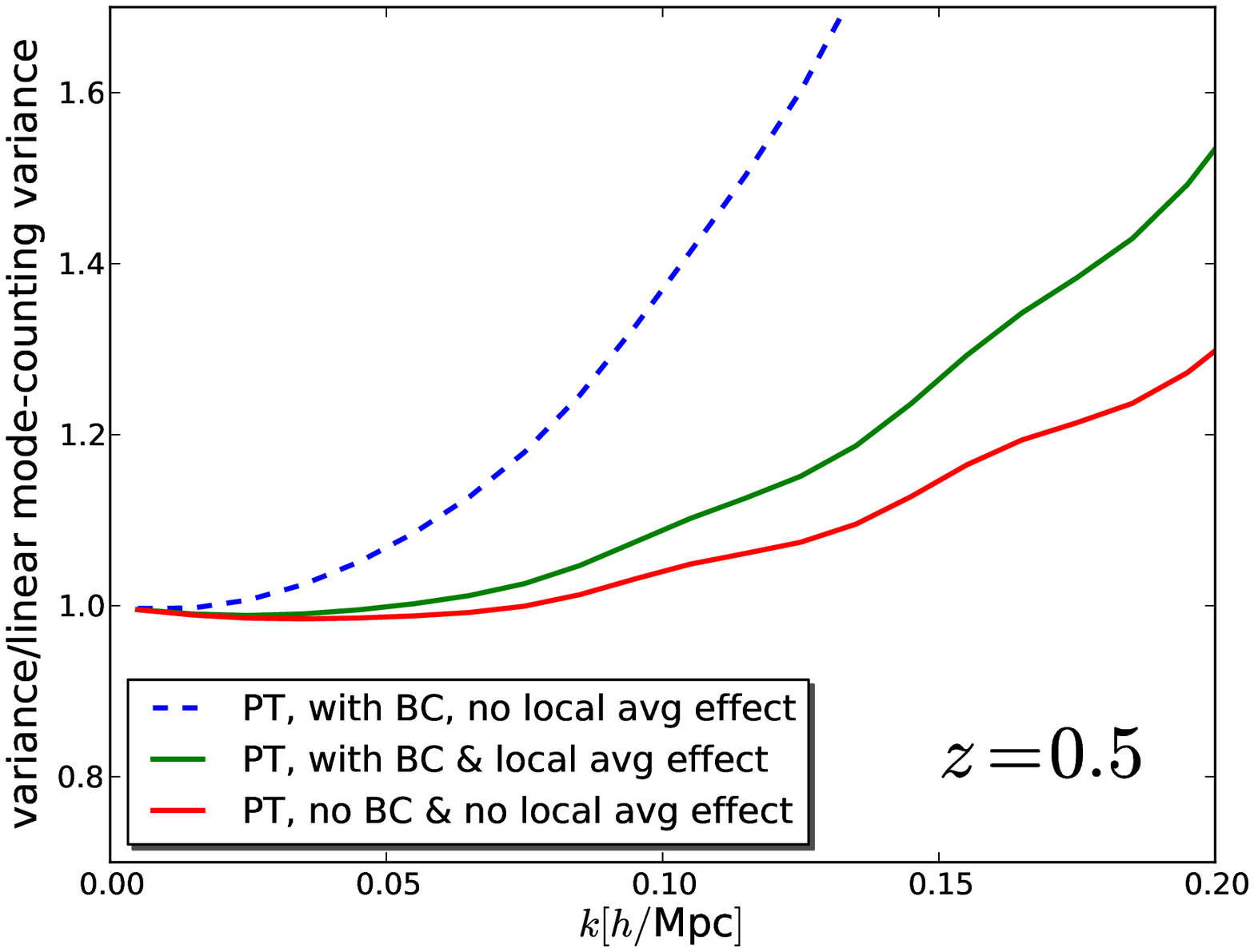}
  \includegraphics[width=0.48\columnwidth]{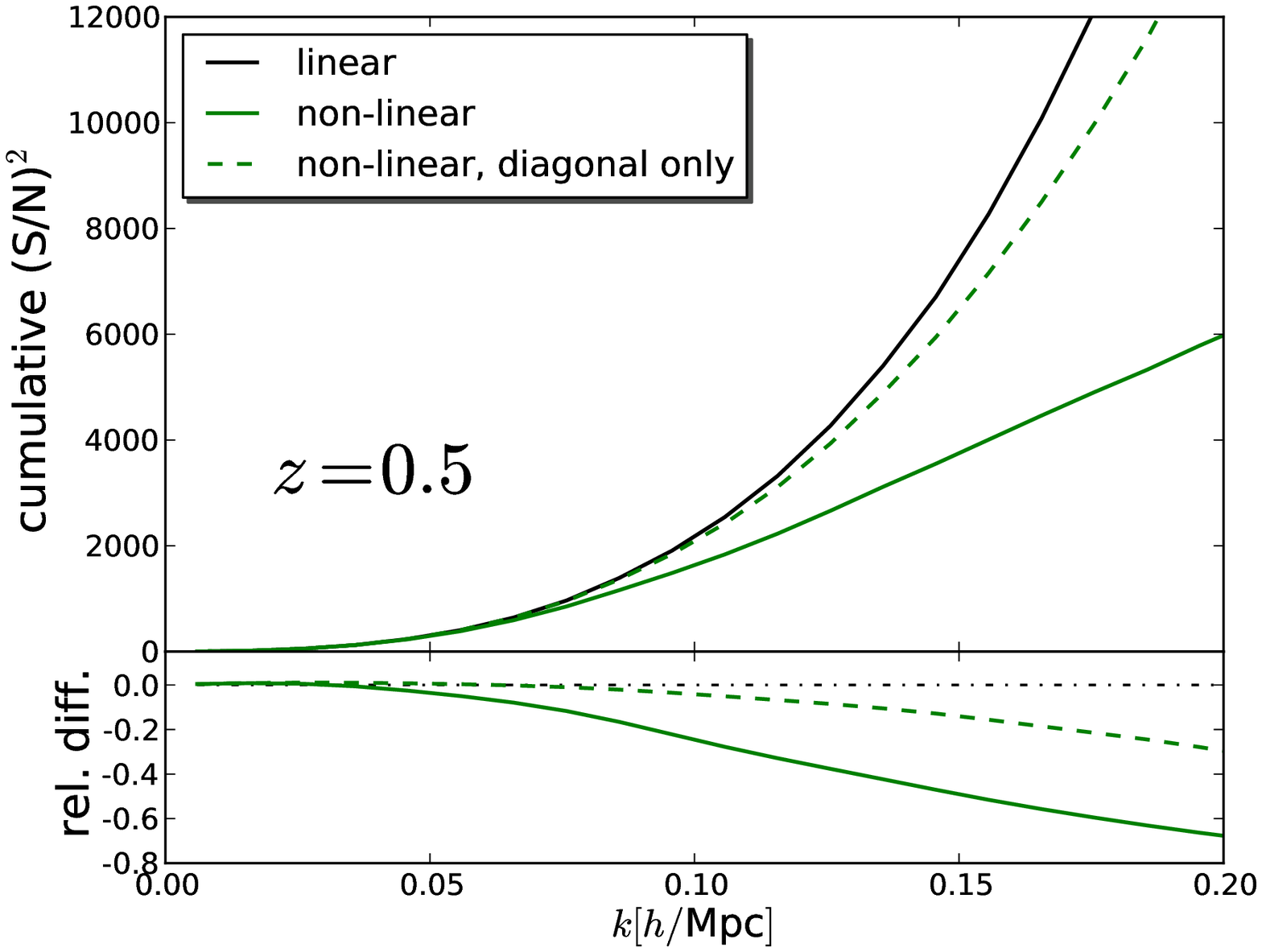}
  }
  \end{center}
  \caption{{\it Left Panel:} Power spectrum variance relative to variance based on linear power spectrum and mode
  counting ($2 P_{\rm lin}^2/N_k$ with $N_k$ the number of modes per bin) for redshift
  $z = 0.5$. The red curve represents
  the case where no modes larger than the survey are present (but other non-linear contributions
  are included). The
  most realistic case is depicted in green and includes the beat coupling and local average effects.
  Non-linear effects cause an increase of variance of up to $50 \%$ on scales $k < 0.2 h/$Mpc.
  The dashed blue curve shows the variance when the local average effect is not taken into account.
  {\it Right Panel:} Cumulative signal-to-noise squared in the amplitude of the non-linear
  power spectrum (see text for details) as a function of largest $k$-bin included (top; bottom
  shows fractional difference relative to the linear matrix).
  The non-linear, off-diagonal terms in the covariance matrix give the strongest contribution
  to the decrease in signal-to-noise at large $k$.
  }
  \label{fig:linvsnonlin}
\end{figure*}

To conclude this section, we use our analytic expressions to quantify the importance of non-linear
corrections to the covariance matrix on a range of scales relevant to large scale structure surveys.
In the left panel of Fig.~\ref{fig:linvsnonlin}, we show the theory based variance, at redshift zero, taking into account
all non-linear effects (green), and taking into account everything except for the beat coupling and local average
effects (red). The figure is the same as in Fig.~\ref{fig:vars}, except that the focus is on the theory curves
in the range $k = 0 - 0.2 h/$Mpc. We see that the full non-linear corrections change the variance at the
$25 \%$ level at $k = 0.15 h/$Mpc and by $\sim 50 \%$ at $k=0.2 h/$Mpc. For comparison, we also show in dashed blue
the variance if we do not take into account the reduction in excess variance due to the local
average effect. Neglecting to take the latter effect into account would lead to a gross overestimate of the non-linear
variance.

To quantify the effect of not only the non-linear corrections to the
diagonal of the covariance matrix, but also of the non-diagonal elements,
we next calculate
the squared signal-to-noise ratio,
\beq
\label{eq:sn2}
(S/N)^2(k_{\rm max}) \equiv \sum_{k_i, k_j < k_{\rm max}} {\bf C}^{-1}_{k_i, k_j} \, P(k_i) \, P(k_j).
\eeq
The sum here is over all pairs of power spectrum bins with both central $k$ values below $k_{\rm max}$.
For $P(k_i)$, we use the (non-linear) power spectrum in the fiducial model, so that $S/N$ can be
thought of as a {\it detection} (or amplitude) signal-to-noise.
The top figure of the right panel of Fig.~\ref{fig:linvsnonlin} shows this statistic using a (diagonal)
linear covariance matrix (based on mode counting; black curve), the full non-linear matrix (green),
and the non-linear matrix with the off-diagonal elements set to zero (dashed green). The bottom figure
shows the relative difference with the linear signal-to-noise squared. Compared to the linear case,
the non-linear, but diagonal covariance matrix indeed decreases the signal-to-noise, as expected from
the increase in variance seen in the right panel. However, the off-diagonal elements have
a significantly stronger effect and decrease the signal-to-noise even further.
The signal-to-noise gives an idea of how much Fisher matrix elements in a parameter forecast,
or $\chi^2$ values in a Monte Carlo chain, are affected by the non-linear corrections
to the covariance matrix.
This means that when a power spectrum study, whether with real data or in a Fisher forecast,
includes $k$ modes well into the non-linear regime, it is not enough to just include
non-linear corrections into the variance of the power spectrum. Instead, the full non-linear
covariance matrix needs to be taken into account.

\section{General Survey Geometry}
\label{sec:gen}

So far, we have found it convenient to describe the statistics of the density field
in terms of a discrete set of overdensity modes. Whereas this is exact when considering the modes
of a full periodic box, this should be considered an effective description in the case of subboxes,
justified by their simple geometry.
In this section, we work out the formalism for a general survey geometry in terms of the
continuum of Fourier modes that exist in an infinite universe. We closely follow the
notation and results of \cite{FKP}, but will add to this the trispectrum contributions arising
in perturbation theory. The discussion in this section will provide justification
for the discrete description of the previous sections. One issue that we will pay particular attention to
is the motivation for using the diagonal mode-counting expression for the disconnected part of the covariance matrix
in Eq.~(\ref{eq:covgen}). As we will see, in the FKP formalism, the window function correlates power spectrum
estimators in different bins (even in the Gaussian case), and the mode counting expression is not accurate.
We will explain why mode counting was justified in the previous sections, argue that for a realistic survey
the FKP description including cross correlations is the relevant one, and finally test the latter description against
simulations.

We first define our Fourier convention as
\beq
\delta({\bf x}) = \int \frac{d^3 \k}{(2 \pi)^3} \, e^{\imag \k \cdot {\bf x}} \, \delta(\k).
\eeq
The statistics of the continuum of Fourier modes are given by
\beqa
\langle \delta(\k) \, \delta(\kp) \rangle &=& (2\pi)^3 \, P(k) \, \delta^D(\k + \kp) \nonumber \\
\langle \delta(\k) \, \delta(\kp) \, \delta(\kpp) \rangle &=& (2\pi)^3 \, B(\k, \kp, \kpp) \, \delta^D(\k + \kp + \kpp) \nonumber \\
\langle \delta(\k) \, \delta(\kp) \, \delta(\kpp) \, \delta({\bf k'''}) \rangle_c &=& (2\pi)^3 \, T(\k, \kp, \kpp, {\bf k'''}) \, \delta^D(\k + \kp + \kpp + {\bf k'''}) \nonumber \\
&\dots&
\eeqa

For a survey with background number density $\bar{n}({\bf x})$, consider then the weighted density field
\beq
F({\bf x}) = \frac{\bar{n}({\bf x}) \, w({\bf x}) \, \delta({\bf x})}{\left[\int d^3 {\bf x} \, \bar{n}^2({\bf x}) \, w^2({\bf x})\right]^{\ha}} \equiv G({\bf x}) \, \delta({\bf x}),
\eeq
where $w({\bf x})$ is a weight function that can be chosen to maximize signal to noise. We will ignore shot noise in the following,
but it is straightforward to include it in the Gaussian approximation (see FKP).
The Fourier transform is given by $F(\k) = (G * \delta)(\k)$, where ``$*$'' indicates a convolution, and
the power spectrum can be estimated by $\hat{P}(\k) = |F(\k)|^2$, so that
\beq
\label{eq:pkexp}
\langle \hat{P}(\k)\rangle = \int \frac{d^3 {\bf k}'}{(2 \pi)^3} \, P(k') \, |G(\k - \kp)|^2.
\eeq
The measured power spectrum is thus a weighted average of the true spectrum,
with a weight function of width of order the fundamental mode of the survey,
i.e.~$\Delta k \sim \pi/L$ if $L$ is the typical scale of the survey.
As a relevant example, the window function for a cubic box of
side $L$ ($\bar{n} \equiv {\rm const}$, $w \equiv 1$ inside the box and zero outside) is given by
\beq
G(\k) = L^{3/2} \, j_0\left( L k_x/2 \right) \, j_0\left( L k_y/2 \right) \, j_0\left( L k_z/2 \right),
\eeq
with $j_0(x) = \sin x/x$ the zeroth spherical Bessel function.
If the true spectrum varies little across this range of scales, we get
\beq
\langle \hat{P}(\k) \rangle \approx P(k).
\eeq

\subsection{Covariances - disconnected (Gaussian) contribution}

We first consider the contribution to the covariance between FKP power spectrum estimators arising
from products of two-point functions. This is the only contribution for a Gaussian density field and we
can follow FKP for its description.

We start from the two-point function
\beq
\langle F(\k) \, F^*(\kp)\rangle = \int \frac{\kpp}{(2 \pi)^3} \, P(k'') \, G(\k - \kpp) \, G^*(\kp - \kpp) \approx P(k) \, Q(\kp - \kp),
\eeq
where the second equality is true in the same limit where the (expectation value of the) power spectrum estimator equals the true spectrum,
and $Q(\k)$ is the Fourier transform of the (normalized) squared window function,
\beq
Q({\bf x}) = \frac{\bar{n}^2({\bf x}) \, w^2({\bf x})}{\int d^3 {\bf x} \, \bar{n}^2({\bf x}) \, w^2({\bf x})}.
\eeq
Hence, the covariance
\beq
\langle \delta \hat{P}(\k) \, \delta \hat{P}(\kp) \rangle \approx |P(k) \, Q(\k - \kp)|^2 + (\kp \to -\kp).
\eeq
In FKP, the power spectrum in a bin $i$ is given by
\beq
\label{eq:fkpbin}
\hat{P}_i = \int_i \frac{d^3 \k}{V_{k,i}} \, \hat{P}(\k),
\eeq
where $V_{k,i}$ is the $k$-volume of bin $i$,
so that the covariance between (isotropic) bins is
\beq
\label{eq:fkpgen}
{\bf C}_{ij} = 2 \int_i \frac{d^3 \k}{V_{k,i}} \, \int_j \frac{d^3 \kp}{V_{k,j}} \, P^2(k) \, |Q(\k - \kp)|^2.
\eeq
Because of the extended nature of the window function $|G|^2$, there will thus be correlations between different bins.
In the limit that the bin width is much larger than the width of this window function, these correlations are negligible and one
can make a further simplification by integrating out $|G|^2$. Using
\beq
\int \frac{d^3 \k}{(2 \pi)^3} \, |Q(\k)|^2 = \frac{\int d^3 {\bf x} \, \bar{n}^4({\bf x}) \, w^4({\bf x})}{\left[ \int d^3 {\bf x} \, \bar{n}^2({\bf x}) \, w^2({\bf x}) \right]^2} \equiv V^{-1}_{\rm eff}
\eeq
one ends up with
\beq
{\bf C}_{ij} = \frac{2 P^2(k_i)}{N_k} \, \delta^K_{ij}, \quad \quad N_k = \frac{V_{k,i}}{(2\pi)^3/V_{\rm eff}},
\eeq
where $k_i$ is a typical mode in the $i$-th bin and we have further assumed that the power spectrum varies little across a bin.
This is the well know mode-counting result. The expression for the effective volume is further simplified in FKP by choosing an
optimal weight function, but we will stick to general $w({\bf x})$.
Note that for our cubic box (or any other geometry with constant background number density and weighting),
$V_{\rm eff} = V$.

\begin{figure*}
  \includegraphics[width=0.48\columnwidth]{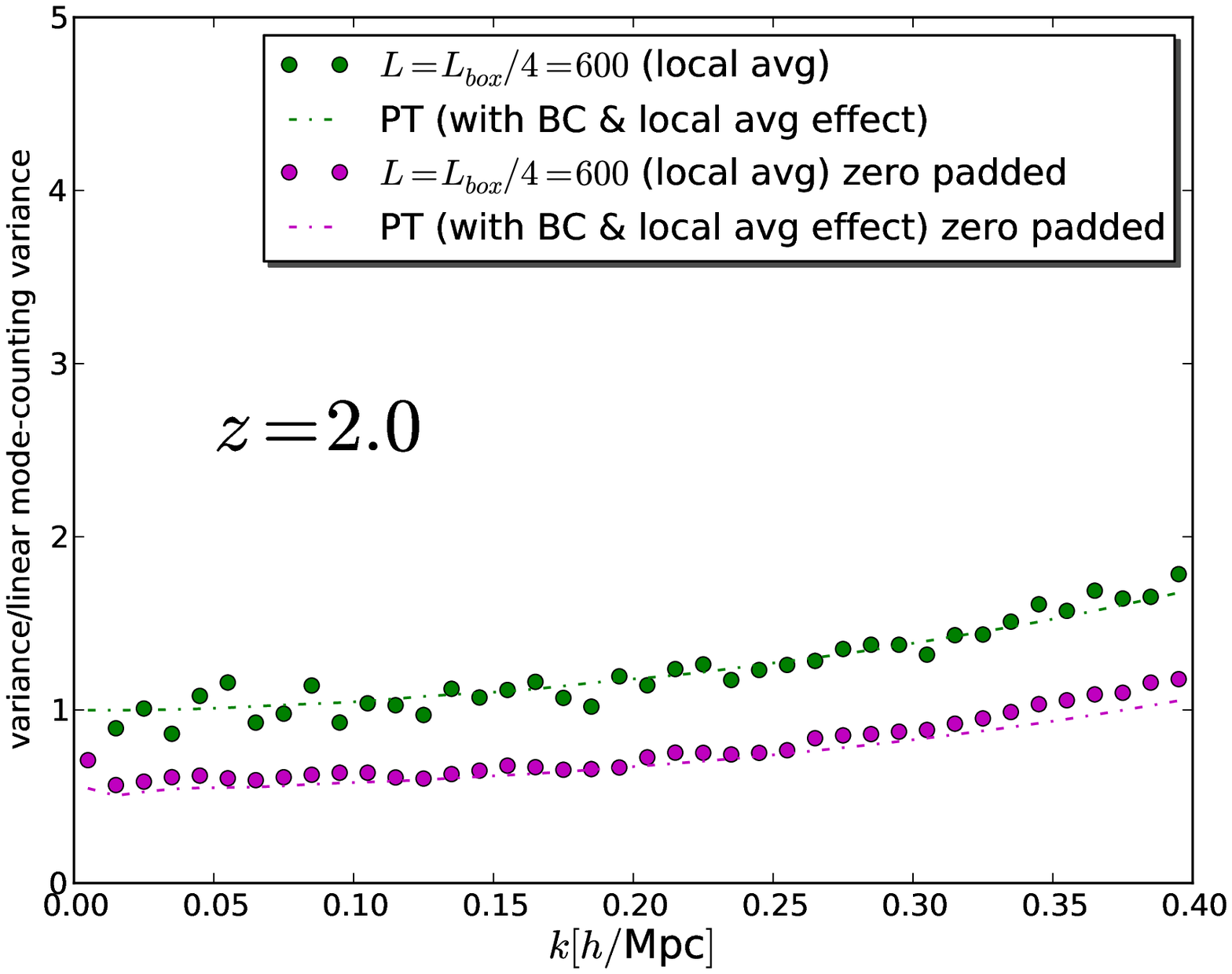}
  \includegraphics[width=0.48\columnwidth]{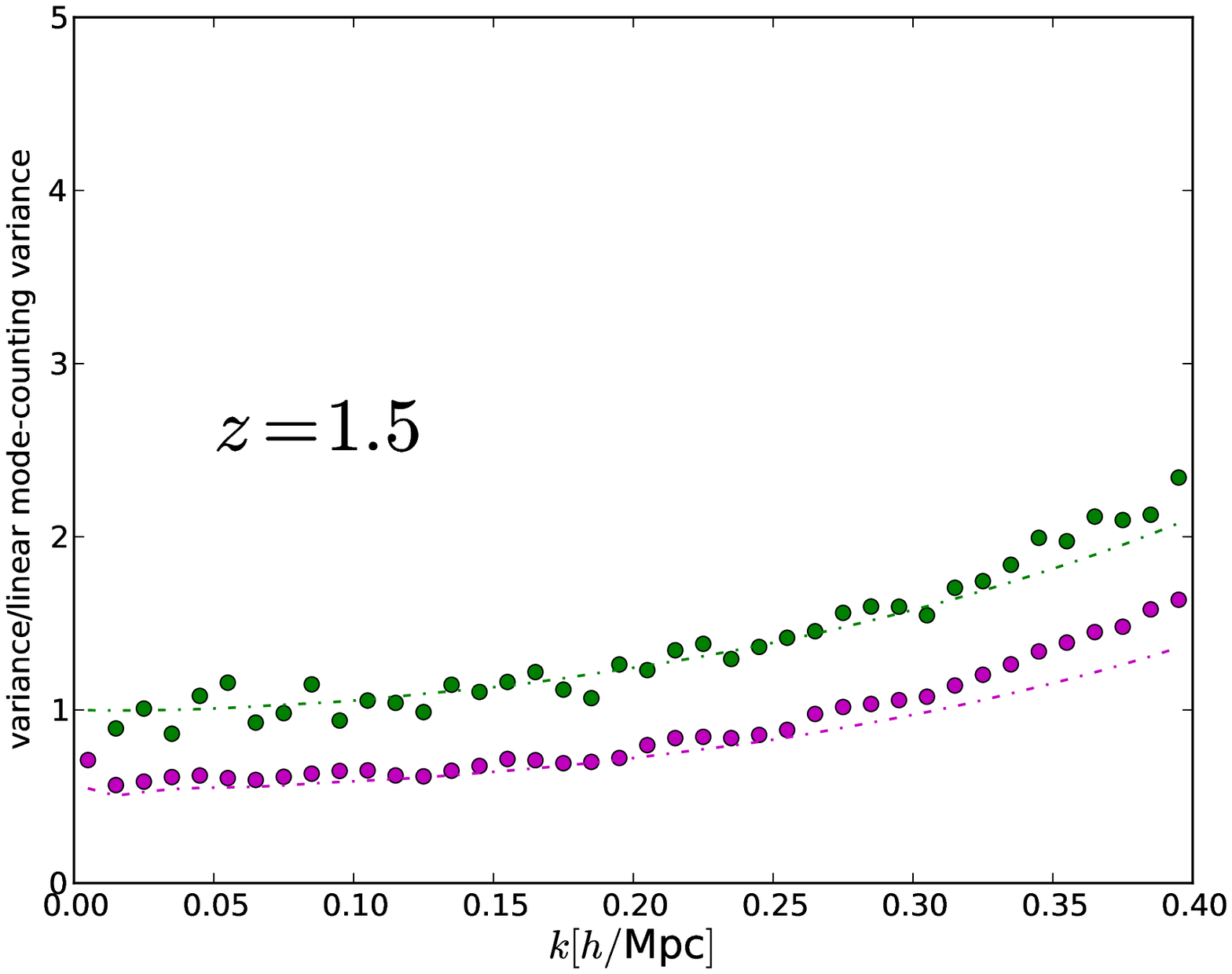}
  \includegraphics[width=0.48\columnwidth]{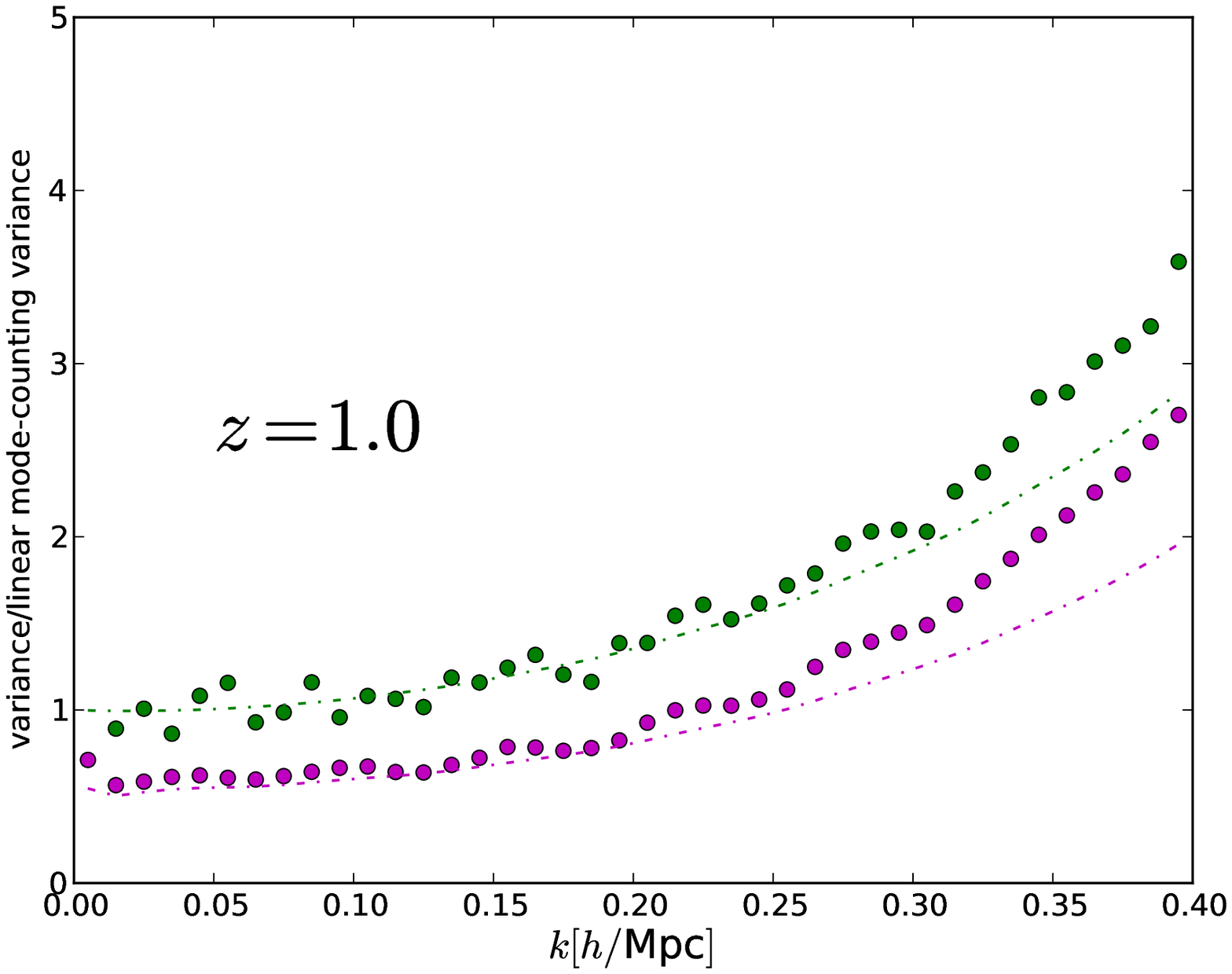}
  \includegraphics[width=0.48\columnwidth]{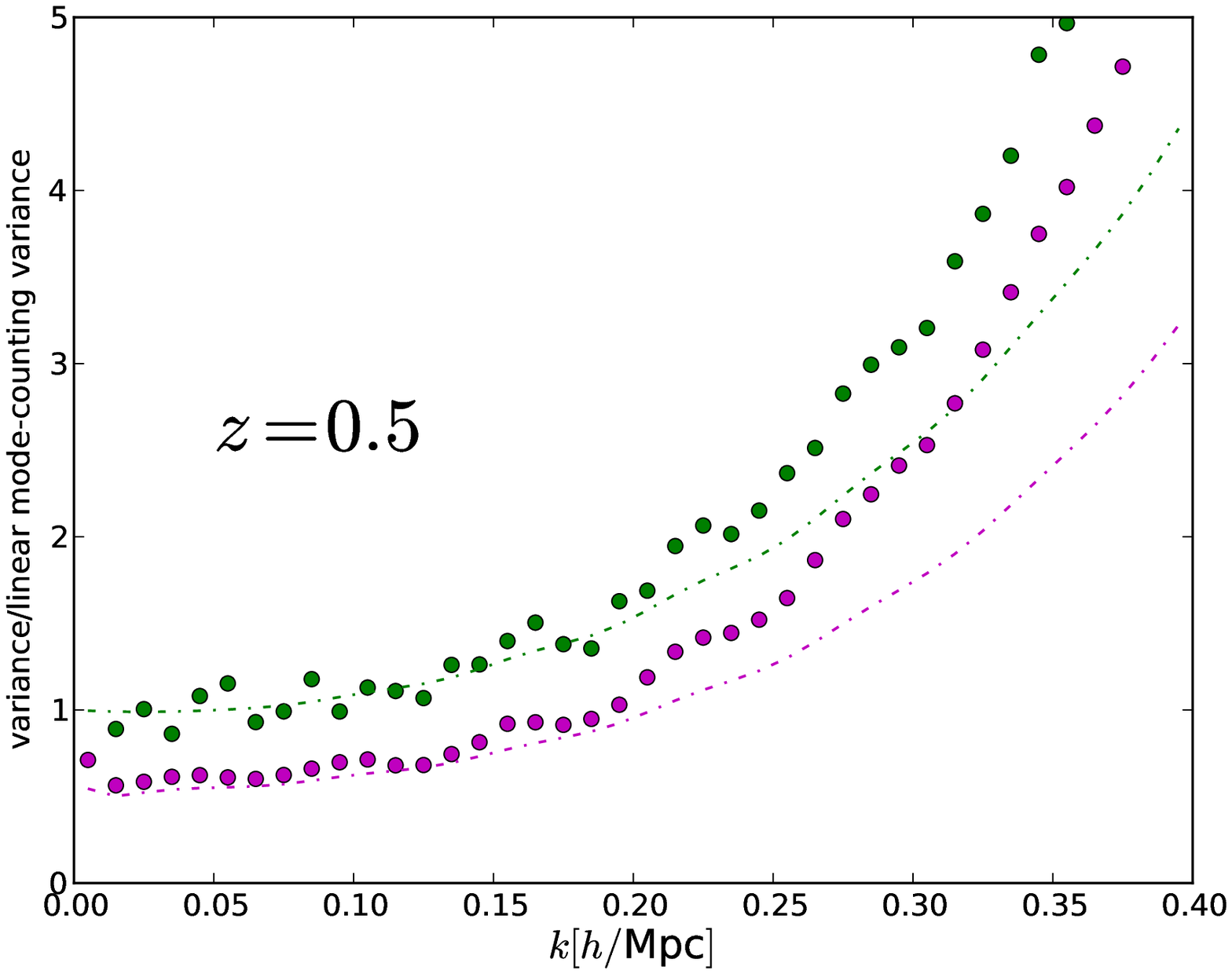}
  \includegraphics[width=0.48\columnwidth]{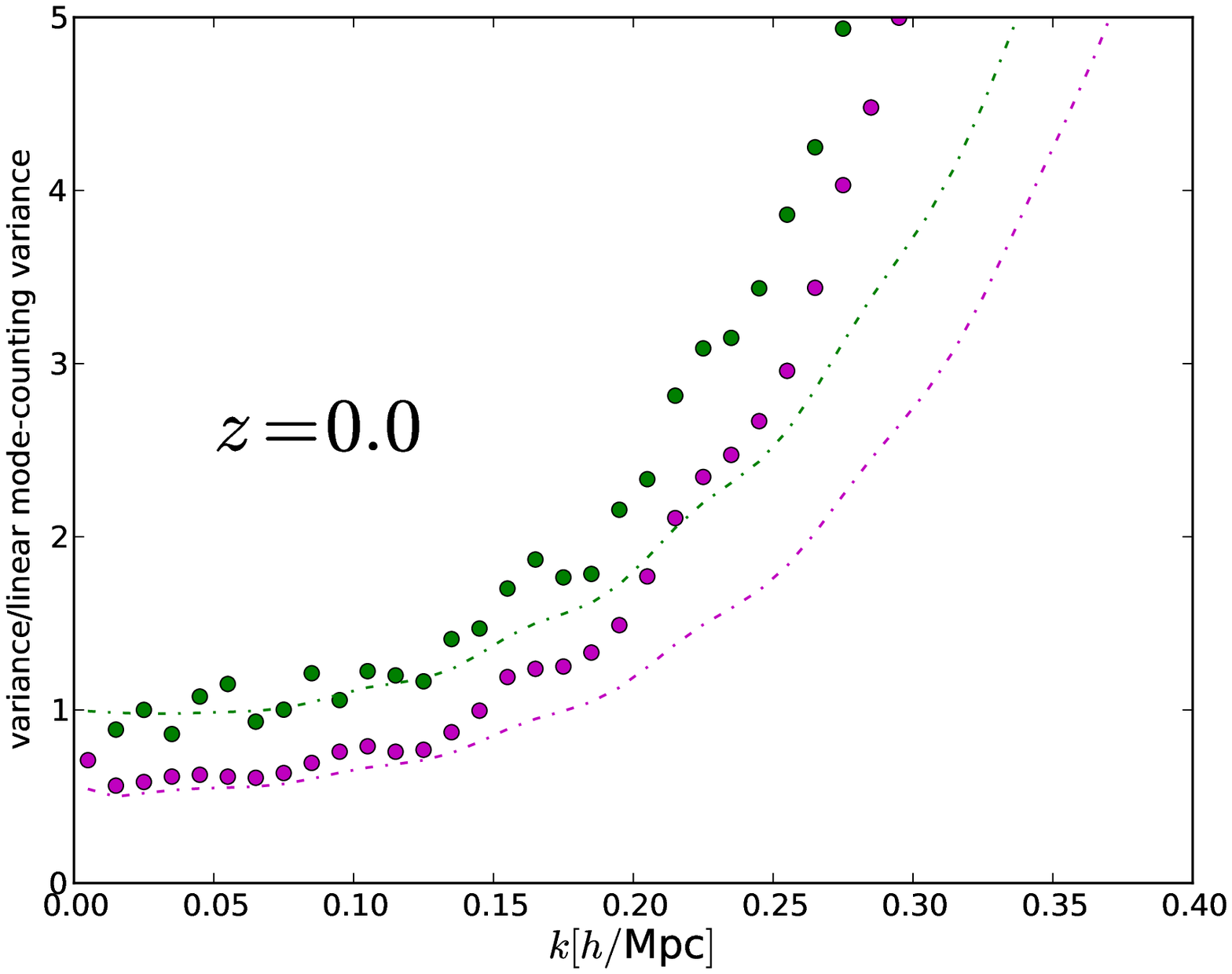}
  \caption{Power spectrum variance relative to (linear) mode-counting variance at redshifts
  $z = 0 - 2$ for Case 3, i.e.~with both beat coupling
  {\it and} the local average effect included. Comparison is between variances in spectrum
  averaged over bins (of width $\Delta k = 0.01 h$Mpc$^{-1}$) with fine sampling (as in FKP, Eq.~(\ref{eq:sparse avg})),
  which has mode mixing due to the window function (magenta dots),
  and average over modes that are multiples of fundamental mode of subbox (Eq.~(\ref{eq:fkpbin})),
  which decorrelates the power spectrum estimator in bins (green dots).
  We also refer to the former case as the ``zero padding'' covariance matrix, as the power is estimated from
  simulations by using the full $2400 h^{-1}$Mpc simulation box to do the Fourier transform, but zero padding
  the region outside the $600 h^{-1}$Mpc subbox.
  Dashed lines show theory predictions from Eq.~(\ref{eq:c3}) (green)
  and Eq.~(\ref{eq:c3 geom}) (magenta).
  The theory covariance matrix thus successfully includes window function effects
  and non-linear effects.
  On large scales (small $k$), the green curve and dots approach unity because
  the bin averaging scheme causes the variance to be given by mode counting and because non-linear effects vanish.
  The more realistic zero padding case however has significantly lower variances,
  but strong correlations between neighboring bins (even on linear scales) so that the total information
  content is the same.
  }
  \label{fig:varszp}
\end{figure*}

\begin{figure*}
  \begin{center}{
  \includegraphics[width=0.48\columnwidth]{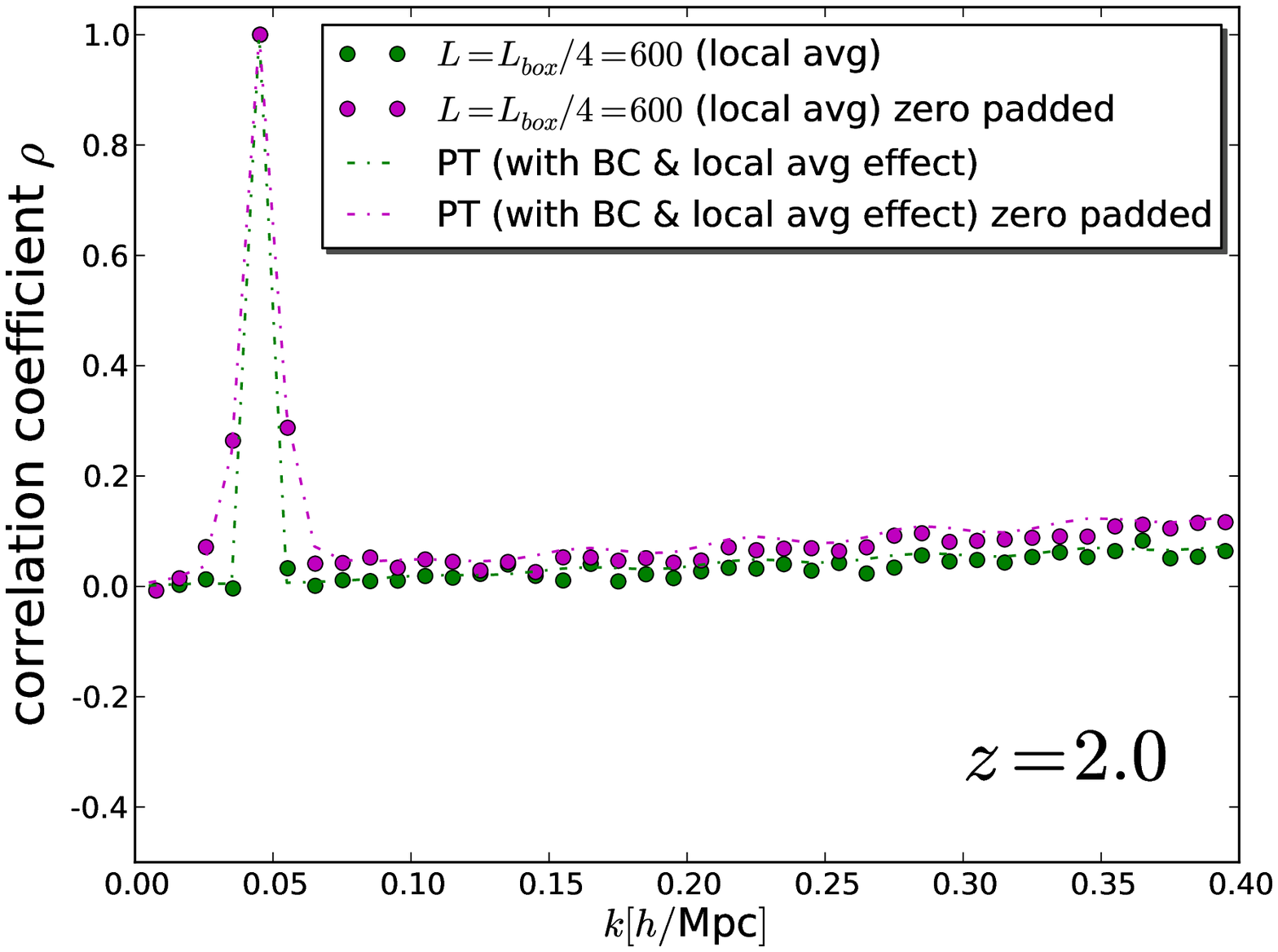}
  \includegraphics[width=0.48\columnwidth]{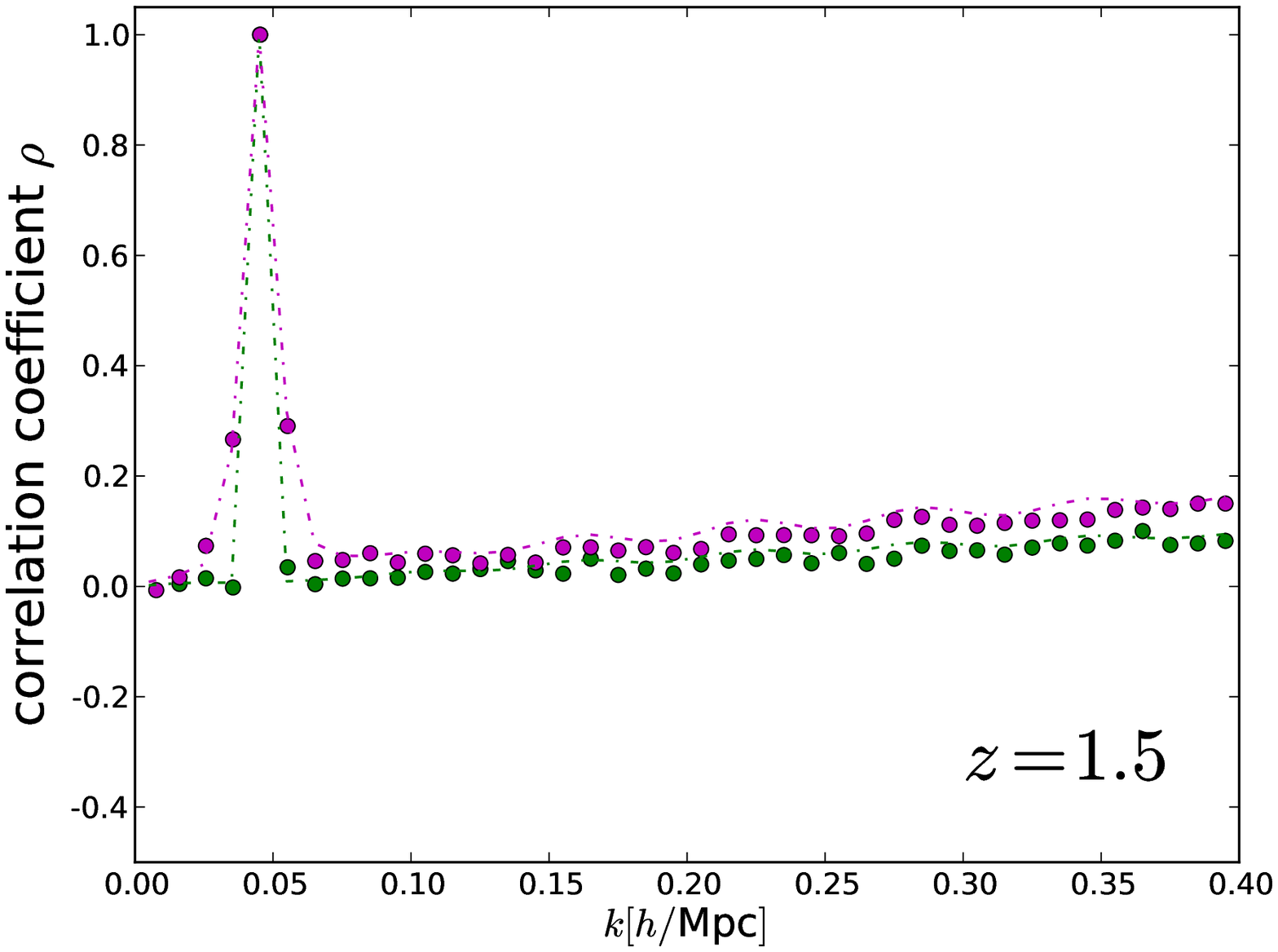}
  \includegraphics[width=0.48\columnwidth]{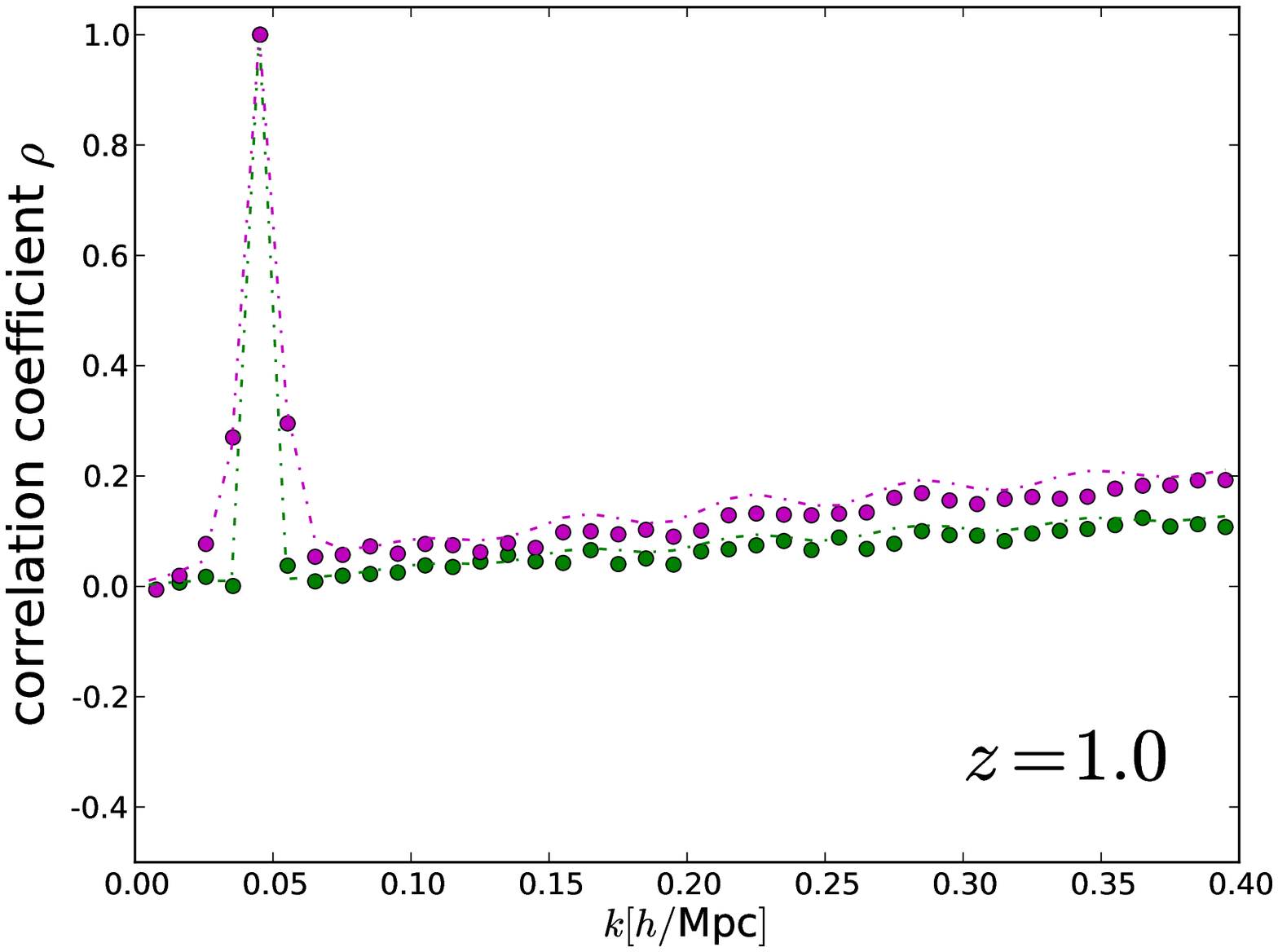}
  \includegraphics[width=0.48\columnwidth]{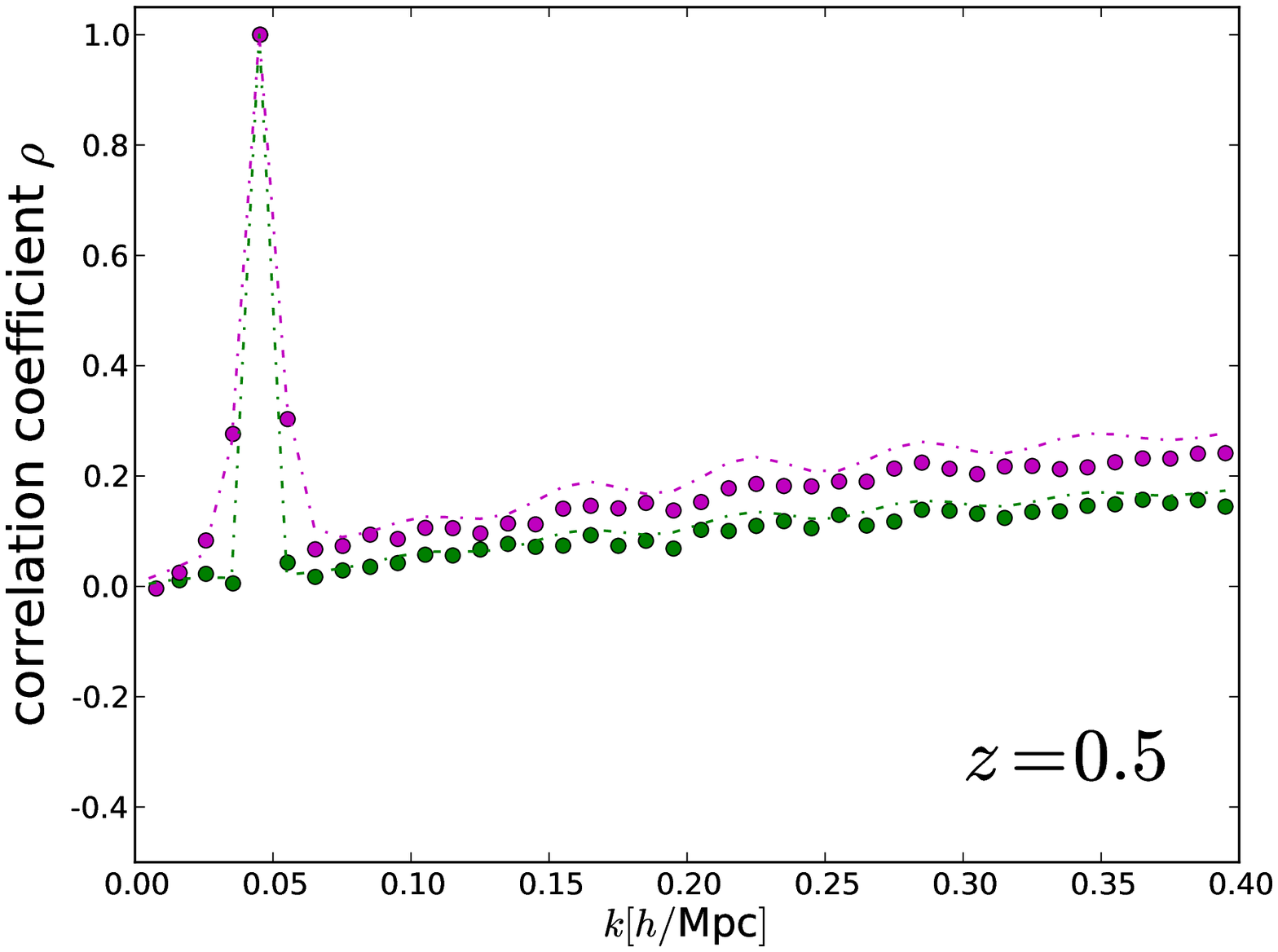}
  \includegraphics[width=0.48\columnwidth]{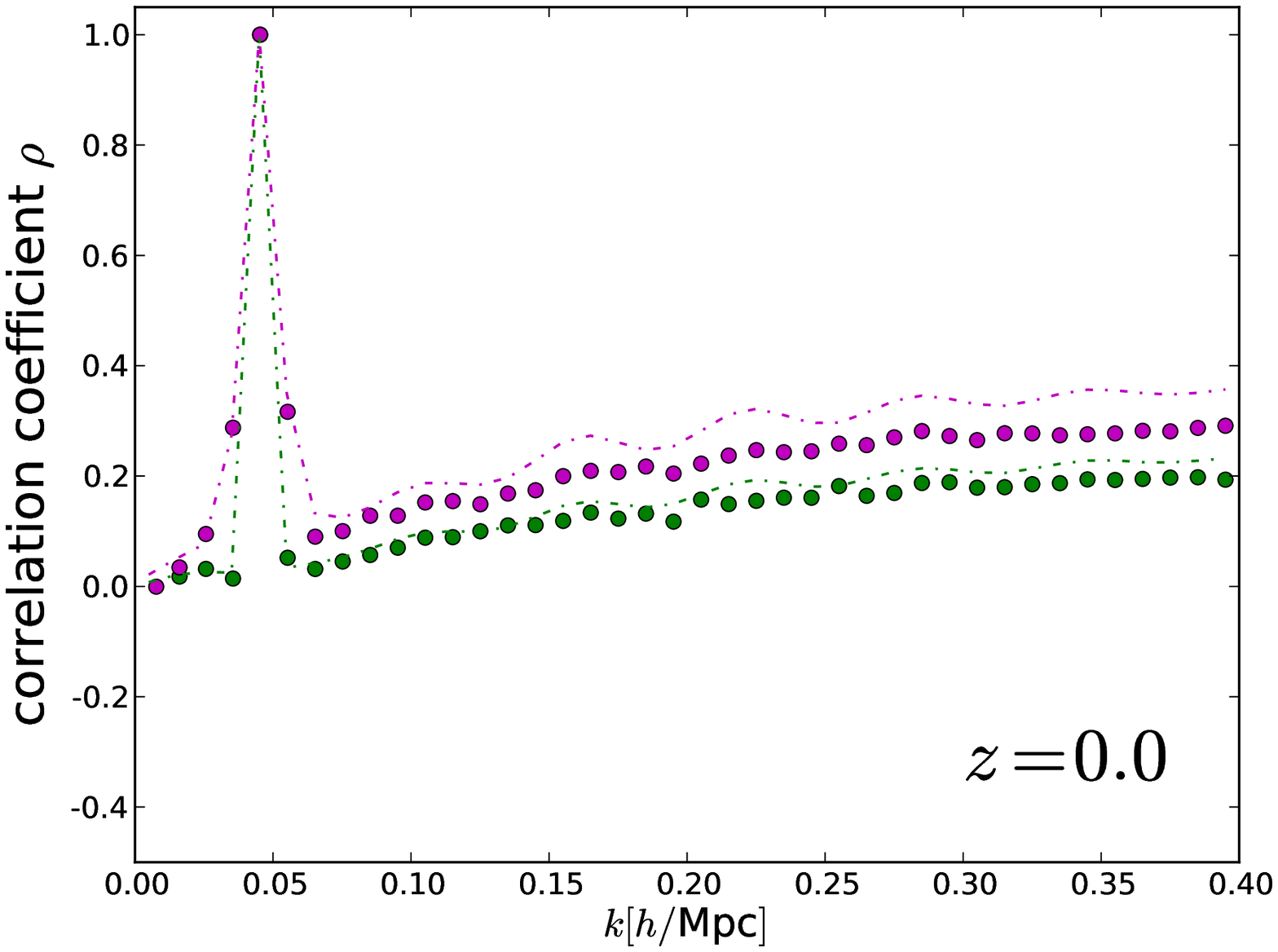}
  }
  \end{center}
  \caption{As Fig.~\ref{fig:varszp}, but here showing the correlation coefficients
  $\rho_{ij} \equiv {\bf C}_{ij}/\sqrt{{\bf C}_{ii} \, {\bf C}_{jj}}$
  of $i$-th bin at $k$ on x-axis, relative to $j$-th bin at $k=0.04-0.05 h$Mpc$^{-1}$.
  The theory predictions again agree quite well with the simulation results
  although more so for correlations relative to bins at larger $k$, as
  shown in Figs \ref{fig:corrik9zp} and \ref{fig:corrik14zp}.
  Note in particular the large correlation between neighboring bins
  in the case where the bin average is based on fine sampling/zero padding (Eq.~(\ref{eq:fkpbin}), magenta).
  This is due to mode mixing by the window function and is separate from the cross
  correlations due to non-linear evolution. It is absent when estimators are averaged
  using the sparse bin sampling (green), Eq.~(\ref{eq:sparse avg}).
  }
  \label{fig:corrik4zp}
\end{figure*}

\begin{figure*}
  \begin{center}{
  \includegraphics[width=0.48\columnwidth]{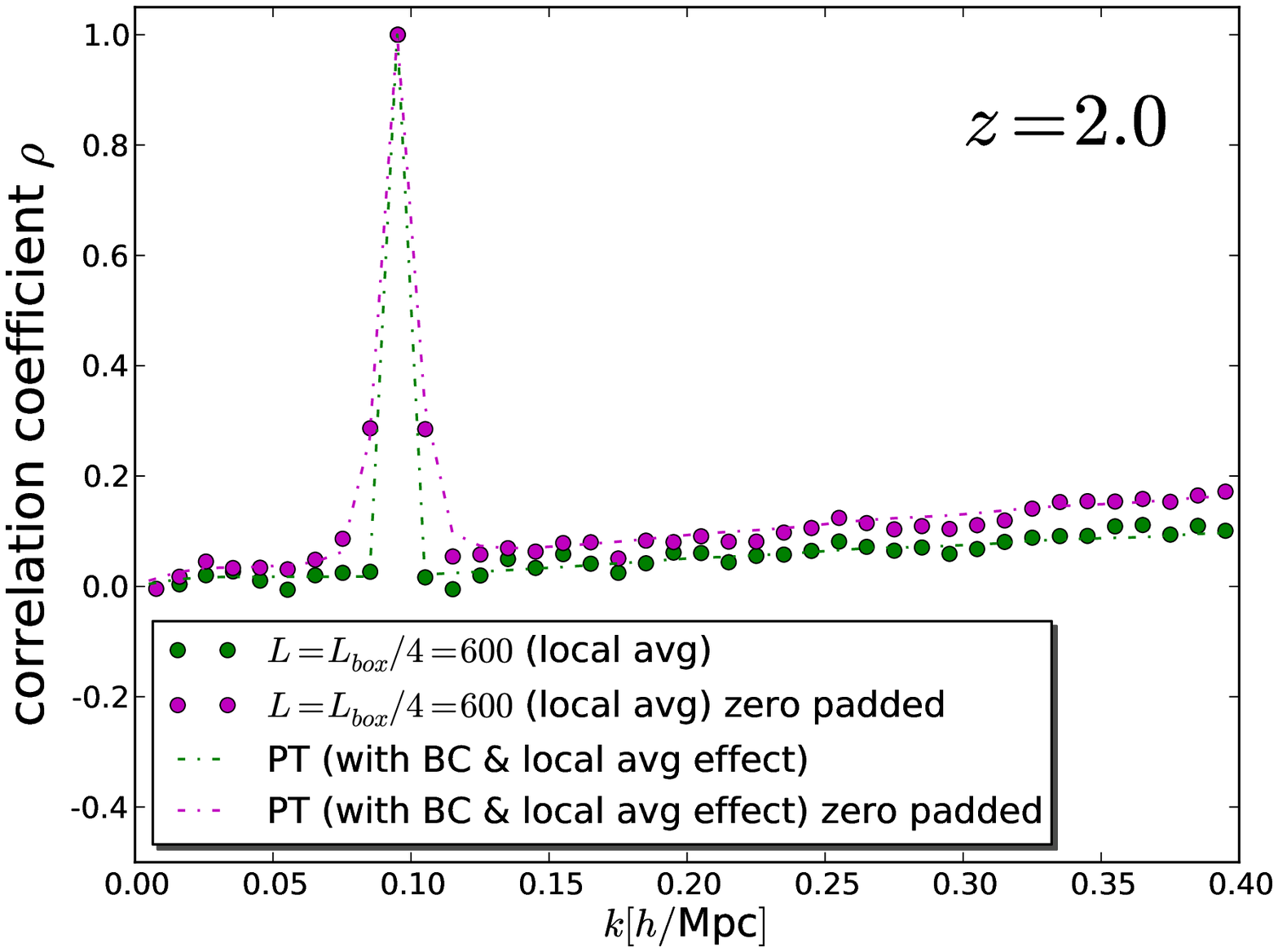}
  \includegraphics[width=0.48\columnwidth]{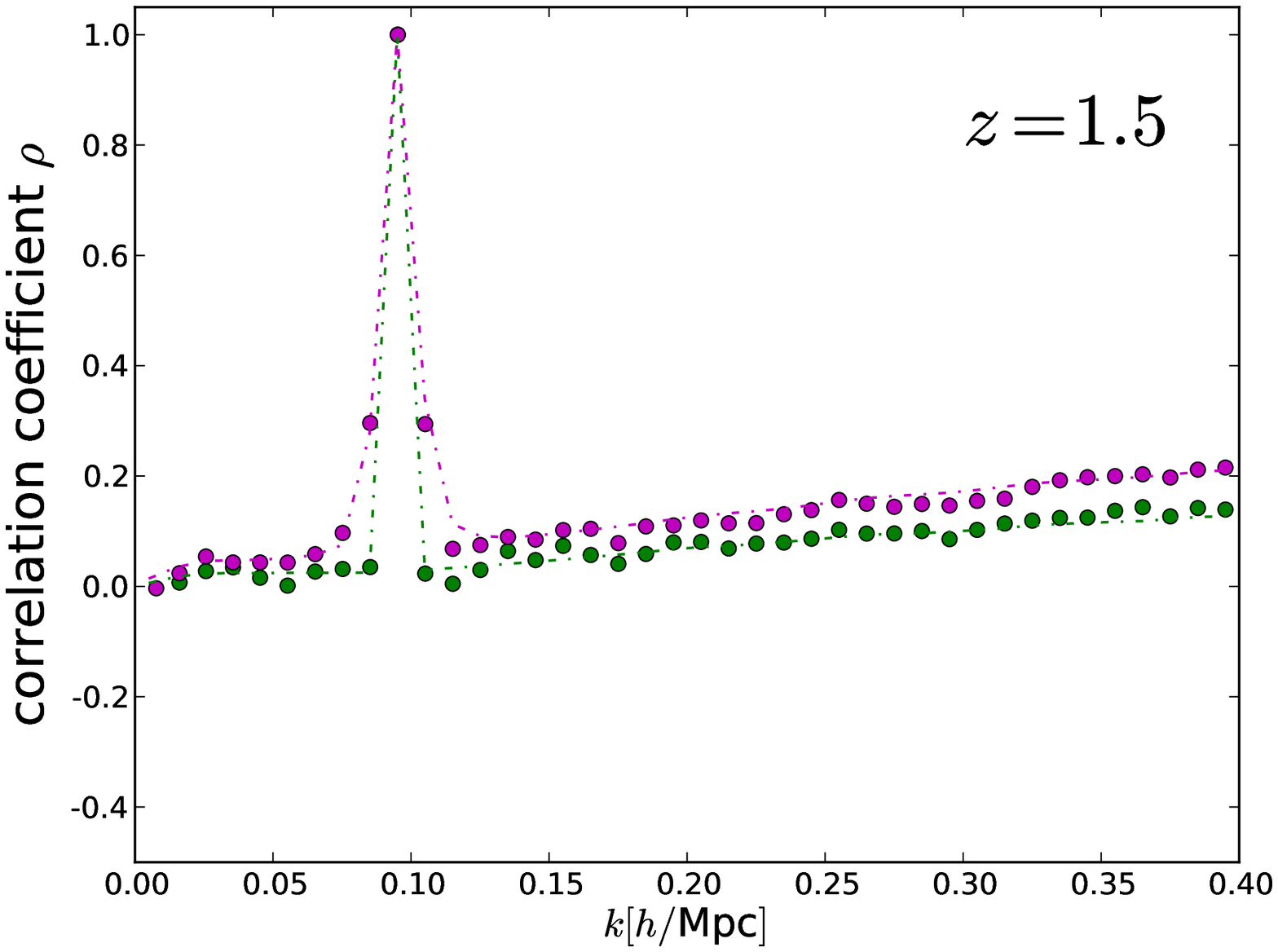}
  \includegraphics[width=0.48\columnwidth]{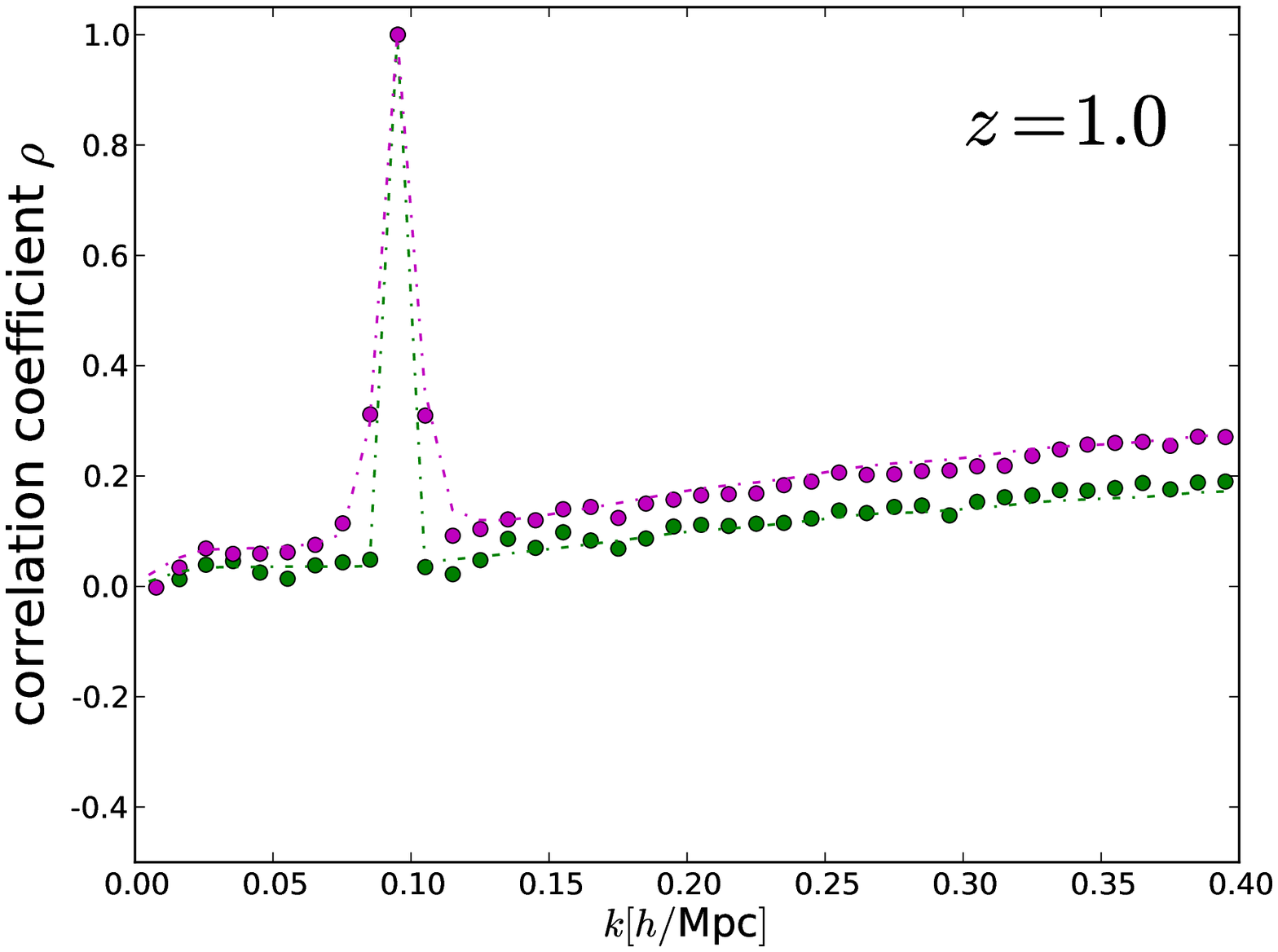}
  \includegraphics[width=0.48\columnwidth]{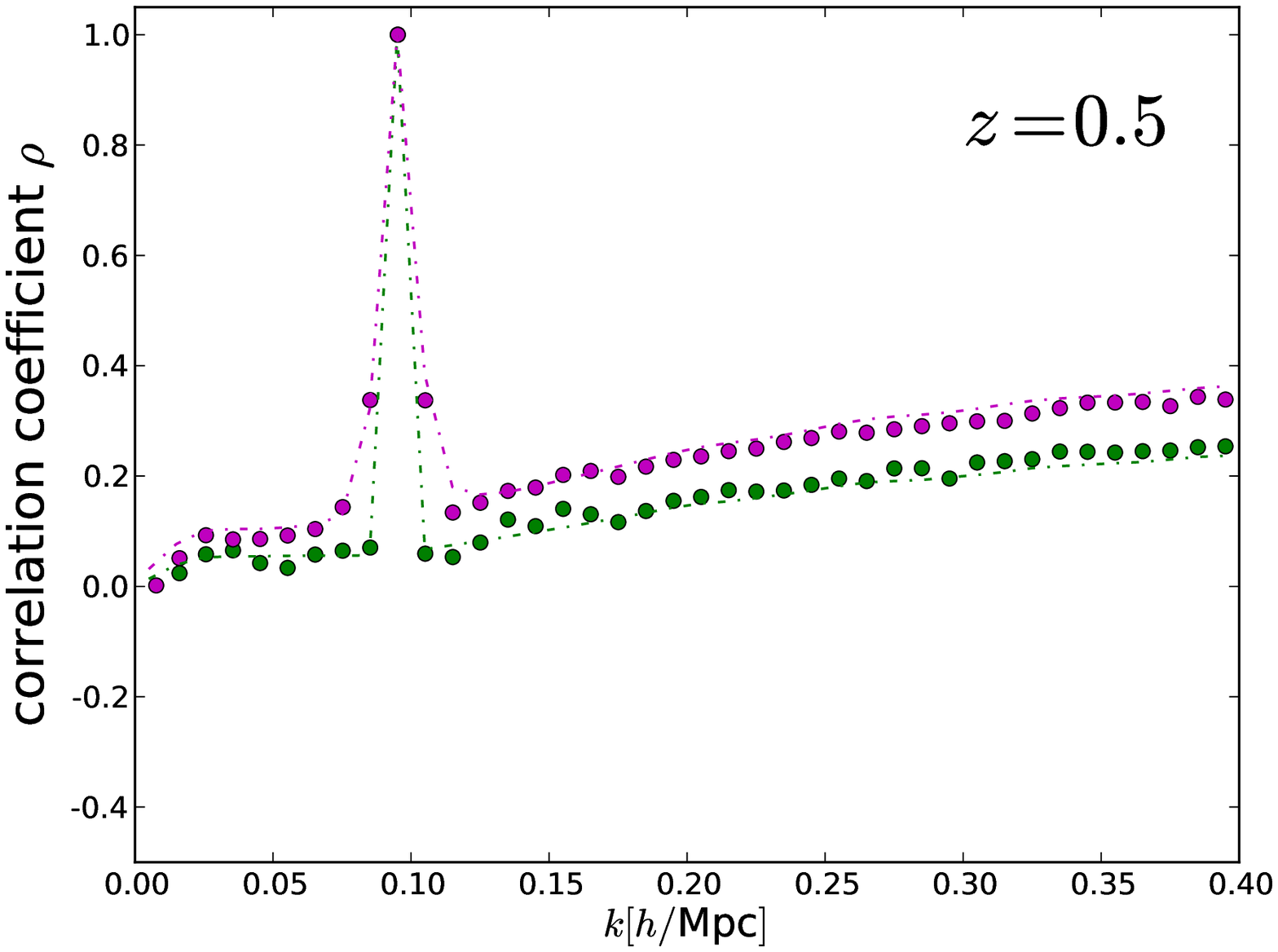}
  \includegraphics[width=0.48\columnwidth]{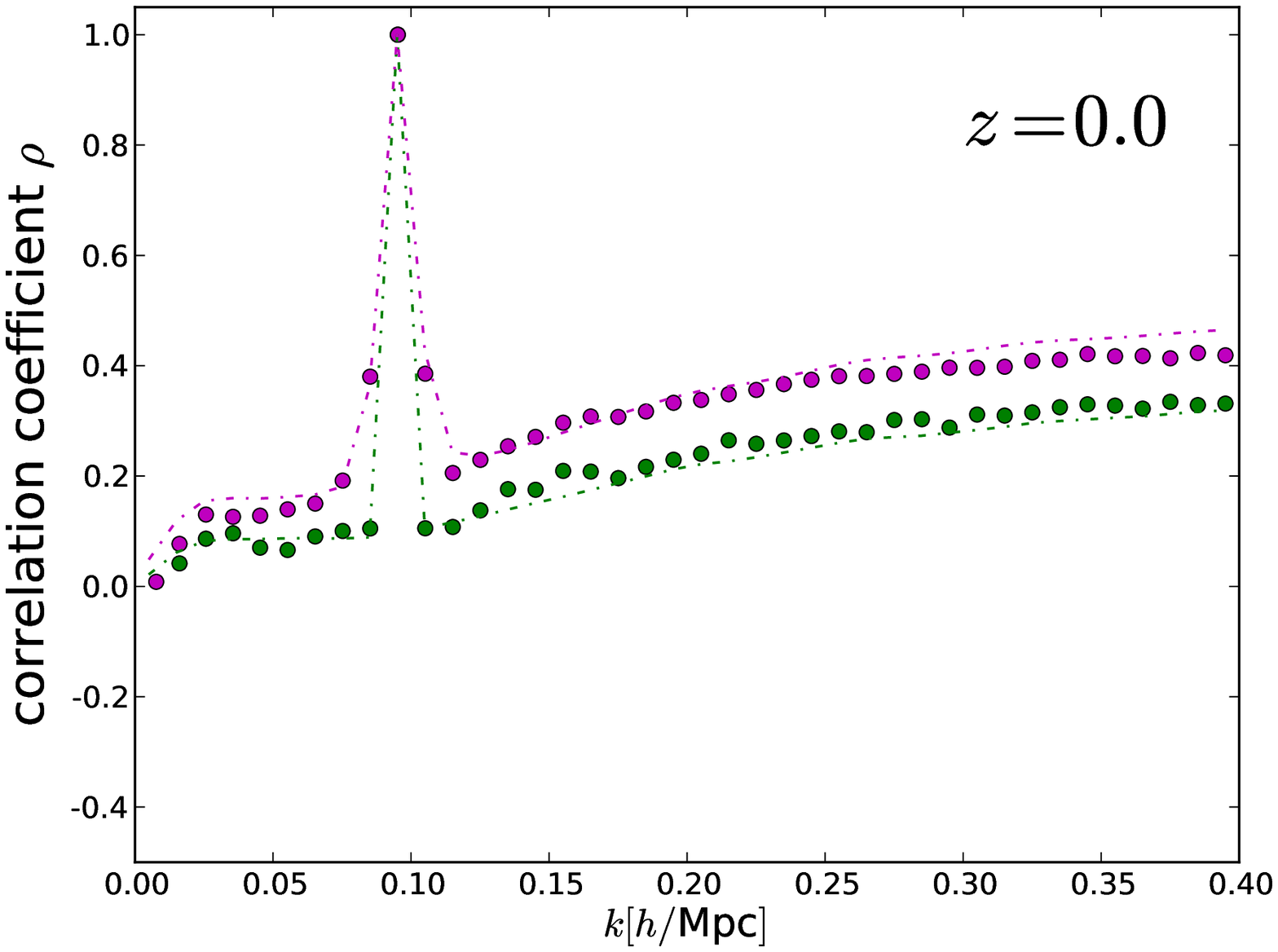}
  }
  \end{center}
  \caption{As Fig.~\ref{fig:corrik4zp}, but correlations relative
  to bin at $k=0.09-0.1 h$Mpc$^{-1}$.
  }
  \label{fig:corrik9zp}
\end{figure*}

\begin{figure*}
  \begin{center}{
  \includegraphics[width=0.48\columnwidth]{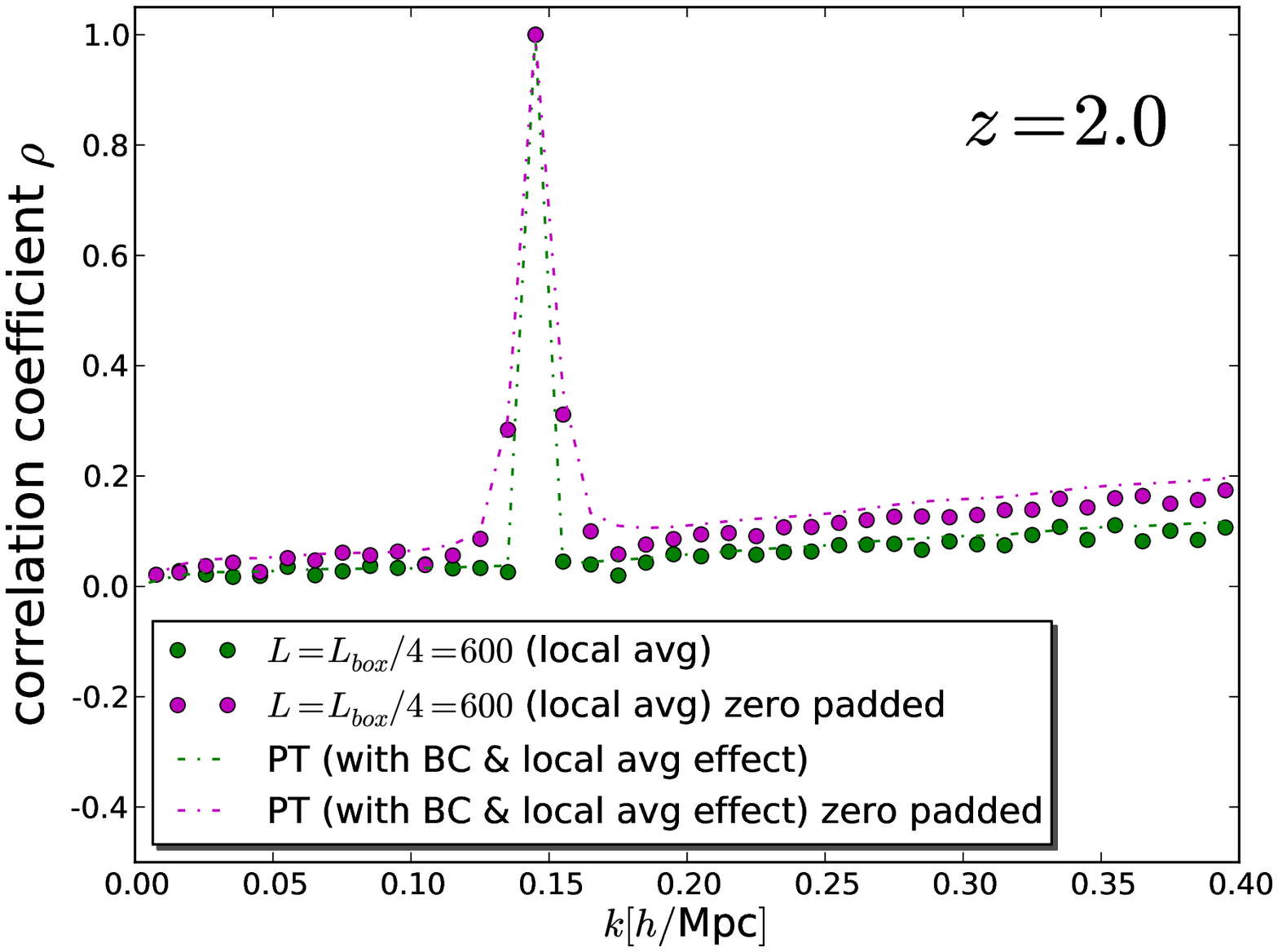}
  \includegraphics[width=0.48\columnwidth]{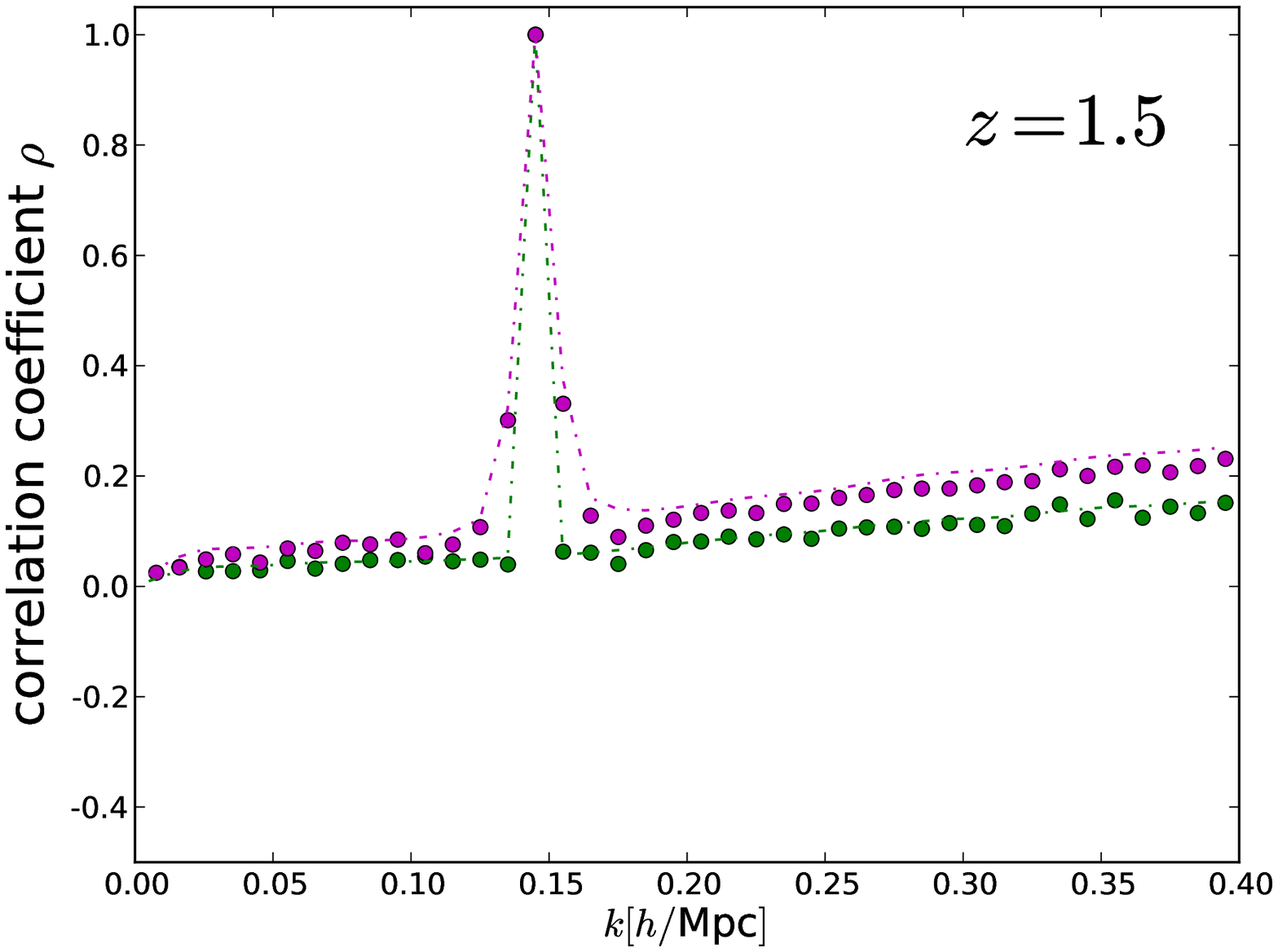}
  \includegraphics[width=0.48\columnwidth]{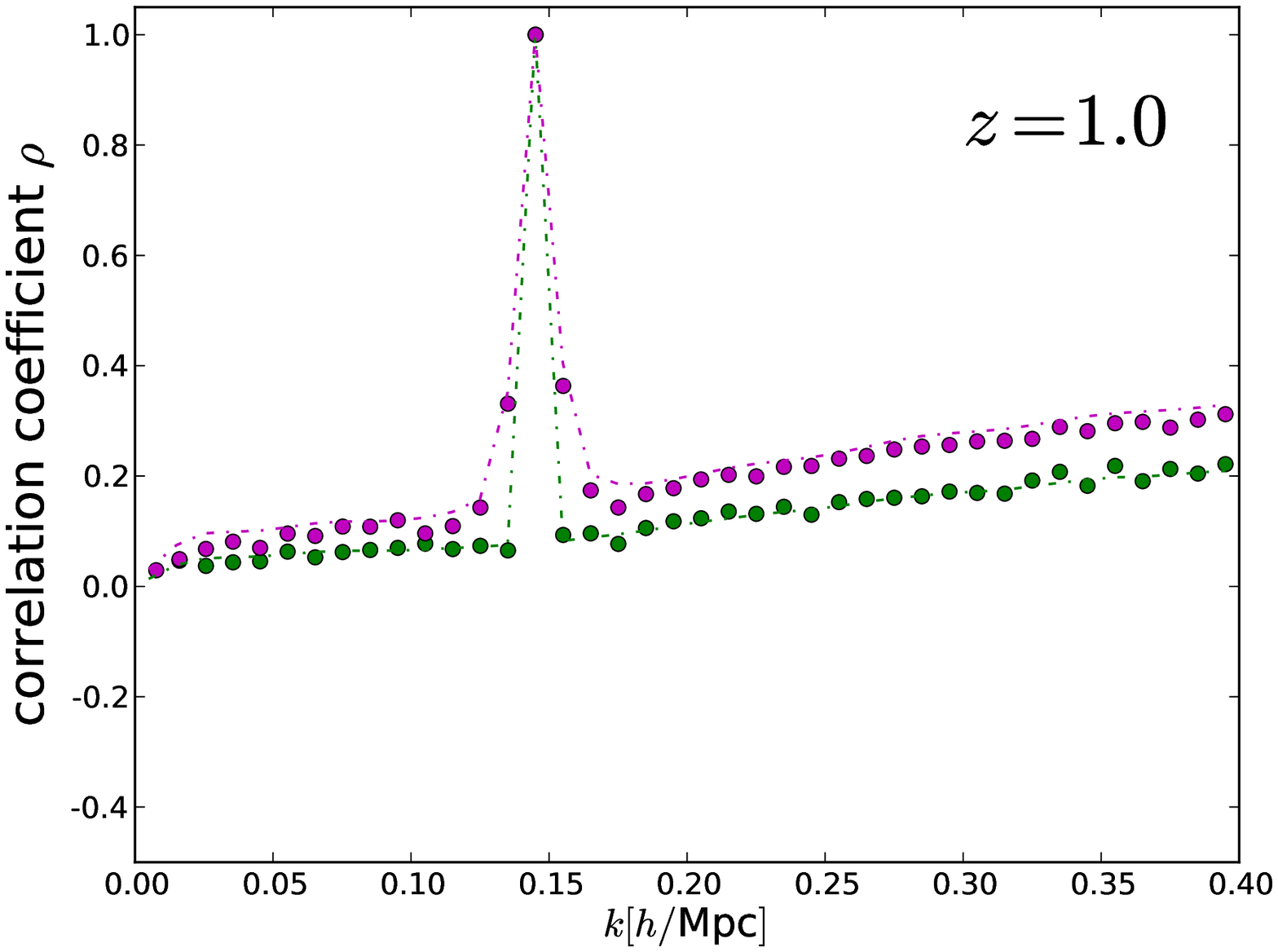}
  \includegraphics[width=0.48\columnwidth]{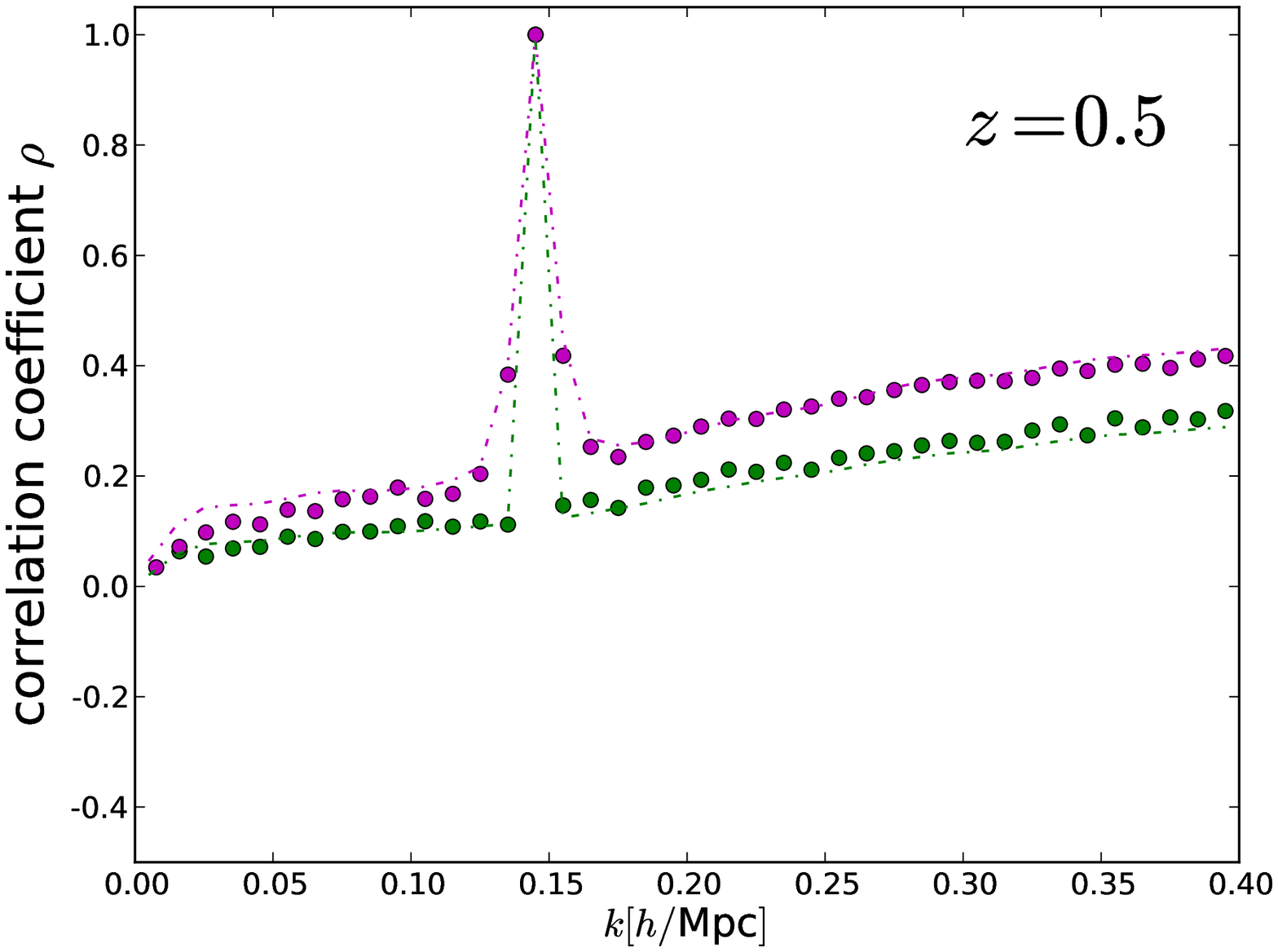}
  \includegraphics[width=0.48\columnwidth]{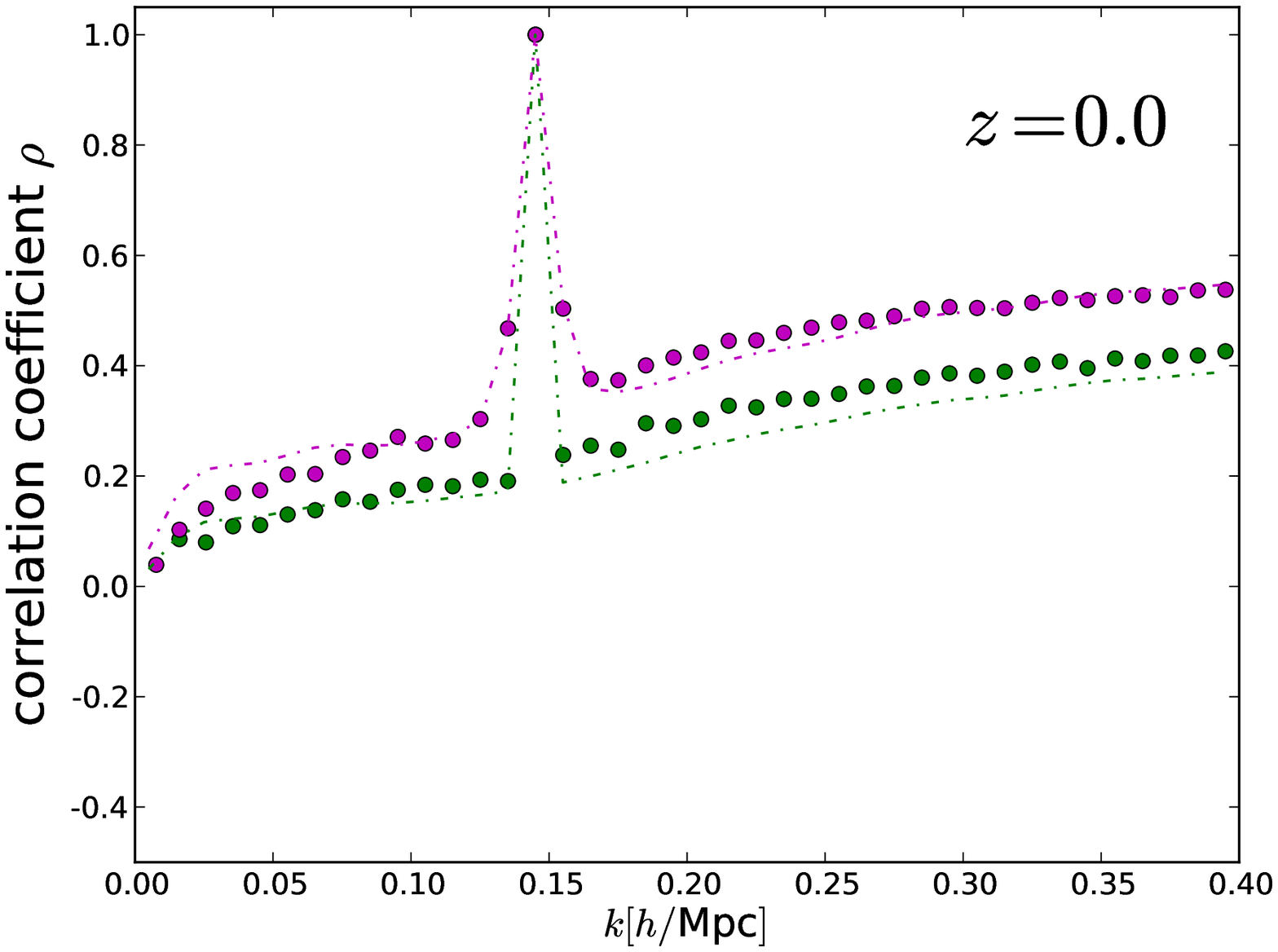}
  }
  \end{center}
  \caption{As Fig.~\ref{fig:corrik4zp}, but correlations relative
  to bin at $k=0.14-0.15 h$Mpc$^{-1}$.
  }
  \label{fig:corrik14zp}
\end{figure*}

Assuming that embedding our $L_{\rm sub} = 600 h^{-1}$Mpc cubed box in a $L = 2400 h^{-1}$Mpc cubed periodic box is a good approximation
to embedding it in an infinite universe with a continuum of Fourier modes, we can apply the above to our simulated scenario and find that
for our bin width $\Delta k = 0.01 h$Mpc$^{-1}$, there should be significant correlations between bins and the mode-counting argument is
not valid, in fact overpredicting the variances. We show this in figures \ref{fig:varszp} - \ref{fig:corrik14zp}, dashed lines, which we will
discuss in more detail below.

Why then was it correct to use the diagonal, mode-counting expression in the previous sections? The reason is that,
in keeping with our effective periodic box description, our bin average only includes modes which are multiples
of the {\it subbox}'s fundamental mode, i.e.~$\k = 2 \pi/L_{\rm sub} \, {\bf n}$ with the components of ${\bf n}$
integers,
\beq
\label{eq:sparse avg}
\hat{P}_i = \frac{1}{N_i} \sum_{\k \in i} \hat{P}(\k).
\eeq
This decorrelates the binned power spectrum estimators as the mode mixing kernel
\beq
Q(\k) = j_0\left( L_{\rm sub} k_x/2 \right) \, j_0\left( L_{\rm sub} k_y/2 \right) \, j_0\left( L_{\rm sub} k_z/2 \right)
\eeq
vanishes for separations that are a non-zero multiple of $2 \pi/L_{\rm sub}$. Since the kernel equals unity for zero
separation, the covariance in this averaging scheme is given by the mode-mixing result,
\beq
{\bf C}_{ij} = 2 \frac{1}{N_i} \frac{1}{N_j} \sum_{\k \in i} \sum_{\kp \in j} \, P^2(k) \, |Q(\k - \kp)|^2
= \delta^K_{ij}\, 2 \frac{1}{N_i} \frac{1}{N_j} \sum_{\k \in i}  \, P^2(k) = \frac{2 P^2(k_i)}{N_i} \, \delta^K_{ij}.
\eeq

This begs the question if a similar binning scheme
can be applied to an actual survey, in order to decorrelate power spectrum estimates in different bins.
Unfortunately, for a realistic survey, the geometry will be much more complicated than a simple cube,
making this rather difficult.
In practice, therefore, one would typically embed the survey in a much larger cubic volume (zero-padding the part not covered
by the survey),
apply a Fourier transform, estimate the power spectrum for each $\k$ on the grid by the embedding box (i.e.~modulo $2\pi/L$ with
$L$ the size of the large box), and finally average to obtain the binned spectrum. This is thus a much denser sampling than
the one we applied previously and approaches the infinitely dense FKP bin average of Eq.~(\ref{eq:fkpbin}). In a more realistic scenario therefore,
mode counting would not be sufficient and the off diagonal covariances are significant.

For this reason, it is useful to numerically test
the mode mixing in the covariance matrix due to the window function as given in Eq.~(\ref{eq:fkpgen}). We do this by
again calculating the covariance matrix for our $L_{\rm sub} = 600 h^{-1}$Mpc subbox, but this time using a
binned spectrum averaged over all multiples of the fundamental mode of the large $L = 2400 h^{-1}$Mpc simulation
box, thus increasing the sampling density by a factor of $4^3=64$ and approaching the FKP bin average. In practice,
we zero-pad the exterior of the subbox and then apply the Fourier transform to the full box.
The results are shown in figures \ref{fig:varszp} - \ref{fig:corrik14zp} and compared to the analytic expression (\ref{eq:fkpgen}). To model the
inevitable non-linear effects, we add trispectrum contributions to the theory prediction, which we will describe in the next
subsection. Even without these contributions however, we can already see from the linear regime that the FKP expression
works very well and that indeed with the more realistic averaging scheme, the off diagonal covariances are considerable and the variances
are reduced accordingly.

To conclude this subsection, we briefly compare our beat coupling investigation to
that of \cite{takahashietal09}.
They too study the beat coupling effect by considering a subbox of a larger simulation volume (although in their case
the subbox is only a factor $2^3=8$ smaller than that of the simulation, $L_{\rm sub}=500 h^{-3}$Mpc$^3$). They use two approaches for estimating the power spectrum.
Their ``zero-padding'' treatment corresponds to the FKP-like method described in the preceding paragraphs, while the other
approach is equivalent to the one we used consistently before this section. While they do not comment on the origin of the observed differences,
it is reassuring that the simulation results in their Fig.(10) agree with our explanation and with our own simulation results.
Interestingly, in \cite{takahashietal09}, the excess covariance due to large modes is significantly smaller than
expected based on beat coupling only. However, they use the local (subbox) average
to calculate the overdensity. The lower variance can thus be explained very well by the local average affect discussed in
this article.

\subsection{General Survey Geometry: Perturbation Theory}
\label{subsec:pt}

The discussion in the previous section covers the linear regime where the underlying density field
is to a good approximation Gaussian. Including non-linear effects, it still describes the disconnected
contribution to the covariance matrix provided that the non-linear power spectrum is used. However, additional terms
are now needed as the covariance matrix receives contributions from the connected part of the four-point function.
These terms were already included in the figures \ref{fig:varszp} - \ref{fig:corrik14zp} (and were
discussed in a simpler framework in section \ref{sec:form}). Here we discuss the details of
the inclusion of the trispectrum terms for arbitrary window function.

Including these terms, the covariance between single-mode estimators now becomes
\beqa
\label{eq:fkpnonlin}
\langle \delta \hat{P}(\k) \, \delta \hat{P}(\kp) \rangle &=& |P(k) \, Q(\k - \kp)|^2 + (\kp \to -\kp) \nonumber \\
&+& \frac{1}{(2\pi)^9}\,\int d^3 {\bf q}_1 d^3 {\bf q}_2 d^3 {\bf q}_3 d^3 {\bf q}_4 \,
\delta^D({\bf q}_1-{\bf q}_2+{\bf q}_3-{\bf q}_4)\,G({\bf q}_1) G^*({\bf q}_2) G({\bf q}_3) G^*({\bf q}_4) \nonumber \\
&\times& \, T(\k - {\bf q}_1, -(\k - {\bf q}_2), \kp - {\bf q}_3,-(\kp - {\bf q}_4)).
\eeqa
Note that this is the generalization of Eq.~(\ref{eq:covindiv}), which
we used to describe the covariance in a cubic box.
To leading order in perturbation theory, the trispectrum
is given by\footnote{Unlike in section \ref{sec:form}, but consistent
with perturbation theory, we now immediately express
perturbation theory quantities in terms of the non-linear power spectrum $P$ and not $P^{\rm lin}$.} (see \cite{scoccetal99})
\beq
T({\bf k}_1, {\bf k}_2, {\bf k}_3, {\bf k}_4) = 4 \left[ F_2({\bf k}_{12}, -\k_1) \, F_2({\bf k}_{12}, -\k_3) \, P(k_1) \, P(k_{12})\, P(k_3)
+ {\rm perm.} \right] + 6 \left[F_3(\k_1, \k_2, \k_3) \, P(k_1)\, P(k_2) \,P(k_3) + {\rm cyc.}\right],
\eeq
with 12 distinct permutations for the first term, 3 cyclic permutations for the second,
and where $\k_{ij} \equiv \k_i + \k_j$.
Inserting the arguments from Eq.~(\ref{eq:fkpnonlin}) into this perturbation theory expression,
and omitting the ${\bf q}_i$ dependence whenever ${\bf q}_i$ appears in an argument as a correction to a larger vector of
order $k$, $k'$, $|\k + \kp|$, etc,
the trispectrum contribution becomes
\beqa
&T&(\k - {\bf q}_1, -(\k - {\bf q}_2), \kp - {\bf q}_3,-(\kp - {\bf q}_4)) =
\left[ 4 \, P(|\k + \kp|)\, \left( F_2(\k + \kp, -\k) \,P(k) + (\k \to \kp) \right)^2 + (\kp \to -\kp) \right] \nonumber \\
&+& 4 \, P(k) \, P(k') \, P(|{\bf q}_2 - {\bf q}_1|) \, \left( F_2({\bf q}_2 - {\bf q}_1, \k) + F_2({\bf q}_2 - {\bf q}_1, -\k) \right)
\, \left( F_2({\bf q}_2 - {\bf q}_1, \kp) + F_2({\bf q}_2 - {\bf q}_1, -\kp) \right) \nonumber \\
&+& 12 \, P(k) \, P(k') \, \left[F_3(\k,-\k,\kp) \, P(k) + (\k \leftrightarrow \kp) \right]
\eeqa
where we have used that the trispectrum arguments
sum up to zero. The terms that are independent of the ${\bf q}_i$'s can be pulled out of the
integral in Eq.~(\ref{eq:fkpnonlin}) so that the integral over mode mixing kernels simply
gives a factor of $V_{\rm eff}^{-1}$. The other terms only depend on the combination ${\bf q}_1 - {\bf q}_2$
so that three of the four Fourier integrals can be carried out analytically and one remains, giving the result
\beqa
\label{eq:fkpnlfull}
\langle \delta \hat{P}(\k) \, \delta \hat{P}(\kp) \rangle &=& |P(k) \, Q(\k - \kp)|^2 + (\kp \to -\kp) \nonumber \\
&+& \frac{1}{V_{\rm eff}}  \left[ 4 \, P(|\k + \kp|)\, \left( F_2(\k + \kp, -\k) \,P(k) + (\k \leftrightarrow \kp) \right)^2 + (\kp \to -\kp) \right] \nonumber \\
&+& \frac{1}{V_{\rm eff}} 12 \, P(k) \, P(k') \, \left[F_3(\k,-\k,\kp) \, P(k) + (\k \leftrightarrow \kp) \right] \nonumber \\
&+& \frac{1}{V_{\rm eff}} 4 \, P(k) \, P(k') \, \left(\int \frac{d^3 {\bf u}}{(2\pi)^3}\, |Q|^2({\bf u})\right)^{-1}\, \nonumber \\
&\times& \int \frac{d^3 {\bf u}}{(2\pi)^3}\,
|Q|^2({\bf u}) P(u) \, \left( F_2({\bf u}, \k) + F_2({\bf u}, -\k) \right)
\, \left( F_2({\bf u}, \kp) + F_2({\bf u}, -\kp) \right)\, ).
\eeqa

The final step towards a covariance matrix is to apply the bin average to Eq.~(\ref{eq:fkpnlfull}), but for the trispectrum
terms
the subtle, averaging related effects described in the previous subsection are not important because the correlations from the trispectrum
are not as narrow as those for the disconnected (or Gaussian) terms. The angle averaging can be done analytically for the last term
(the beat coupling term) and has to
be carried out numerically for the remaining trispectrum contributions. The final result is
\beqa
{\bf C}_{ij} &=& 2 \int \frac{d^3 \k}{V_{k,i}} \, \int \frac{d^3 \kp}{V_{k,j}} \, P^2(k) \, |Q(\k - \kp)|^2 \nonumber \\
&+& \frac{1}{V_{\rm eff}} \int \frac{d^3 \k}{V_{k,i}} \, \int \frac{d^3 \k}{V_{k,i}} \, T^0(\k, -\k, \kp, -\kp) \nonumber \\
&+& \frac{1}{V_{\rm eff}} \, 16 \, \left( \frac{17}{21}\right)^2 \,
\left(\int \frac{d^3 {\bf u}}{(2\pi)^3}\, |Q|^2({\bf u})\right)^{-1}\, \int \frac{d^3 {\bf u}}{(2\pi)^3}\,
|Q|^2({\bf u}) P(u),
\eeqa
where $T^0(\k, -\k, \kp, -\kp)$ represents the second and third lines of Eq.~(\ref{eq:fkpnlfull}).
The last term is the beat coupling term, which can now explicitly be seen to be proportional
to a weighted average of the power spectrum over large modes. The integral describing this average
can be compared to that for the expectation value of the zero-mode power spectrum, Eq.~(\ref{eq:pkexp}).
The only difference is that the latter is given in terms of $G$ while the former is given in terms of
$Q$ (note that the normalization integral appearing in the beat coupling term is equal to one when $Q$
is replaced by $G$). However, for a cubic box (with $\bar{n} \equiv {\rm const}, w \equiv 1$ inside the box), the two quantities
are exactly equal and the average appearing in the beat coupling expression above can be replaced by the zero-mode 
power spectrum, thus justifying our use of $P^{\rm lin}_0$ in Eq.~(\ref{eq:cov}) in section \ref{subsec:PT}.

We will not rederive the correction due to the local average effect for the case of arbitrary geometry,
but in analogy with the cubic subbox case, we will use the following expression for Case 3 in
Figs \ref{fig:varszp}-\ref{fig:corrik14zp}:
\beqa
\label{eq:c3 geom}
{\bf C}_{ij} &=& 2 \int \frac{d^3 \k}{V_{k,i}} \, \int \frac{d^3 \kp}{V_{k,j}} \, P^2(k) \, |Q(\k - \kp)|^2 \nonumber \\
&+& \frac{1}{V_{\rm eff}} \int \frac{d^3 \k}{V_{k,i}} \, \int \frac{d^3 \k}{V_{k,i}} \, T^0(\k, -\k, \kp, -\kp) \nonumber \\
&+& \frac{1}{V_{\rm eff}} \, \frac{676}{441} \,
\left(\int \frac{d^3 {\bf u}}{(2\pi)^3}\, |Q|^2({\bf u})\right)^{-1}\, \int \frac{d^3 {\bf u}}{(2\pi)^3}\,
|Q|^2({\bf u}) P(u).
\eeqa
Note that for a varying background density, the estimate of the background density $\bar{\rho}({\bf x})$ used in
$\delta({\bf x}) = (\rho({\bf x}) - \bar{\rho}({\bf x}))/\bar{\rho}({\bf x})$ is not only affected by the effective
zero mode, but also by smaller modes. These will also affect the covariance matrix, but we will not go into this effect
here, as it is beyond the topic we set out to study.

\section{Summary and Discussion}
\label{sec:disc}

In this work, we have studied the effects of modes larger than the survey
on the (dark matter) power spectrum covariance matrix.
We have built an analytic description that includes the beat coupling effect of \cite{HRS06},
but also the previously overlooked (at least in analytic studies) effect of
large modes on the estimated average density $\bar{\rho}$ which enters the overdensity through $\delta=(\rho-\bar{\rho})/\bar{\rho}$
(the local average effect).
We have confirmed this model and the role of individual contributions by comparing to covariance matrices
obtained from N-body simulations. To study the role of super-survey modes in the simulations, we
estimated the power spectrum from a subvolume embedded in a considerably larger simulation volume.\\

{\bf We summarize our main results below:}
\begin{itemize}
\item
We build a model based on perturbation theory for the matter covariance matrix that includes the effects
of modes larger than the survey. For the variances, we find excellent agreement with simulations
for $k < 0.4 h$Mpc$^{-1}$ (or larger) at $z=2$ and for $k < 0.2 (0.15) h$Mpc$^{-1}$ at $z=0.5 (0)$.
Agreement is even better for the correlation coefficients, with our model predicting the correct
coefficients for at least $k < 0.4 h$Mpc$^{-1}$ at all redshifts.
\item
When isolated, the beat coupling effect from \cite{HRS06,rimesham06}, can indeed be described by the last term in
Eq.~(\ref{eq:c2}), as shown by the blue points and curves of Figs \ref{fig:vars}-\ref{fig:corrik14}.
\item
In a more realistic approach, the {\it local average effect} needs to be taken into account as well.
This has previously been overlooked in analytic studies and we derive its effect using perturbation theory,
leading to Eq.~(\ref{eq:c3}). It reduces the covariance, leaving only $10 \%$ of the original beat coupling excess covariance,
as shown by the green lines and points in Figs \ref{fig:vars}-\ref{fig:corrik14}. This also explains
the disagreement found between the beat coupling-only expression and simulations in \cite{takahashietal09}.
We conclude that the beat coupling excess covariance is not as important as previously thought.
\item
Eq.~(\ref{eq:c3 geom}) gives the final result for
the matter covariance matrix for arbitrary survey geometry,
including not only the above mentioned effects, but also the correlations between neighboring
power spectrum bins and the related reduced variance
due to the survey's window function. It is depicted
in Figs \ref{fig:varszp}-\ref{fig:corrik14zp} and again agrees well with simulations.
It can be used as a first step towards a covariance matrix for the galaxy (or other tracer's) power spectrum.

\end{itemize}

In a real survey of large scale structure, the effects discussed in this paper are relevant
because there are always modes larger than the survey volume.
Our results are thus important for large scale structure surveys and in particular galaxy surveys, 
as we quantify the expected excess covariance due to these modes, as well as
the more standard covariance contributions.
With a complete description of the covariance matrix now available, it is also possible to
study the cosmology dependence of the covariance matrix in an efficient manner. We refer to
the Appendix for a first look into this issue.

To build a complete covariance matrix
for the galaxy power spectrum, one needs to include the effects of shot noise, galaxy bias
and redshift space distortions. While the effect of shot noise is easy to incorporate within the FKP formalism (at least in the limit
where the shot noise can be treated as Gaussian),
the other effects are more complicated and clearly beyond the scope of our paper. However, a very rough first
approximation to a galaxy covariance matrix could be obtained by multiplying the dark matter
matrix by the galaxy bias to the fourth power, including an angle
averaged Kaiser factor (as in \cite{takahashietal09})
to account for redshift space distortions, and using FKP to include the effect of shot noise.

In addition to the complications arising from observing galaxies (or other tracers) instead
of dark matter, one also needs to be able to describe the covariance for a realistic survey geometry
(not many surveys have a cubic footprint). For this reason, we presented in section \ref{sec:gen}
a full description of the model for arbitrary survey geometry, assuming the FKP
estimator is used. By comparing to simulations, we showed that the expressions from
\cite{FKP} (with trispectrum terms added) can be used to accurately describe the
mode mixing due to the survey window function and the resulting
correlations between neighboring power spectrum bins. While section \ref{sec:gen}
does not present any completely new results relative to the previous sections and
what is in the literature, its main purpose is to present a complete set of equations
to describe the matter power spectrum for arbitrary survey geometry. This can then serve
as a stepping stone towards a full galaxy covariance matrix.

\acknowledgments 
We thank Ryuichi Takahashi, Masahiro Takada and Shun Saito
for useful discussion and for sharing their simulation outputs,
which helped motivate our investigation.
We also thank Beth Reid for insightful discussions.
RdP and CW are supported by FP7-IDEAS-Phys.LSS 240117.
OM is supported by AYA2008-03531 and the Consolider Ingenio project CSD2007-00060.
LV acknowledges support of FP7-IDEAS-Phys.LSS 240117.
N-body simulations and calculations on N-body outputs were
done on the cluster of computers {\it Hipatia}
(UB computing facilities and ERC grant FP7- IDEAS Phys.LSS 240117).

\appendix 

\section{Cosmology Dependence of Covariance Matrix}

One major advantage of the analytic expressions presented in this work,
is that they allow for a quick estimate of the covariance matrix, especially
compared to methods using N-body simulations. The ease with which covariance
matrices can be calculated makes the analytic method perfectly suited for studying
the cosmology dependence of the covariance matrix.
A full study of this cosmology dependence would consider how constraints on cosmological parameters
are affected and would quantify the error induced by ignoring the cosmology dependence.
Such an investigation is beyond the scope of this article and we will leave it for future work.
In this appendix, we will simply quantify how much the covariance matrix changes as we vary
individual cosmological parameters.

For simplicity, we again consider the covariance in the matter power spectrum,
as estimated from a cubic volume with $V = L^3 = (600 h^{-1}$Mpc$)^3$. We imagine the
spectrum is ``measured'' at redshift $z = 0.5$ and include the smearing effect due to the window
function that arises when the FKP estimator is used. In other words, we will consider
the covariance matrix
given by Eq.~(\ref{eq:c3 geom}), with the power spectrum evaluated at
$z = 0.5$.

In practice, galaxy surveys typically measure the power spectrum
relative to a fixed, fiducial background cosmology, see e.g.~\cite{Tegetal2006}.
Therefore, when a Monte Carlo chain is run, the theoretical power spectrum
at each point in parameter space needs to be rescaled to account for the effect of
using the fiducial background cosmology as opposed to the actual background cosmology
at that point in parameter space. Only after this rescaling can it be compared to the observed spectrum.
For consistency, we therefore also rescale the covariance
to the fiducial background cosmology,
\beq
{\bf C}_f(\Delta k_i, \Delta k_j) =  A^{-6} \, {\bf C}(\Delta k_i/A, \Delta k_j/A)|_{V = V_f \, A^3},
\eeq
where ${\bf C}_f$ is the covariance matrix of the power spectrum estimator relative to the fiducial background cosmology,
${\bf C}$ the true covariance matrix
given by Eq.~(\ref{eq:c3 geom}) (evaluated at the actual volume $V=V_f \, A^3$, with $V_f$ the survey volume
calculated in the fiducial cosmology),
$\Delta k_i$ the bin widths, and $A \equiv d_V(z)/d_V^f(z)$ the dilation factor (see \cite{Tegetal2006}).

Since the different terms in the covariance matrix scale either like the second or third power of the power spectrum,
we expect a significant dependence on $\sigma_8$ and other parameters
changing the overall normalization of the spectrum. The beat coupling/local average term
is particularly sensitive to the power at $k \lesssim 2\pi/L$ so that it should depend
on any parameter
that affects the power on very large scales.
To make this more quantitative, we consider the $\Lambda$CDM fiducial cosmology described in section
\ref{sec:nbody} and change each parameter $p$ by a step $\Delta p$, one by one, keeping the remaining parameters
fixed.
For the step sizes, we choose $\Delta p = 2 \, \sigma_p$, where $\sigma_p$
is the parameter uncertainty from
\cite{reidetal10} (Table 3), obtained from combining the SDSS-II Data Release 7 \cite{DR7} halo power spectrum
with WMAP5 \cite{wmap5} cosmic microwave background (CMB) data.
These step sizes are indicative of
the relevant parameter range in a Markov Chain Monte Carlo (MCMC) analysis with power spectrum data.
We show the fiducial parameters and their step sizes in Table \ref{tab:steps}.

\begin{table*}[hbt!]
\begin{center}
\small
\begin{tabular}{c|ccccc}
\hline\hline
 & $\omega_b$ & $\omega_m$ & $\Omega_m$ & $\sigma_8$ & $n_s$\\
\hline\hline
fiducial & 0.023 & 0.1323 & 0.27 & 0.79 & 0.96\\
\hline
step size & 0.00116 & 0.008 & 0.038 & 0.05 & 0.026 \\
\hline\hline
\end{tabular}
\caption{Fiducial values and step sizes used to test cosmology dependence of matter covariance matrix.
}
\label{tab:steps}
\end{center}
\end{table*}

\begin{figure*}
  \begin{center}{
  \includegraphics[width=0.48\columnwidth]{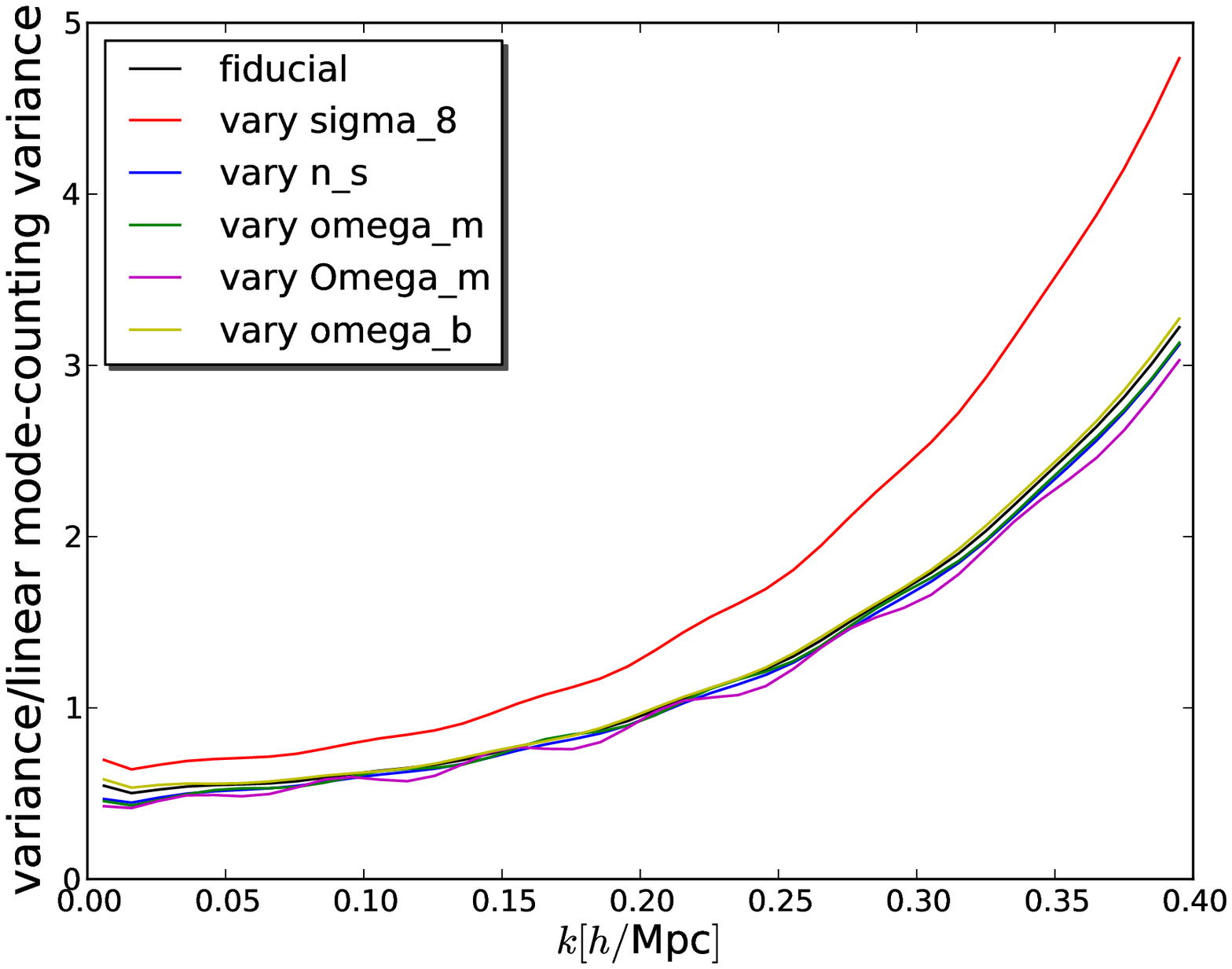}
  \includegraphics[width=0.48\columnwidth]{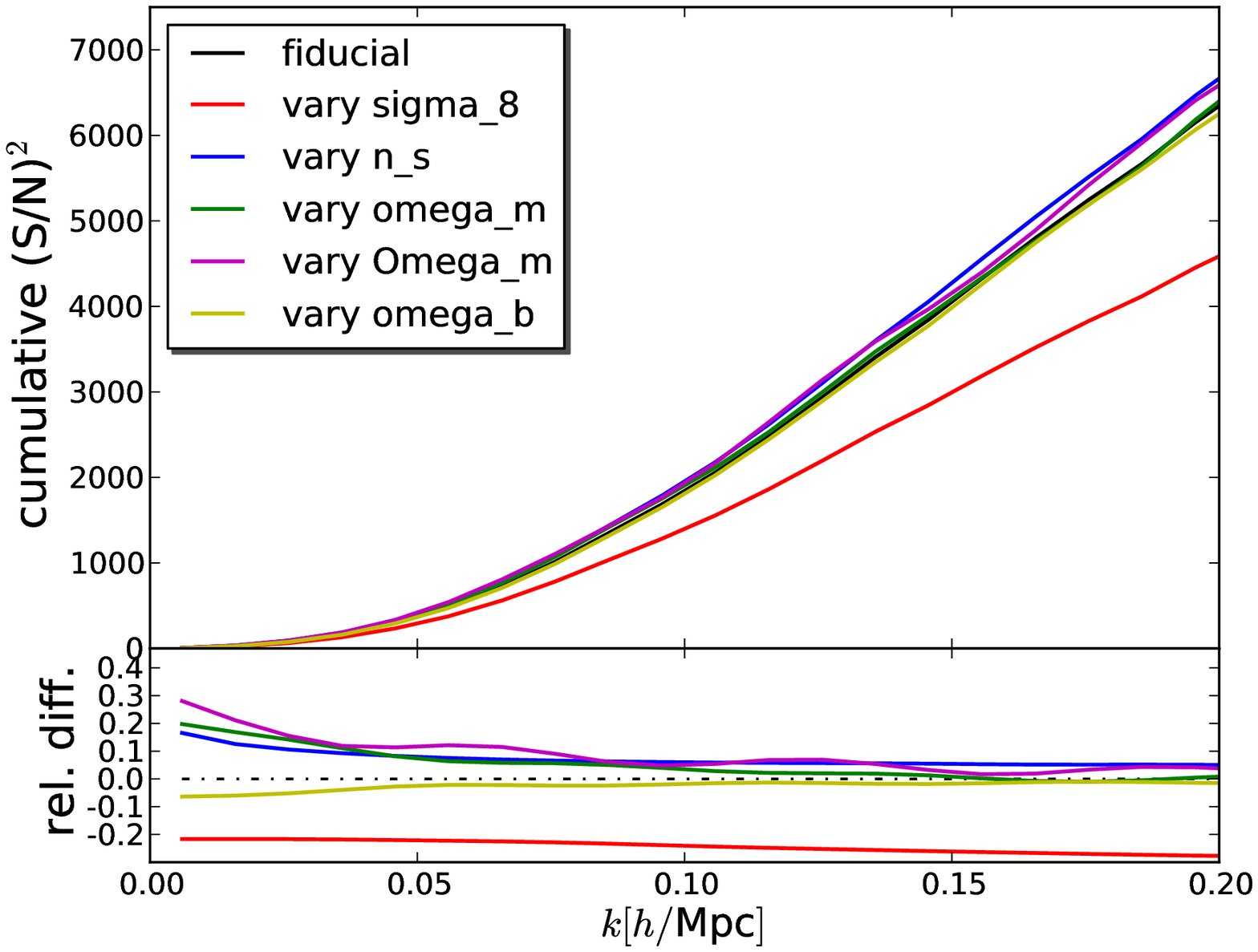}
  }
  \end{center}
  \caption{{\it Left Panel:} Variance relative to linear, mode counting based variance
  as a function of cosmology. The black curve shows the fiducial cosmology, and the other curves
  depict the effect of varying each $\Lambda$CDM parameter by approximately twice its error bar
  expected from a large scale structure plus CMB measurement (see text, and Table \ref{tab:steps}).
  {\it Right Panel:} The signal-to-noise squared in the detection/amplitude of the power spectrum as a function
  of the largest included mode $k$ (main figure). The bottom inset shows the relative difference with the
  fiducial cosmology.
  Both panels show that the strongest parameter dependence is on $\sigma_8$ and $\Omega_m$.
  }
  \label{fig:cosmodep}
\end{figure*}

In the left panel of figure \ref{fig:cosmodep}, we show the effect on the variance of
each parameter. As before, we normalize the variance by the variance based on mode counting
and the linear power spectrum (as calculated in the fiducial cosmology). The units on the horizontal
axis are $[h_f/$Mpc$]$, where $h_f = 0.7$ is the dimensionless Hubble parameter in the fiducial model.
The variance in the fiducial model is given by the black curve. As expected, the largest variation
is obtained when varying $\sigma_8$, causing a significant increase in variance.
The effect of the other parameters is much smaller.


The effect of the parameter variations on the correlation coefficients is always below $|\Delta \rho_{i j}| \lesssim 0.05$,
with $\omega_m$ decreasing the coefficients most, and $\sigma_8$ increasing them by the largest amount. Finally, to
incorporate the properties of the full covariance matrix in a single statistic, we calculate
the squared signal-to-noise ratio in the (non-linear) power spectrum amplitude as a function of $k_{\rm max}$,
see Eq.~(\ref{eq:sn2}).
In a more thorough analysis, one could replace the power spectrum
in the expression for $(S/N)^2$ by derivatives with respect to cosmological parameters in order to create a full Fisher matrix.

The right panel of figure \ref{fig:cosmodep} depicts this signal-to-noise squared as a function of $k_{\rm max}$
for the different cosmologies, with the bottom inset showing the relative difference with respect to $(S/N)^2$
in the fiducial model. We focus on the range $k = 0 - 0.2 \, h_f/$Mpc, as this is a more realistic range for
a galaxy survey (due to strong non-linearities and shot noise on smaller scales),
and because we have seen that our approach loses accuracy on smaller scales.
Consistent with the picture arising from the left panel, we see again that $\sigma_8$ has by far the largest effect,
decreasing the signal-to-noise by $\sim 30 \%$ (with only a weak dependence of the relative change on $k_{\rm max}$).
The other parameters have a much more modest effect ($\ls 10 \%$ for $k_{\rm max} \sim 0.1 h/$Mpc).
We conclude that reasonable variations
in cosmic parameters can cause $\sim 30 \%$ changes in the covariance matrix and thus in $\Delta \chi^2$ values
in an MCMC chain.

Finally, we note that, for a galaxy survey, the effect of galaxy bias on the covariance matrix
will be very important as well, and will look similar to the effect of $\sigma_8$ discussed above.

\bibliography{refs}

\end{document}